\documentclass[nofootinbib,reprint,amsmath,amssymb,showkeys,showpacs,aps]{revtex4-1}
\usepackage{graphicx}
\usepackage[T1]{fontenc}
\usepackage{dcolumn}
\usepackage{xcolor,colortbl}
\usepackage{multirow}
\usepackage{stackengine}
\usepackage{tikz-cd}
\usepackage{bm}
\usepackage{mathtools}
\usepackage{tikz}
\usepackage{wrapfig}
\newcommand{\qm}[1]{``#1''}
\usepackage{hyperref}
\hypersetup{colorlinks, linkcolor={black},citecolor={blue},urlcolor={blue}}

\usepackage{accents}
\usepackage{scalerel}
\usepackage{bbm}
\usepackage{tabularx,siunitx}
\usepackage{wasysym}
\usepackage{ mathrsfs }

\newcommand{\tg}[1]{\accentset{\land}{#1}}
\newcommand{\stg}[1]{\accentset{\scaleto{\diamond}{2.5pt}}{#1}}
\begin{document}

\preprint{APS/123-QED}

\title[Extended Geometric Trinity of Gravity]{ Extended Geometric Trinity of Gravity}

\author{Salvatore Capozziello$^{1,2,3}$}\email{capozziello@unina.it}
\author{Sara Cesare$^{2,3}$}\email{s.cesare@ssmeridionale.it}
\author{Carmen Ferrara$^{2,3}$}\email{carmen.ferrara-ssm@unina.it}
\affiliation{$^1$ Dipartimento di Fisica "E. Pancini", 
Universit\`a degli Studi di Napoli ``Federico II'', Via Cinthia Edificio 6, 80126 Napoli, Italy\\
$^2$ Scuola Superiore Meridionale, Largo San Marcellino 10, I-80138 Napoli, Italy,\\
$^3$ Istituto Nazionale di Fisica Nucleare, Sezione di Napoli, Complesso Universitario di Monte S. Angelo, Via Cinthia Edificio 6, I-80126 Napoli, Italy}

\begin{abstract}
Extensions of equivalent representations of gravity are discussed in the metric-affine framework. First, we focus on: (i) General Relativity, based upon the metric tensor whose dynamics is given by the Ricci curvature scalar  $R$; (ii) the Teleparallel Equivalent of General Relativity, based on tetrads and spin connection  whose dynamics is given by  the  torsion scalar  $T$; (iii) the Symmetric Teleparallel Equivalent of General Relativity, formulated with respect to both the metric tensor and the affine connection and characterized by the non-metric scalar $Q$ with the role of  gravitational  field. They represent the so-called Geometric Trinity of Gravity, because, even if based on different frameworks and different dynamical variables, such as curvature, torsion, and non-metricity, they express the same gravitational dynamics. Starting from this framework, we construct their extensions with the aim to study possible equivalence.  We discuss the straightforward  extension of General Relativity, the  $f(R)$ gravity, where $f(R)$ is an arbitrary function of the Ricci scalar. With this paradigm in mind, we take into account $f(T)$ and $f(Q)$ extensions showing that they are not equivalent to $f(R)$. Dynamical equivalence is achieved if boundary terms  are considered, that is  $f(T-\tilde{B})$ and $f(Q-B)$ theories. The latter are the extensions of  Teleparallel Equivalent of General Relativity and Symmetric Teleparallel of General Relativity, respectively.  We obtain that $f(R)$, $f(T-\tilde{B})$, and $f(Q-B)$ form the Extended Geometric Trinity of Gravity. The aim is to show that also if dynamics are equivalent, foundations of theories of gravity can be very different.
\end{abstract}

\maketitle

\tableofcontents{}

\section{Introduction}\label{sec:Intro}

General Relativity (GR) is a fundamental theory from several points of view \cite{Wald, Straumann, CDL, FC}. First, it explains  a wide class of   phenomena that cannot be properly described in the framework of Newtonian Gravity (NG) as the  Mercury precession and the deflection of light rays by gravitational sources \cite{Will1993, Ni2016, DeMarchi2020}. Then, it makes  predictions on   phenomena unexpected in classical physics like  gravitational waves \cite{Abbott2016, Abbott2022, Sathyaprakash2009, Bailes2021, LIGOScientific2021psn} and  black holes \cite{EHC20191, EHC20192, EHC20193, EHC20194, EHC20195, EHC20196}.

It is also worth mentioning that GR has led to significant changes in the conceptual basis of NG. In particular, the idea of  absolute and fixed space and time, which are also incompatible with the \qm{Mach Principle} \cite{Mach},
has been  superseded by the idea that spacetime can be a dynamical quantity described by field equations.
Nevertheless, NG successfully predicts various gravitational observations, related both to terrestrial experiments and to planetary motion \cite{Will1993}.
This implies that NG cannot be entirely abandoned in favor of GR but it has to be correctly reproduced by the latter in regimes in which its validity is experimentally proved (i.e. in the Weak Field Limit). The fact that GR indeed reduces to NG in the Weak Field Limit is once again a positive trait of GR itself, as it does not entirely overthrow a consistent and successful theory as NG.
\\
However, Einstein's theory of gravity is far from being unattached by drawbacks and shortcomings on both a theoretical and an experimental level. In fact, it is a non-renormalizable theory, it does not explain the nature of dark matter and dark energy, and it leaves the cosmological constant problems, i.e. the magnitude and the coincidence problems, unsolved \cite{Capozziello:2011, oneloop1974, Obukhov2017}. 
\\
For these reasons, in the last century there have been many attempts to modify and/or extend the GR framework, relaxing the Riemaniann constraints and giving rise to the so-called Metric-Affine theories of gravity \cite{Kible1961, Sciama1962, Hehl1995, JHK,ComparingEG}. In this framework, the affine connection, together with the metric tensor, acquires the role of fundamental field and the former is no longer chosen \textit{a-priori} to be the Levi-Civita connection, as in the GR case. As a result, several theories have been investigated, such as the \emph{Extended} and the \emph{Alternative Theories of Gravity} (see e.g., \cite{Clifton2006,Capozziello2008,Sergei,Vasilis, Nojiri,Sotiriou2010,FC,CDL,Cai2016}). Clearly, the main problem is related to the effective foundation of gravitational interaction which is matter of intense debate in the scientific community. See e.g. Ref. \cite{Mancini:2025asp} and references therein.\\
The transition to the Extended Theories of Gravity (ETGs) can be done by correcting and extending Einstein's theory of gravity through the addition of:
\begin{itemize}
	\item higher order terms in the curvature invariants, as $R^2$, $R_{\alpha\beta\mu\nu}R^{\alpha\beta\mu\nu}$, $R_{\mu\nu}R^{\mu\nu}$, $R\thinspace\Box^{n}R$, etc.;
	\item terms containing either minimally or non-minimally coupled scalar fields
\end{itemize}
in the Einstein-Hilbert action
\begin{equation}\label{eq: 3.1}
		S_{\text{EH}}=\int d\Omega \sqrt{-g} R\nonumber.
\end{equation}
The importance of such a transition lies in the recovery of Mach's principle (e.g. in the Brans-Dicke theory), which can be considered as referred to the assumption of a varying gravitational coupling constant, and with the possibility of solving Einstein's theory flaws both at infrared and ultraviolet scales.
Finally, the addition of higher order terms in the curvature invariants can be considered as a reasonable choice in black hole physics due to the strong gravity regime or, equivalently, to the high curvature in the region nearby the considered black hole.

Taking the path of Metric Affine Geometries (MAGs), spacetime is described by a four-dimensional differentiable manifold $M$, a symmetric metric $g_{\mu \nu}$, and a linear affine connection $\Gamma^{\lambda}{}_{\mu \nu}$: $\mathcal{M} = (M, g_{\mu\nu}, \Gamma^{\lambda}{}_{\mu \nu})$.  The metric encodes distances and angles, while the connection independently defines parallel transport and covariant derivatives. Moreover, the linear affine connection leads to the definition of two additional tensors, which, along with the curvature tensor, represent the dynamical variables of MAGs: the torsion and the non-metricity tensor. It is also possible to restrict the generic MAG to one of the following subclasses, fixing the affine connection:

\begin{itemize}
    \item the \emph{Riemannian geometry} represents the Einstein scaffold for its theory and the gravitational effects are encoded in the curvature tensor \cite{Misner1973,Romano2019}.
    \item the \emph{Teleparallel geometries} are based on a trivial curvature and on the concept of \emph{Fernparallelismus} or \emph{parallelism at distance}, since the parallel transport of vectors becomes independent of the path \cite{ATG, Martin2019}. 
    Among these, one can itemize:
\begin{enumerate}
    \item[1)] \emph{metric teleparallel theories} where the gravitational field is described by torsion. In particular, if gravitational action is given by the torsion scalar $T$, we deal with Teleparallel Equivalent to General Relativity (TEGR);
    \item[2)] \emph{symmetric teleparallel theories} where the gravitational field is described by non-metricity. In particular, if the gravitational action is given by the non-metricity scalar $Q$, we deal with Symmetric Teleparallel Equivalent to General Relativity (STEGR);
    \end{enumerate}
\item The \emph{Riemann-Cartan geometry} is given by metric compatible curvature and torsion tensors. It is also known as $U_4$ or Einstein-Cartan-Sciama-Kibble theory \cite{Hehl1971,Hehl1973b,Hehl1974b,Hehl1976}.
    \item The \emph{Weyl geometry} is defined with respect to the non-metricity. This theory is the base of the $U(1)$ gauge theory \cite{Wheeler2018}.
\item The \emph{Minkowski geometry} is obtained by setting curvature, torsion, and non-metricity to zero, since the flat metric $\eta_{\mu\nu}$ and the null affine connections are used. This represents the framework of Special Relativity \cite{Misner1973,Romano2019}.
\end{itemize}

In this work, we focus our attention on MAGs, in particular on the teleparallel and Riemann theories, analyzing GR and its dynamically equivalent formulations and also their respective extensions, highlighting their similarities and differences. The idea is to realize if the results of the so called \textit{Geometric Trinity of Gravity} hold or not for extensions of  GR, TEGR and STEGR.\\

The layout of the paper is the following. In Sec. \ref{Section3.2}, we describe the general framework of  Metric-Affine theories of Gravity, where equivalent theories of GR and metric $f(R)$ gravity can be considered.  Moreover, the  mathematical tools necessary for the formulation of teleparallel theories are contained in Sec. \ref{sec: Teleparallel framework}. We focus on the so-called {\it Geometric Trinity of Gravity} (GTG) \cite{JHK} and the \textit{Extended Geometric Trinity of Gravity} (EGTG). In Sec. \ref{sec: Trinity} we infer the dynamical equivalence of GR, TEGR, and STEGR; in Sec. \ref{Section 3.3} the same analysis is also performed  for the respective extended theories, i.e. $f(R)$, $f(T-B)$, and $f(Q-B)$. The cases of $f(T)$ and $f(Q)$ gravity are also considered as particular cases. 
Moreover, in Sec. \ref{sec: discussion} we briefly analyze the conservation laws for the Geometric Trinity of Gravity and the Extended Geometric Trinity of Gravity pointing out that further degrees of freedom (DoFs) with respect to the basic theory can be modeled  out as geometric counter parts of effective stress-energy tensors \cite{Capozziello:2012uv}. 
Finally, we conclude our work in Sec. \ref{sec:end}, outlining the fundamental aspects of the treated topics and considering some future research perspectives to improve our comprehension of gravity theories. In the Appendix A, the field equations of the considered extended gravity theories are derived. 

\emph{Notation.} We adopt the metric signature  $(-,+,+,+)$. 
Greek letters ($\alpha, \beta,..=0,1,2,3$) denote the general manifold indices, whereas Latin letters ($a, b,..=0,1,2,3$) represent the tangent space indices. For instance, spacetime and tangent space coordinates can be respectively indicated as $\{x^{\mu}\}$ and $\{x^{a}\}$ and lead to the definition of local bases for vector and covector fields. 
$\eta^{\alpha \beta }=\eta_{\alpha \beta }={\rm diag}(-1,1,1,1)$ is the flat metric. The determinant of the metric $g_{\mu \nu}$ is denoted by $g$. Finally, unless differently stated, we refer to quantities in metric,  teleparallel, and symmetric teleparallel formalism respectively with a circle, a hat, and a diamond over the symbol defining the desired object (e.g. $\accentset{\circ}{\Gamma}^{\alpha}_{\phantom{\alpha}\mu\nu}$,
$\hat{\Gamma}^{\alpha}_{\phantom{\alpha}\mu\nu}$, $\stg{\Gamma}^{\alpha}_{\phantom{\alpha}{\mu\nu}}$).

\section[Metric-affine Gravity]{Metric-affine  Gravity}\label{Section3.2}

 GR can be  formulated as a  metric-affine theory. 
To understand how the latter is defined and how it differs from the standard  metric formulation, it is important to recall the core of the Einstein picture, which is   \qm{purely metric}. Einstein considered the metric  as the only fundamental field appearing in the action, with the role of completely describe the spacetime dynamics. Assuming  the metric field exclusive role,  the covariant derivative is  the Levi-Civita connection, which is the only connection  satisfying  the  two following properties \cite{Carroll}:
\begin{enumerate}
\item[(i)] it is compatible with the metric tensor
     \begin{equation}\label{eq:metric postulate}
 	  \nabla_{\rho}g_{\mu\nu}=0;
    \end{equation}
\item[(ii)]  it is symmetric under the interchange of  lower indices 
 \begin{equation} \label{eq:symmetric connection}
\Gamma^{\lambda}_{\phantom{\lambda}\mu\nu}=\Gamma^{\lambda}_{\phantom{\lambda}\nu\mu}\,.
\end{equation}
Such a condition  can be equivalently rewritten, introducing the torsion tensor
\begin{equation}\label{eq: 3.3}
 	T^{\lambda}_{\phantom{\lambda}\mu\nu}=\Gamma^{\lambda}_{\phantom{\lambda}\mu\nu}-\Gamma^{\lambda}_{\phantom{\lambda}\nu\mu},
 \end{equation}
 and then
 \begin{equation}\label{eq: 3.4}
 	T^{\lambda}_{\phantom{\lambda}\mu\nu}=0\,,
 \end{equation}
 for the Levi-Civita connection.
\end{enumerate}

 Conditions \eqref{eq:metric postulate} and \eqref{eq:symmetric connection} give 
\begin{equation}
	\nabla_{\rho}g_{\mu\nu}-\nabla_{\mu}g_{\nu\rho}-\nabla_{\nu}g_{\rho\mu}=0,
\end{equation}
and then the Levi-Civita connection is obtained:
\begin{equation}\label{eq: 3.6}
	\accentset{\circ}{\Gamma}^{\lambda}_{\phantom{\lambda}\mu\nu}\equiv
	   \left\{\begin{array}{l}
	       \thinspace\lambda\\
	       \mu\nu
	   \end{array}\right\}
    =\frac{1}{2}g^{\lambda\rho}(g_{\mu\rho,\nu}+g_{\nu\rho,\mu}-g_{\mu\nu, \rho}).
\end{equation}
The validity of  relation \eqref{eq: 3.6} implies that the metric tensor owns the additional role of defining parallel transport instead of being just in charge of distance measurements. This peculiarity  allows to state that the \qm{causal structure}, given by the metric $g_{\mu\nu}$, and the \qm{geodesic structure}, given by geodesics and then $\Gamma^{\lambda}_{\mu\nu}$, coincide in GR. At a foundation level, this statement is related to the validity of the Equivalence Principle. See Refs. \cite{Mancini:2025asp, Tino:2020nla} for a discussion.

However, in general, the metric tensor and the linear affine connection are not necessarily related and may be considered as independent fields.
\\In these cases, the connection is no longer \emph{a-priori}  the Levi-Civita connection, as  in a purely metric theory. Then, it is possible to relax the constraints of Eqs \eqref{eq:metric postulate} and \eqref{eq:symmetric connection} considering that:
\begin{enumerate}
	\item [(I)] the connection does not necessarily conserve  covariantly  the metric. This feature can be outlined through the introduction of the \textit{non-metricity tensor} $Q_{\rho\mu\nu}$:
	\begin{equation}\label{eq: 3.7}
		Q_{\rho\mu\nu}=\nabla_{\rho}g_{\mu\nu};
	\end{equation}	
	\item [(II)] the torsion tensor \eqref{eq: 3.3} is not identically null, 
    
    \begin{equation}\label{eq:nntt}
        T^{\lambda}{}_{\mu\nu} \neq 0\,.
    \end{equation} 
    
\end{enumerate}

Using  conditions (I) and (II), Eqs. \eqref{eq: 3.7} and \eqref{eq:nntt}, and the definition of  torsion tensor in Eq. \eqref{eq: 3.3}, the linear connection coefficients $\Gamma^{\alpha}_{\phantom{\alpha}\mu\nu}$ can be  derived in general.
From the definition of  non-metricity tensor \eqref{eq: 3.7} in (I)
\begin{equation}
	\nabla_{\rho}\thinspace g_{\mu\nu}-\nabla_{\mu}\thinspace g_{\nu\rho}-\nabla_{\nu}\thinspace g_{\rho\mu}-Q_{\rho\mu\nu}+Q_{\mu\nu\rho}+Q_{\nu\rho\mu}=0,
\end{equation}
and the expression of the covariant derivative of the metric tensor
\begin{equation}
	\nabla_{\rho}g_{\mu\nu}=\partial_{\rho}g_{\mu\nu}-\Gamma^{\lambda}_{\phantom{\lambda}\rho\mu}g_{\lambda\nu}-\Gamma^{\lambda}_{\phantom{\lambda}\rho\nu}g_{\mu\lambda},
\end{equation}
after some  calculations, one obtains 
\begin{align}\label{eq: intermediate}
	&\partial_{\rho}\thinspace g_{\mu\nu}-\partial_{\mu}\thinspace g_{\nu\rho}-\partial_{\nu}\thinspace g_{\rho\mu}+g_{\nu\lambda}T^{\lambda}_{\phantom{\lambda}\mu\rho}
 +g_{\lambda\mu}T^{\lambda}_{\phantom{\lambda}\nu\rho}+\nonumber\\
 &+g_{\rho\lambda}(\Gamma^{\lambda}_{\phantom{\lambda}\mu\nu}+\Gamma^{\lambda}_{\nu\mu})
	-Q_{\rho\mu\nu}+Q_{\mu\nu\rho}+Q_{\nu\rho\mu}=0.
\end{align}

By summing and subtracting $g_{\rho\lambda}\Gamma^{\lambda}_{\phantom{\lambda}\mu\nu}$ in the LHS of Eq.$\thinspace$\eqref{eq: intermediate}, one gets the most general linear affine connection:
\begin{align}\label{eq: 3.10}
	\Gamma^{\alpha}_{\phantom{\alpha}\mu\nu}=&-\frac{1}{2}g^{\rho\alpha}(g_{\mu\nu,\rho}-g_{\nu\rho, \mu}-g_{\rho\mu,\nu})+\frac{1}{2}(T^{\alpha}_{\phantom{\alpha}\mu\nu}
	-T_{\mu\nu}^{\phantom{\mu\nu}\alpha}+\nonumber\\
    &-T_{\nu\mu}^{\phantom{\nu\mu}\alpha})+ \frac{1}{2}(Q^{\alpha}_{\phantom{\alpha}\mu\nu}-Q_{\mu\phantom{\alpha}\nu}^{\phantom{\mu}\alpha}-Q_{\nu\phantom{\alpha}\mu}^{\phantom{\nu}\alpha})=
	\accentset{\circ}{\Gamma}^{\alpha}_{\phantom{\alpha}\mu\nu}+\nonumber\\
	&+\frac{1}{2}(T^{\alpha}_{\phantom{\alpha}\mu\nu}+T_{\mu\phantom{\alpha}\nu}^{\phantom{\mu}\alpha}+T_{\nu\phantom{\alpha}\mu}^{\phantom{\nu}\alpha})+\frac{1}{2}(Q^{\alpha}_{\phantom{\alpha}\mu\nu}-Q_{\mu\phantom{\alpha}\nu}^{\phantom{\mu}\alpha}+\nonumber\\
	&-Q_{\nu\phantom{\alpha}\mu}^{\phantom{\nu}\alpha})=
	\accentset{\circ}{\Gamma}^{\alpha}_{\phantom{\alpha}\mu\nu}+K^{\alpha}_{\phantom{\alpha}\mu\nu}+L^{\alpha}_{\phantom{\alpha}\mu\nu}.
\end{align}
Torsion is antisymmetric under the interchange of the lower indices:

\begin{equation}\label{proprietà torsione}
    T^{\alpha}_{\phantom{\alpha}\mu\nu}=-T^{\alpha}_{\phantom{\alpha}\nu\mu},
\end{equation}
while the non-metricity tensor is such that:
\begin{equation}\label{proprietà non-metricità}
    Q_{\alpha\mu\nu}=Q_{\alpha\nu\mu}.
\end{equation}
In Eq.$\thinspace$\eqref{eq: 3.10}, we introduced the following new objects, which have tensorial transformation properties under changes of coordinates: 
\begin{itemize}
	\item the \textit{contorsion tensor} 
	\begin{equation}\label{eq: 3.12}
		K^{\alpha}_{\phantom{\alpha}\mu\nu}=\frac{1}{2}(T^{\alpha}_{\phantom{\alpha}\mu\nu}+T_{\mu\phantom{\alpha}\nu}^{\phantom{\mu}\alpha}+T_{\nu\phantom{\alpha}\mu}^{\phantom{\nu}\alpha})
	\end{equation}
    which is clearly antisymmetric under the interchange of the first and third index
    \begin{equation}\label{eq: proprietà contorsione}
    K^{\alpha}_{\phantom{\alpha}\mu\nu}=-K^{\phantom{\nu\mu}\alpha}_{\nu\mu\phantom{\alpha}},
    \end{equation}
    as one can infer from Eq. \eqref{proprietà torsione};
	\item the \textit{disformation tensor}
	\begin{equation}\label{eq: 3.13}
		L^{\alpha}_{\phantom{\alpha}\mu\nu}=\frac{1}{2}(Q^{\alpha}_{\phantom{\alpha}\mu\nu}-Q_{\mu\phantom{\alpha}\nu}^{\phantom{\mu}\alpha}-Q_{\nu\phantom{\alpha}\mu}^{\phantom{\nu}\alpha})
	\end{equation}
    which is instead symmetric under the interchange of the last two indices 
    \begin{equation}
    L^{\alpha}_{\phantom{\alpha}\mu\nu}=L^{\alpha}_{\phantom{\alpha}\nu\mu}
    \end{equation}
    because of Eq. \eqref{proprietà non-metricità}.
\end{itemize}

Moreover, the Riemann tensor can be defined analogously to the curvature tensor in GR as
\begin{align}\label{eq: 3.14}
	&R^{\alpha}_{\phantom{\alpha}\beta\mu\nu}=
	\begin{vmatrix}
	\partial_{\mu} & \partial_{\nu}\\
	\Gamma^{\alpha}_{\phantom{\alpha}\mu\beta} & \Gamma^{\alpha}_{\phantom{\alpha}\nu\beta}
	\end{vmatrix}
	\thinspace+\thinspace
	\begin{vmatrix}
		\Gamma^{\alpha}_{\phantom{\alpha}\mu\sigma} & \Gamma^{\alpha}_{\phantom{\alpha}\nu\sigma}\\
		\Gamma^{\sigma}_{\phantom{\sigma}\mu\beta} & \Gamma^{\sigma}_{\phantom{\sigma}\nu\beta}
	\end{vmatrix}\nonumber\\
	\nonumber\\
	&=\Gamma^{\alpha}_{\phantom{\alpha}\nu\beta,\mu}-\Gamma^{\alpha}_{\phantom{\alpha}\mu\beta,\nu}+\Gamma^{\alpha}_{\phantom{\alpha}\mu\sigma}\Gamma^{\sigma}_{\phantom{\sigma}      
	\nu\beta}-\Gamma^{\alpha}_{\phantom{\alpha}\nu\sigma}\Gamma^{\sigma}_{\phantom{\sigma}\mu\beta}
\end{align}
and it is such that 
\begin{equation}
    R^{\alpha}_{\phantom{\alpha}\beta\mu\nu}=-R^{\alpha}_{\phantom{\alpha}\beta\nu\mu}.
\end{equation}

\begin{figure}[ht]
\renewcommand{\thefigure}{1}
   \begin{center}
	 \includegraphics[width=90 mm]{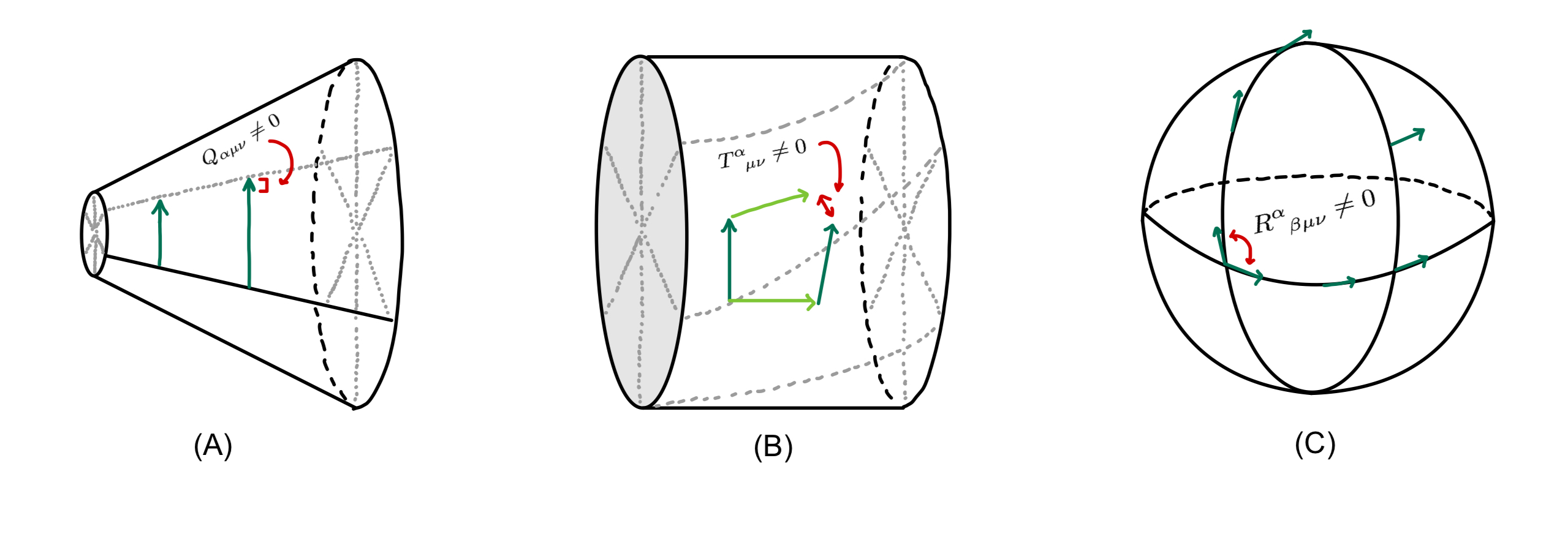}
 	\caption{A pictorial view of  (A) non-metricity, (B) torsion and (C) curvature for parallel-transported vectors. See also \cite{Teleparallel}.}
  	\label{Parallel_transport}
  \end{center}
\end{figure}

In addition, the curvature, torsion, and non-metricity differently affect the parallel transport of a vector on a manifold (see Fig.$\thinspace$\ref{Parallel_transport}).

We have that:
\begin{itemize}
    \item \emph{curvature} manifests its presence when a vector is parallel-transported along a closed curve on a non-flat background and comes back to its starting point forming a non-null angle with its initial position; 
    \item \emph{torsion} entails a rotational geometry, where the parallel transport of two vectors is antisymmetric by exchanging the transported vectors and the direction of transport. This property results in the non-closure of parallelograms; 
    \item \emph{non-metricity} is responsible for altering the length of the vectors along the transport. 
\end{itemize}

In a generic metric-affine theory, all these effects can coexist and they correspond to physical quantities (e.g. the torsion tensor is linked to the spin in the Einstein-Cartan theory \cite{Hehl1976} but it can have also a more general meaning \cite{Capozziello:2001mq}).
A useful alternative to the definition of the Riemann tensor in \eqref{eq: 3.14} can be derived from the decomposition of  linear affine connection \eqref{eq: 3.10}:

\begin{align}\label{eq: 3.15}
	R^{\alpha}_{\phantom{\alpha}\beta\mu\nu}=&\accentset{\circ}{R}^{\alpha}_{\phantom{\alpha}\beta\mu\nu}+\accentset{\circ}{\nabla}_{\mu}D^{\alpha}_{\phantom{\alpha}\nu\beta}-\accentset{\circ}{\nabla}_{\nu}D^{\alpha}_{\phantom{\alpha}\mu\beta}+\nonumber\\	&+D^{\sigma}_{\phantom{\sigma}\nu\beta}D^{\alpha}_{\phantom{\alpha}\mu\sigma}-D^{\sigma}_{\phantom{\sigma}\mu\beta}D^{\alpha}_{\phantom{\alpha}\nu\sigma},
\end{align}
where $D^{\alpha}_{\phantom{\alpha}\mu\nu}$ is defined as the sum of the contorsion \eqref{eq: 3.12} and disformation tensors \eqref{eq: 3.13}
\begin{equation}
	D^{\alpha}_{\phantom{\alpha}\mu\nu}=K^{\alpha}_{\phantom{\alpha}\mu\nu}+L^{\alpha}_{\phantom{\alpha}\mu\nu}.
\end{equation}

Finally, from  the tensorial quantities $T^{\alpha}_{\phantom{\alpha}\mu\nu}$ and $Q_{\alpha\mu\nu}$, we can introduce the torsion scalar $\mathbb{T}$ 

\begin{equation}\label{eq: torsion scalar}
	\mathbb{T}\equiv \frac{a_1}{4}T_{\alpha\mu\nu}T^{\alpha\mu\nu}+\frac{a_2}{2}T_{\alpha\mu\nu}T^{\mu\alpha\nu}-a_3 T_{\alpha}T^{\alpha}, 
\end{equation}
with $T_{\alpha}=T^{\mu}_{\phantom{\mu}\alpha\mu}$, and the non-metricity scalar $\mathbb{Q}$
\begin{align}\label{eq: non-metricity scalar}
	\mathbb{Q}\equiv -\frac{b_1}{4}&Q_{\alpha\mu\nu}Q^{\alpha\mu\nu}+\frac{b_2}{2}Q_{\alpha\mu\nu}Q^{\mu\alpha\nu}+\frac{b_3}{4}Q_{\alpha}Q^{\alpha}+\nonumber\\
    &-(b_4-1)\tilde{Q}_{\alpha}\tilde{Q}^{\alpha}-\frac{b_5}{2}Q_{\alpha}\tilde{Q}^{\alpha}
\end{align}
with $Q_{\alpha}=Q_{\alpha\mu}^{\phantom{\alpha\mu}\mu}$ and $\tilde{Q}_{\alpha}=Q_{\mu\alpha}^{\phantom{\mu\alpha}\mu}$.
In a metric-affine framework, there are several classes of theories whose dynamics can be related to the tensors 
$R^{\alpha}_{\phantom{\alpha}\beta\mu\nu}$, $T^{\alpha}_{\phantom{\alpha}\mu\nu}$, and $Q_{\alpha\mu\nu}$.
As stated in Sec. \ref{sec:Intro}, in this work, we focus our attention on the \textit{flat connection} subclass. In particular, we will consider \textit{metric teleparallel} and \textit{symmetric teleparallel} theories and their dynamical equivalence with respect to GR and its extension $f(\accentset{\circ}{R})$. 

\begin{figure}[h]
\renewcommand{\thefigure}{2}
	 \includegraphics[width=90mm]{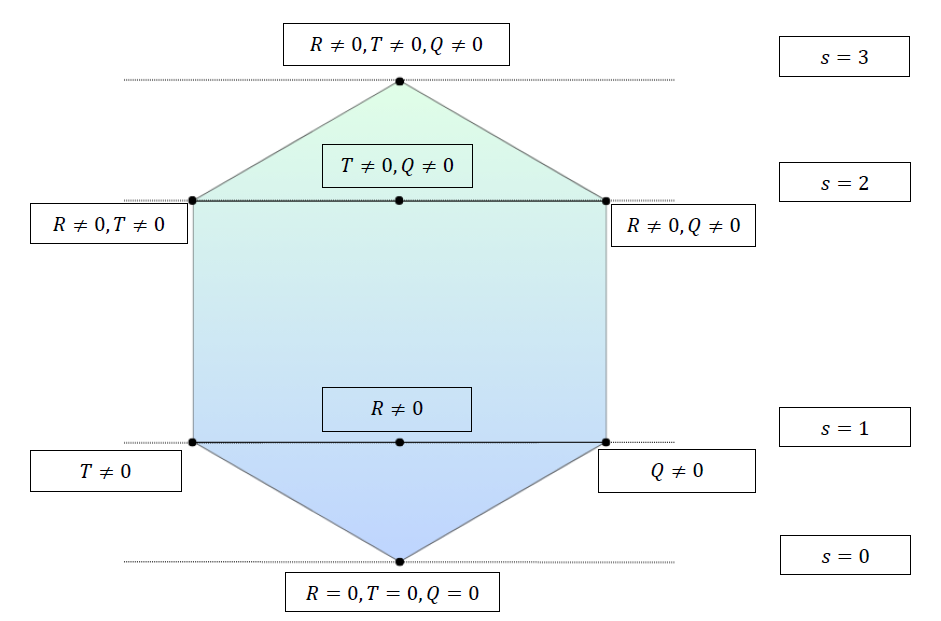}
 	\caption{The Eightfold Way of Gravity. The number of non-trivial geometric invariants is given by $s$. Starting from $s=0$, the Minkowsky case, going up the hexagon, it is possible to recover all the metric-affine theories.}
  	\label{Figure Multiplet of gravity}
\end{figure}

To conclude this preliminary discussion, it is worth noticing that in a very general theory of gravity, geometric information can come from curvature, torsion and non-metricity. If all these geometric invariants are zero,  we are dealing with Minkowski spacetime where no gravitational dynamics is present. On the other hand,  we can establish a sort of {\it Multiplet of Gravity} where the various geometric invariants are the entries. In   Fig.\ref{Figure Multiplet of gravity}, the {\it quantum number} $s$ indicates how many invariants are different from zero in a given gravitational theory. 

Specifically, $s=0$ is Minkowki; $s=1$ corresponds to GR, TEGR, and STEGR (the Trinity Gravity); $s=2$ means that two invariants can be considered; $s=3$ is the full theory where  the whole geometric budget is taken into account.

In analogy with particle physics, we can  consider a sort of {\it Eigthfold Way of Gravity}. Also in this case, we have an {\it octet}.  This picture could be very useful in view of a possible quantization procedure where the various  DoFs are taken into account.

\section{The Teleparallel framework}\label{sec: Teleparallel framework}

Most of the fundamental interactions, i.e, electromagnetism, weak and strong interactions, occurs in the spacetime and are gauge theories, defined by point-dependent transformations in internal spaces  corresponding to different points of the external manifold. In fact, any gauge theory is defined on the fiber bundles, in which the gauge group is connected to each point of the spacetime \cite{Nakahara}. 
On the other hand, gravity is strictly connected with the notion of spacetime. In particular, any gravitational theory is formulated on the tangent bundle: the spacetime is the base space and the tangent vector space, attached at each point of it, is the fiber bundle. The spacetime manifold and the tangent bundle are soldered, being both 4-dimensional, in a such way that the spacetime metric $g_{\mu\nu}$ and the Minkowski metric $\eta_{ab}$ are connected by the tetrad field $e^{a}{}_{\mu}$ : 
\begin{equation}\label{eq: metric}
	g_{\mu\nu}=e^{a}_{\phantom{a}\mu}e^{b}_{\phantom{b}\nu}\eta_{ab},
\end{equation}
Teleparallel gravity is a gauge theory for the translation group and it is defined with respect to the tedrad formalism. Moreover, GR itself can be formulated in this framework, obtaining the above mentioned TEGR.
On the fiber bundle, i.e. the Minkowski tangent space, the gauge transformations represent local translations; thus, we have to respect the invariance under both general coordinate transformations, performed in the spacetime, and local Lorentz transformations, performed in the tangent space.
In this framework, the inertial effects are represented by a purely inertial connection, the \textit{spin connection} or \textit{Lorentz connection}, which depends on local Lorentz transformations and allows to define different classes of frame. The equivalent inertial frames are related by a global Lorentz transformation, and the class of frames, in which inertial effects are not present, is characterized by a trivial inertial Lorentz connection. 
In all other classes of frames, the inertial spin connection is non-vanishing \cite{Martin2019, Pereira2012, ATG}.\\
In Sec. \ref{subsec: Tetrad} we develop the tetrad formalism, while in Sec. \ref{subsec: spin connection} we focus on the local Lorentz transformations with the respective spin connections. Finally in Sec. \ref{subsec:dynamical variables}, we consider the three dynamical variable, curvature, torsion, and non-metricity, in the tetrad formalism.

\subsection{The tetrad formalism}\label{subsec: Tetrad}

Let us consider a 4-dimensional differentiable manifold $\mathcal{M}$, whose tangent space at each point $p\in \mathcal{M}$ is the Minkowski spacetime $T_{p}\mathcal{M}$. A natural differentiable basis of $T_{p}\mathcal{M}$ is given by the sets of gradients
\begin{equation}
    \left\{\partial_{\mu}\right\}:=\left\{\frac{\partial}{\partial_{\mu}}\right\}
\end{equation}
as well as on the cotangent space $T^*_{p}\mathcal{M}$ the covector basis is $\left\{dx^{\mu}\right\}$. Both $\left\{\partial_{\mu}\right\}$ and $\left\{dx^{\mu}\right\}$ satisfy the orthonormality condition:
\begin{align}\label{eq: ortho condition}
    dx^{\mu}\partial_{\nu}&=\delta^{\mu}_{\nu}\\
    dx^a \partial_b&=\delta^a_b
\end{align}
The entire set of such bases represents a basis for the vectors on $T_p\mathcal{M}$, at each point $p\in\mathcal{M}$. Thus, on the
common domains they are defined, we can express each orthonormal vector and covector with respect to the other \cite{Martin2019, Pereira2012, ComparingEG}. A general linear frame is expressed by tetrads or \qm{vierbein} (German for \qm{many legs}, vierbein = four legs):
\begin{align}\label{eq: Tetrad}
	e_{a}&=e_{a}^{\phantom{a}\mu}\partial_{\mu}\\
	e^{a}&=e^{a}_{\phantom{a}\mu}dx^{\mu}.
\end{align}

Tetrads are fundamental dynamical objects, introduced in substitution for the metric field and are defined as linear bases that connect the general spacetime metric $g_{\mu\nu}$ to the Minkowski metric $\eta_{ab}$ according to Eq. \eqref{eq: metric} and the reverse equation 

\begin{equation}\label{eq: A1}
	\eta_{ab}=e_{a}^{\phantom{a}\mu}e_{b}^{\phantom{b}\nu}g_{\mu\nu}.
\end{equation}

The frame $e^{a}_{\phantom{a}\mu}$ is such that respects the orthornormality condition given in Eq. \eqref{eq: ortho condition}:

\begin{align}
&e^{a}_{\phantom{a}\nu}e_{a}^{\phantom{a}\mu}=\delta^{\mu}_{\nu}\label{eq: orthogonality relation}\\
	&e^{a}_{\phantom{a}\mu}e_{b}^{\phantom{b}\mu}=\delta_{b}^{a}.
\end{align}

It is possible to define also the inverse metric components 
\begin{equation}\label{eq: gcontravariant}
	g^{\mu\nu}=e_{a}^{\phantom{a}\mu}e_{b}^{\phantom{b}\nu}\eta^{ab}
\end{equation}
and the determinant of the tetrad fields $e=\textrm{det}(e^{a}_{\phantom{a}\mu})$
\begin{equation}\label{eq: e}
	e=\sqrt{-g}.
\end{equation}

It is worth noticing that the frame and the associated bundle are characteristic of spacetime, therefore if the spacetime is differentiable, they are automatically present \cite{Pereira2012}. \\
A general tetrad basis $\left\{e_a \right\}$, Eq. \eqref{eq: Tetrad}, satisfies the commutation relations \cite{Pereira2012}:
\begin{equation}\label{eq: commutatori tetradi}
    [e_a, e_b ] := f^{c}{}_{ab} e_c,
\end{equation}
where $f^{c}{}_{ab}$ are the structure coefficients or the anholonomy
of frame and are defined by:
\begin{equation}\label{eq: coeff struttura}
     f^{c}{}_{ab} = e^{\mu}_{a}e^{\nu}_{b}(\partial_{\nu}e^{c}_{\mu} - \partial_{\mu}e^{c}_{\nu}).
\end{equation}
The structure coefficients $f^{c}{}_{ab}$ are functions of the spacetime points and measure the not-closure of the parallelogram formed by the vectors $e_a$ and $e_b$ \cite{ComparingEG}. If $f^{c}{}_{ab} \neq 0$, the tedrad basis is said to be \textit{anholonomic} or \textit{non-trivial} tetrads.
However, in a class of frame $e^{'a}$ it is possible to set 
\begin{equation}\label{eq: olonomia}
    f^{c}{}_{ab}=0,
\end{equation}
locally. This is the class of inertial frame, in which the holonomy of the tetrads is restored. In fact, Eq. \eqref{eq: olonomia} means that $e'_{a}$ is locally a closed differential form and there exists a neighborhood around the point $p$ on which:
\begin{equation}
    e'^{a}=dx^{a}.
\end{equation}

In absence of gravitation, the anholonomy is only caused by the inertial forces which are present in those frames: the metric $g_{\mu\nu}$ represents the Minkowski metric in a general coordinate system $\eta_{\mu\nu}$. 
In absence of inertial forces, the class of inertial frames is characterized by vanishing structure coefficients, since all coordinate bases are holonomic. This property is valid everywhere for frames belonging to this inertial class \cite{Aldrovandi03, ATG}. In fact, a closed differential form is always locally integrable, or exact. Thus, locally inertial frames are always holonomic. In these frames, inertial effects locally compensate gravitation. If we consider the Minkowski metric in a holonomic basis, i.e. in absence of gravitation, in any other coordinates, $\eta_{\mu\nu}$ will be a function of the spacetime coordinates, regardless the nature of tetrads field, i.e. holomic (inertial) or not. Tedrads always relate the tangent Minkowski space to a Minkowski spacetime:
\begin{equation}
    \eta_{ab} = \eta_{\mu\nu} e^{\mu}_{a} e^{\nu}_{b}.
\end{equation}
These are the frames appearing in Special Relativity and are usually called trivial frames, or trivial tetrads. Different classes of frames are obtained by performing local Lorentz transformations and in each class, the different frames are related to the others through global Lorentz transformations.

\subsection{The Lorentz connection}\label{subsec: spin connection}

The spin connection, or \textit{Lorentz connection} $\omega_{\mu}$ is a 1-form acting in the Lorentz algebra:
\begin{equation}
	\omega_{\mu}=\frac{1}{2}\omega^{ab}_{\phantom{ab}\mu}L_{ab},
\end{equation}
where $L_{ab}$ identifies a representation of the Lorentz generators and $\omega^{ab}_{\mu}$ are the spin connection coefficients, which are antsymmetric in the algebraic indices, i.e. $\omega^{ab}{}_{\mu} = -\omega^{ba}{}_{\mu}$, because of the antisymmetry of  generators $L_{ab}$. In fact, in order to respect the covariance of the Einstein equations, the connection $\omega_{\mu}$  behaves like vectors in the spacetime indices and it has a non-tensorial character in the Lorentz indices to compensate the non-tensoriality of ordinary derivatives. These connections are gauge potentials, introduced to produce derivatives which behave covariantly under gauge transformations, and are related to the linear group $GL(4,\mathbb{R})$.
The spin-connection represents the inertial effects in the considered frame.
\\As in GR, the Levi-Civita connection is defined so to obtain the notion of a derivative with tensorial transformation properties, the spin-connection guarantees the introduction of the \textit{Fock-Ivanenko covariant derivative}
\begin{equation}
	D_{\mu}=\partial_{\mu}-\frac{i}{2}\omega^{ab}_{\phantom{ab}\mu}L_{ab}
\end{equation}
whose second part acts only on the algebraic, or tangent space indices \cite{Capozziello:2009zza}.
Thus, a tetrad field relates tangent space (or internal) tensors with those related to the spacetime (or external). It is possible to relate the connection 1-form $\omega$ and the linear affine connection:
\begin{equation}\label{eq: fock derivata}
    \Gamma^{\lambda}_{\mu \nu} = e^{\lambda}_{a}e^{b}_{\nu}\omega^{a}_{b\mu} + e^{\lambda}_{a}\partial_{\mu}e^{a}_{\nu} \equiv e^{\lambda}_{a}D_{\mu} e^{a}_{\nu},
\end{equation}
\begin{equation}\label{eq: cov derivata}
    \omega^{a}_{b\mu} = e^{a}_{\lambda} e^{\nu}_{b} \Gamma^{\lambda}_{\mu\nu} + e^{a}_{\sigma}\partial_{\mu}e^{\sigma}_{b} \equiv e^{a}_{\nu} \nabla_{\mu} e^{\nu}_{b}.
\end{equation} 

We stress again that $\nabla_{\mu}$ is the covariant derivative defined with respect to the connection $\Gamma^{\lambda}_{\mu\nu}$,  acting  on external indices, and it can be defined for tensorial fields. On the other hand, the Fock-Ivanenko derivative $D_{\mu}$ acts on internal indices and can be defined for all tensorial and spinorial fields.
However, both Eqs \eqref{eq: fock derivata} and \eqref{eq: cov derivata} encode the \emph{tedrad postulate} \cite{ComparingEG}: the total covariant derivative of the tetrad fields, expressed in terms of  connection for both internal and external indices, vanishes identically, namely
\begin{equation}\label{eq: tetrad postulate}
    \nabla_{\mu}e^{a}_{\nu} = \partial_{\mu}e^{a}_{\nu}-\Gamma^{\lambda}_{\mu\nu}e^{a}_{\lambda}+\omega^{a}_{b\mu}e^{b}_{\nu} = 0.
\end{equation}

Let us introduce the object
\begin{equation}\label{eq: A12}
	W^{a}_{\phantom{a}\mu\nu}=\partial_{\mu}e^{a}_{\phantom{a}\nu}
\end{equation}
for future convenience.
Thus, the metric compatibility given in Eq. \eqref{eq:metric postulate} in the tetrad formalism can be expressed through the spin connection, which has to be Lorentzian, i.e. $\omega_{ab\mu} = -\omega_{ba\mu}$ \cite{ComparingEG}.
This means that if the metric postulate Eq. \eqref{eq:metric postulate} is not valid, the corresponding spin connection cannot assume values in the Lie algebra of the Lorentz group, because it is not a Lorentz connection.
The metric compatibility holds if and only if we choose a Lorentz connection. \\
Let us first consider the inertial frames of Sec. \ref{subsec: Tetrad} $\left\{e'^{a}_\mu \right\}$, written in a general coordinate system $\left\{x'^{\mu}\right\}$, whose holomic form is $e'^{a}_{\mu} = \partial_{\mu}x'^{a}$, where $x'^{a}$ is a point–dependent Lorentz vector: $x'^{a}= x'^{a}(x'^{\mu})$. \\ A local Lorentz transformation $x^a= \Lambda^a_b x'^{b}$ transforms the holomic frame in the new frame
\begin{equation}
    e'^{a}_{\mu}=\partial_{\mu}x'^{a}=\Lambda^{a}_{b}(x) x^{b}_{\mu}
\end{equation}
whose explicit form is:
\begin{equation}\label{eq:newframe}
    e^{a}_{\mu} = \partial_{\mu}x^{a} + \omega^{a}_{b\mu}x^{b} = D_{\mu}x^{a},
\end{equation}
where 
\begin{equation}\label{eq: omega inerziale}
    \omega^{a}_{b\mu} = \Lambda^{a}_{c}\partial_{\mu}\Lambda^{c}_{b},
\end{equation}
is a Lorentz connection that represents the inertial effects present in the Lorentz rotated frame $e^{a}_{\mu}$ and $D_{\mu}$ is the associated covariant derivative \cite{ComparingEG, Martin2019}.
Under a local Lorentz transformation $\Lambda^{a}_{b}(x)$, the spin connection becomes:
\begin{equation}\label{eq: omega spuria}
    \omega^{a}_{b\mu} = \Lambda^{a}_{c}(x)\omega'^{c}_{d\mu}\Lambda^{d}_{c}+\Lambda^{a}_{c}\partial_{\mu}\Lambda^{c}_{b}.
\end{equation}
Thus, in the RHS of Eq. \eqref{eq: omega spuria}, we find two contributions: the first term represents  non-inertial effects caused by the changed frame, while the second term represents the inertial effects to the rotation of the new frame with respect to the previous one. Therefore, the inertial connection of Eq. \eqref{eq: omega inerziale} is the connection obtained from a Lorentz transformation, expressed in Eq. \eqref{eq: omega spuria} considering a vanishing spin connection $\omega'^{c}_{d\mu}$. Thus, starting from an inertial frame, in which the inertial spin connection vanishes, through a local Lorentz transformation $\Lambda^{a}_{b}(x^{\mu})$, it is always possible to find different classes of frames and, in each class, the infinitely many frames are related to the others through a global Lorentz transformation, given by $\Lambda^{a}_{b} = \mbox{const}$. In the inertial frames (i.e. $e'^{a}_{\mu} = \partial_{\mu}x'^{a}$), the inertial effects are absent since the Lorentz connections vanish. The structure coefficients  \eqref{eq: coeff struttura} can be written in terms of the spin connection, considering both Eqs \eqref{eq: tetrad postulate} and \eqref{eq: omega inerziale}:
\begin{equation}
    f^{c}_{ab} = - (\omega^{c}_{ab}-\omega^{c}_{ba}).
\end{equation}
Therefore, the definition $\omega^{c}_{ab}$ becomes:
\begin{equation}
    \omega^{c}_{ab} = \frac{1}{2} (f_{a}^{\text{ c }}{}_{b} + f_{b}^{\text{ c }}{}_a + f^{c}_{ab}). 
\end{equation}

It is worth remarking that, in Teleparallel Gravity, we have always to consider tetrads with the related spin connections, thus the couple $\left\{ e^a_{\mu}, \omega^a_{b \mu}\right\}$. However, according to Eq. \eqref{eq: omega inerziale}, there are special frames in which the spin connections are assumed vanishing; in this case, we consider the couple $\left\{e^a_{\mu},0\right\}$, since, in this case, the tetrad and the spin connection are treated as independent variables \cite{Martin2019}. This gauge choice of  spin connection is the so-called the Weitzenb\"ock gauge; its definition is equivalent to the tedrad postulate $\nabla_{\mu}e^{a}_{\nu} = 0$. In the class of frames in which $\omega^{a}_{b\mu}$ vanishes, it becomes:
\begin{equation}\label{eq: weitzenbock}
     \partial_{\mu}e^{a}_{\nu}-\Gamma^{\lambda}_{\mu\nu}e^{a}_{\lambda}=0.
\end{equation}
This is the \textit{distant parallelism condition}, from where Teleparallel Gravity takes its name. In this gauge, the corresponding Weitzenb\"ock connection and torsion can be written as follows{}
\begin{align}
    \hat{\Gamma}^{\alpha}_{\phantom{\alpha}\mu\nu}&=e_{a}^{\phantom{a}\alpha}\partial_{\mu}e^{a}_{\phantom{a}\nu},\label{eq: A16} \\
    \hat{T}^{\alpha}_{\phantom{\alpha}\mu\nu}&=e_{a}^{\phantom{a}\alpha}\hat{T}^{a}_{\phantom{a}\mu\nu}=e_{a}^{\phantom{a}\alpha}[\partial_{\mu}e^{a}_{\phantom{a}\nu}-\partial_{\nu}e^{a}_{\phantom{a}\mu}] \label{eq: tttn}.
\end{align}

\subsection{The dynamical variables in  tetrad formalism}\label{subsec:dynamical variables}

Now one has at disposal all the ingredients to define  torsion, curvature, and non-metricity tensors associated to a generic affine connection. Indeed, these tensorial quantities can be formulated in the context of a coordinate-free framework \cite{GE,Nakahara}
\begin{align}
	&T(X, Y)=\nabla_XY-\nabla_YX-[X,Y],\\
	&R(X,Y,Z)=\nabla_{X}\nabla_{Y}Z-\nabla_{Y}\nabla_{X}Z-\nabla_{[X,Y]}Z,\\
	&Q(X,Y, Z)=\nabla_Z[g(X,Y)]-g(\nabla_ZX,Y)-g(X,\nabla_Z Y),
\end{align}
and indicating the coordinate basis $\{\partial_{\mu}\}$ as $\{e_{\mu}\}$, one easily derives
\begin{align}\label{eq: A17}
	T^{a}{}_{\mu\nu}&=\langle e^{a}, T(e_{\mu},e_{\nu})\rangle \notag \\ &=\langle e^{a},\nabla_{e_{\mu}}e_{\nu}-\nabla_{e_{\nu}}e_{\mu}-[e_{\mu},e_{\nu}] \rangle \nonumber\\
	&=\langle e^{a}{}_{\rho}dx^{\rho},(\Gamma^{\lambda}{}_{\mu\nu}-\Gamma^{\lambda}{}_{\nu\mu})e_{\lambda}\rangle\nonumber \\ &=
	e^{a}{}_{\lambda}(\Gamma^{\lambda}{}_{\mu\nu}-\Gamma^{\lambda}{}_{\nu\mu}) \nonumber\\
	&=W^{a}{}_{\mu\nu}-W^{a}{}_{\nu\mu}+\omega^{a}{}_{b\mu}e^{b}{}_{\nu}-\omega^{a}{}_{b\nu}e^{b}{}_{\mu}.
\end{align}
\begin{align}
	R^{a}{}_{b\mu\nu} &= \langle e^{a}, R(e_{\mu},e_{\nu})e_{b} \rangle \notag \\ &=\langle e^{a}, (\nabla_{e_{\mu}}\nabla_{e_{\nu}}-\nabla_{e_{\nu}}\nabla_{e_{\mu}})e_{b}{}^{\rho}e_{\rho} \rangle \nonumber\\
	&=\langle e^{a}_{\phantom{a}\sigma}dx^{\sigma}, (\Gamma^{\lambda}_{\phantom{\lambda}\nu\mu}-\Gamma^{\lambda}_{\phantom{\lambda}\mu\nu})\partial_{\lambda}(e_{b}^{\phantom{b}\rho})e_{\rho} \rangle +e^{a}_{\phantom{a}\sigma}e_{b}{}^{\rho}	R^{\sigma}_{\phantom{\sigma}\rho\mu\nu} \nonumber\\
	&=(\Gamma^{\lambda}_{\phantom{\lambda}\mu\nu}-\Gamma^{\lambda}_{\phantom{\lambda}\nu\mu})e_{b}^{\phantom{b}\rho}W^{a}_{\phantom{a}\lambda\rho}+e^{a}_{\phantom{a}\sigma}e_{b}^{\phantom{b}\rho}R^{\sigma}_{\phantom{\sigma}\rho\mu\nu} \nonumber\\
	&=\partial_{\mu}\omega^{a}_{\phantom{a}b\nu}-\partial_{\nu}\omega^{a}_{\phantom{a}b\mu}+\omega^{a}_{\phantom{a}c\mu}\omega^{c}_{\phantom{c}b\nu}-\omega^{a}_{\phantom{a}c\nu}\omega^{c}_{\phantom{c}b\mu}
\end{align}
\begin{align}\label{eq: A19}
	Q_{\alpha ab}&=\nabla_{e_\alpha}[g(e_a,e_b)]-g(\nabla_{e_{\alpha}}e_a,e_b)-g(e_a,\nabla_{e_{\alpha}e_b}) \nonumber\\
	&=\partial_{\alpha}(g_{\mu\nu})e_{a}^{\phantom{a}\mu}e_{b}^{\phantom{b}\nu}-e_{a}^{\phantom{a}\mu}W^{c}_{\phantom{c}\alpha\mu}\eta_{cb}-\omega^{c}_{\phantom{c}a\alpha}\eta_{cb} \nonumber\\
	&-e_{b}^{\phantom{b}\mu}W^{c}_{\phantom{c}\alpha\mu}\eta_{ac}-	\omega^{c}_{\phantom{c}b\alpha}\eta_{ac} \nonumber \\ &=-\omega^{c}_{\phantom{c}a\alpha}\eta_{cb}-\omega^{c}_{\phantom{c}b\alpha}\eta_{ac}.
\end{align}
where the relation
\begin{equation}
	\nabla_{e_{\mu}}(fX)=e_{\mu}[f]X+f\nabla_{e_{\mu}}X,\quad \textrm{with $f$ being a scalar},\nonumber
\end{equation}
and the definition of the general affine connection \eqref{eq: fock derivata} have been used.
If we consider the inertial spin connection, Eq. \eqref{eq: omega inerziale}, the curvature tensor, Eq \eqref{eq: 3.14}, vanishes, while torsion can be vanishing if one takes into account trivial tetrads, i.e. $e^a_\mu=\partial_\mu x^a$ and $\omega^a_{b \mu}=0$ \cite{Pereira2012, Martin2019}.

\section{The Geometric Trinity of Gravity}\label{sec: Trinity}

In the metric teleparallel framework, the linear affine connection \eqref{eq: 3.10} can be rewritten as follows
\begin{equation}
	\hat{\Gamma}^{\alpha}_{\phantom{\alpha}\mu\nu}=\accentset{\circ}{\Gamma}^{\alpha}_{\phantom{\alpha}\mu\nu}+\hat{K}^{\alpha}_{\phantom{\alpha}\mu\nu}
\end{equation}
and the Riemann tensor \eqref{eq: 3.15} becomes
\begin{align}\label{eq: Riemann TT}{}
	R^{\alpha}_{\phantom{\alpha}\beta\mu\nu}&=\accentset{\circ}{R}^{\alpha}_{\phantom{\alpha}\beta\mu\nu}+\accentset{\circ}{\nabla}_{\mu}\hat{K}^{\alpha}_{\phantom{\alpha}\nu\beta}-\accentset{\circ}{\nabla}_{\nu}\hat{K}^{\alpha}_{\phantom{\alpha}\mu\beta}+\nonumber\\
&+\hat{K}^{\sigma}_{\phantom{\sigma}\nu\beta}\hat{K}^{\alpha}_{\phantom{\alpha}\mu\sigma}-\hat{K}^{\sigma}_{\phantom{\sigma}\mu\beta}\hat{K}^{\alpha}_{\phantom{\alpha}\nu\sigma}\equiv 0
\end{align}
where the vanishing curvature constraint has been imposed.
\\Starting from the definition of the Riemann tensor in Eq. \eqref{eq: Riemann TT}, it is possible to define the associated Ricci scalar:

\begin{equation}\label{eq: RTT}
	\accentset{\circ}{R}+\accentset{\circ}{\nabla}_{\mu}\hat{K}^{\mu\nu}_{\phantom{\mu\nu}\nu}-\accentset{\circ}{\nabla}_{\nu}\hat{K}_{\mu}^{\phantom{\mu}\mu\nu}+\hat{K}^{\sigma\nu}_{\phantom{\sigma\nu}\nu}\hat{K}^{\mu}_{\phantom{\mu}\mu\sigma}-\hat{K}^{\sigma}_{\phantom{\sigma}\mu\nu}\hat{K}^{\mu\nu}_{\phantom{\mu\nu}\sigma}=0.
\end{equation}

The last four terms in Eq.$\thinspace$\eqref{eq: RTT} may be straightforwardly recast as

\begin{align}\label{eq: T1}
	&\accentset{\circ}{\nabla}_{\mu}\hat{K}^{\mu\nu}_{\phantom{\mu\nu}\nu}-\accentset{\circ}{\nabla}_{\nu}\hat{K}_{\mu}^{\phantom{\mu}\mu\nu}=\accentset{\circ}{\nabla}_{\mu}(\hat{K}^{\mu\nu}_{\phantom{\mu\nu}\nu}-\hat{K}_{\nu}^{\phantom{\nu}\nu\mu})=2\accentset{\circ}{\nabla}_{\mu}\hat{K}^{\mu\nu}_{\phantom{\mu\nu}\nu}=\nonumber\\
	&=2\accentset{\circ}{\nabla}_{\mu}\Big[\frac{1}{2}(\hat{T}^{\mu\nu}_{\phantom{\mu\nu}\nu}+\hat{T}^{\nu\mu}_{\phantom{\nu\mu}\nu}+\hat{T}_{\nu}^{\phantom{\nu}\mu\nu})\Big]=2\accentset{\circ}{\nabla}_{\mu}\hat{T}_{\nu}^{\phantom{\nu}\mu\nu}=2\accentset{\circ}{\nabla}_{\mu}\hat{T}^{\mu},
\end{align}
\begin{align}\label{eq: T2}
	\hat{K}^{\sigma\nu}_{\phantom{\sigma\nu}\nu}\hat{K}^{\mu}_{\phantom{\mu}\mu\sigma}=-\hat{K}^{\sigma\nu}_{\phantom{\sigma\nu}\nu}\hat{K}_{\sigma\mu}^{\phantom{\sigma\mu}\mu}=-\hat{T}^{\alpha}\hat{T}_{\alpha},
\end{align}
\begin{align}\label{eq: T3}
	-&\hat{K}^{\sigma}_{\phantom{\sigma}\mu\nu}\hat{K}^{\mu\nu}_{\phantom{\mu\nu}\sigma}=\hat{K}_{\sigma\mu\nu}\hat{K}^{\sigma\nu\mu}=\frac{1}{4}(\hat{T}_{\sigma\mu\nu}\hat{T}^{\sigma\nu\mu}+\hat{T}_{\sigma\mu\nu}\hat{T}^{\nu\sigma\mu}+\nonumber\\
    &+\hat{T}_{\sigma\mu\nu}\hat{T}^{\mu\sigma\nu}+\hat{T}_{\mu\sigma\nu}\hat{T}^{\sigma\nu\mu}+\hat{T}_{\mu\sigma\nu}\hat{T}^{\nu\sigma\mu}+\hat{T}_{\mu\sigma\nu}\hat{T}^{\mu\sigma\nu}+\nonumber\\
    &+\hat{T}_{\nu\sigma\mu}\hat{T}^{\sigma\nu\mu}
    +\hat{T}_{\nu\sigma\mu}\hat{T}^{\nu\sigma\mu}+\hat{T}_{\nu\sigma\mu}\hat{T}^{\mu\sigma\nu})=\nonumber\\
	&=\frac{1}{4}(\hat{T}^{\alpha\mu\nu}\hat{T}_{\alpha\mu\nu}+2\hat{T}_{\alpha\mu\nu}\hat{T}^{\mu\alpha\nu}),
\end{align}

having used the antisymmetric properties of the torsion and of the contorsion tensor of Eqs \eqref{proprietà torsione} and \eqref{eq: proprietà contorsione}.

Using Eqs$\thinspace$\eqref{eq: T1}-\eqref{eq: T3} in Eq.$\thinspace$\eqref{eq: RTT}, one has
\begin{equation}\label{eq: TEGR1}
	\accentset{\circ}{R}=\frac{1}{4}(-\hat{T}^{\alpha\mu\nu}\hat{T}_{\alpha\mu\nu}-2\hat{T}_{\alpha\mu\nu}\hat{T}^{\mu\alpha\nu})+\hat{T}^{\alpha}\hat{T}_{\alpha}-2\accentset{\circ}{\nabla}_{\mu}\hat{T}^{\mu}
\end{equation}
where the sum of the first three terms in the RHS of Eq.$\thinspace$\eqref{eq: TEGR1} is equal to the torsion scalar of Eq. \eqref{eq: torsion scalar}, if the free parameters $a_i$ (with $i=1, 2, 3$) are equal to 1. Using the following notation to refer to the torsion scalar with the above fixed parameters
\begin{equation}\label{eq: TSFP}
	\hat{T}=\frac{1}{4}(-\hat{T}^{\alpha\mu\nu}\hat{T}_{\alpha\mu\nu}-2\hat{T}_{\alpha\mu\nu}\hat{T}^{\mu\alpha\nu})+\hat{T}^{\alpha}\hat{T}_{\alpha},
\end{equation}
the relation \eqref{eq: TEGR1} can be recast as 

\begin{equation}\label{eq: TEGR}
	\accentset{\circ}{R}=\hat{T}-\tilde{B}
\end{equation}

where $\tilde{B}$ is the boundary term
\begin{equation}\label{eq: TEGR boundary term}
	\tilde{B}=2\accentset{\circ}{\nabla}_{\mu}\hat{T}^{\mu}=\frac{2}{\sqrt{-g}}\partial_{\mu}(\sqrt{-g}\hat{T}^{\mu}).
\end{equation}

Then, Eq. \eqref{eq: TEGR} clearly shows that, in the framework of metric teleparallel geometries, a dynamically equivalent theory to GR, the above defined TEGR, can be formulated. Indeed, the application of the variational principle to the Einstein-Hilbert action
\begin{equation}\label{eq:GR action}
	S_{\text{EH}}=\int d^4x \sqrt{-g} \accentset{\circ}{R}
\end{equation}
leads to the same field equations that are instead obtained by varying the TEGR action
\begin{equation}\label{eq:TEGR action}
	S_{\text{TEGR}}=\int d^4x \sqrt{-g}\hat{T}
\end{equation}
where the boundary term \eqref{eq: TEGR boundary term} has been neglected.
\\\\Besides the TEGR case, one may similarly introduce another equivalent theory to GR, which confers a description solely in terms of  non-metricity tensor, belonging to the class of symmetric teleparallel theories of gravity.
In such theories, the linear affine connection \eqref{eq: 3.10} reduces to the following expression:

\begin{equation}\label{3.18}
	\stg{\Gamma}^{\alpha}_{\phantom{\alpha}\mu\nu}=\accentset{\circ}{\Gamma}^{\alpha}_{\phantom{\alpha}\mu\nu}+\stg{L}^{\alpha}_{\phantom{\alpha}\mu\nu},
\end{equation}
since the torsion tensor is trivial.
\\Using equation \eqref{3.18}, the Riemann tensor \eqref{eq: 3.15} becomes
\begin{align}\label{eq: STT Riemann}
	R^{\alpha}_{\phantom{\alpha}\beta\mu\nu}
	&=\accentset{\circ}{R}^{\alpha}_{\phantom{\alpha}\beta\mu\nu}+\accentset{\circ}{\nabla}_{\mu}\stg{L}^{\alpha}_{\phantom{\alpha}\nu\beta}-\accentset{\circ}{\nabla}_{\nu}\stg{L}^{\alpha}_{\phantom{\alpha}\mu\beta}+\nonumber\\
	&+\stg{L}^{\alpha}_{\phantom{\alpha}\mu\sigma}\stg{L}^{\sigma}_{\phantom{\sigma}\nu\beta}
	-\stg{L}^{\alpha}_{\phantom{\alpha}\nu\sigma}\stg{L}^{\sigma}_{\phantom{\sigma}\mu\beta}\equiv 0,
\end{align}
because of the vanishing curvature constraint.
\\Performing the  contractions in Eq.$\thinspace$\eqref{eq: STT Riemann}, one finds the expression of the Ricci scalar to be
\begin{align}\label{3.19}
	\accentset{\circ}{R}+\accentset{\circ}{\nabla}_{\mu}(\stg{L}^{\mu}-\stg{\tilde{L}}^{\mu})+\stg{\tilde{L}}_{\mu}\stg{L}^{\mu}
	-\stg{L}_{\mu\nu\sigma}\stg{L}^{\sigma\mu\nu}=0
\end{align}
where 
\begin{align}
	\stg{L}^{\mu}=\stg{L}^{\mu}{}_{\nu\nu}&=\frac{1}{2}g^{\nu\rho}(\stg{Q}^{\mu}_{\phantom{\mu}\rho\nu}-\stg{Q}{_{\rho\phantom{\mu}\nu}^{\phantom{\rho}\mu}}
 -\stg{Q}_{\nu\phantom{\mu}\rho}^{\phantom{\nu}\mu})=\nonumber\\
 &=\frac{1}{2}(\stg{Q}^{\mu}-2\stg{\tilde{Q}}^{\mu}),
\end{align}
\begin{align}
	\stg{\tilde{L}}{^{\mu}}&=\stg{L}{^{\nu\mu}_{\phantom{\nu\mu}\nu}}=\stg{L}^{\nu\phantom{\nu}\mu}_{\phantom{\nu}\nu}=\nonumber\\
    &=\frac{1}{2}g^{\mu\rho}(\stg{Q}{^{\nu}_{\phantom{\nu}\nu\rho}}-\stg{Q}{_{\nu\phantom{\nu}\rho}^{\phantom{\nu}\nu}}-\stg{Q}{_{\rho\phantom{\nu}\nu}^{\phantom{\rho}\nu}})=-\frac{1}{2}\stg{Q}^{\mu}.
\end{align}
Finally, rewriting the last three terms in the LHS of Eq.$\thinspace$\eqref{3.19} in terms of non-metricity, one gets:

\begin{equation}\label{3.24}
	\accentset{\circ}{\nabla}_{\mu}(\stg{L}^{\mu}-\stg{\tilde{L}}^{\mu})=\overset{\circ}{\nabla}_{\mu}(\stg{Q}{^{\mu}}-\stg{\tilde{Q}}{^{\mu}})
\end{equation}
\begin{align}
	&\stg{\tilde{L}}{_{\mu}}\stg{L}{^{\mu}}-\stg{L}{_{\mu\nu\sigma}}\stg{L}{^{\sigma\mu\nu}}=
	-\frac{1}{4}\stg{Q}{_{\mu}}(\stg{Q}{^{\mu}}-2\stg{Q}{^{\mu}})
	-\frac{1}{4}(\stg{Q}{_{\mu\nu\sigma}}+\nonumber\\
    &-\stg{Q}{_{\nu\mu\sigma}}-\stg{Q}{_{\sigma\mu\nu}})(\stg{Q}{^{\sigma\mu\nu}}-\stg{Q}{^{\mu\sigma\nu}}-\stg{Q}{^{\nu\sigma\mu}})=\nonumber\\
	&=-\frac{1}{4}\stg{Q}{_{\mu}}\stg{Q}{^{\mu}}+\frac{1}{2}\stg{Q}{_{\mu}}\stg{\tilde{Q}}{^{\mu}}-
	\frac{1}{2}\stg{Q}{_{\mu\nu\sigma}}\stg{Q}{^{\nu\mu\sigma}}+\frac{1}{4}\stg{Q}{_{\mu\nu\sigma}}\stg{Q}{^{\mu\nu\sigma}}.\label{3.25}
\end{align}

In this case, it is possible to recognize Eq.$\thinspace$\eqref{3.25} as the opposite of the non-metricity scalar $\mathbb{Q}$ \eqref{eq: non-metricity scalar}, when the coefficients $b_i, i=1,..., 5$ are equal to 1. Thus, the non-metricity scalar becomes:
\begin{equation}\label{3.26}
	\stg{Q}=-\frac{1}{4}\stg{Q}{_{\mu\nu\sigma}}\stg{Q}{^{\mu\nu\sigma}}+\frac{1}{2}\stg{Q}{_{\mu\nu\sigma}}\stg{Q}{^{\nu\mu\sigma}}
	+\frac{1}{4}\stg{Q}{_{\mu}}\stg{Q}{^{\mu}}-\frac{1}{2}\stg{Q}{_{\mu}}\stg{\tilde{Q}}{^{\mu}}.
\end{equation}
Finally, gathering together Eqs \eqref{3.19}, \eqref{3.24}, and \eqref{3.25} one can express the Ricci tensor in the symmetric teleparallel framework:

\begin{equation}\label{3.27}
	\accentset{\circ}{R}=\stg{Q}-B
\end{equation}
where $B$ is a term of pure divergence (i.e. a boundary term) defined as
\begin{equation}
	B=\overset{\circ}{\nabla}_{\mu}(\stg{Q}{^{\mu}}-\stg{\tilde{Q}}{^{\mu}})=\frac{1}{\sqrt{-g}}\partial_{\mu}[\sqrt{-g}(\stg{Q}{^{\mu}}-\stg{\tilde{Q}}{^{\mu}})].
\end{equation}

As in the TEGR case, Eq.$\thinspace$\eqref{3.27} leads to the dynamical equivalence between GR and the above mentioned  STEGR. In fact, the Einstein-Hilbert action leads to the same field equations as the ones derived from
\begin{equation}\label{STEGR action}
	S_{\textrm{STEGR}}=\int d^4x \sqrt{-g}\thinspace\stg{Q}.
\end{equation}

Both TEGR and STEGR are characterized by a flat connection that leads to a trivial curvature tensor. The simplest example of a flat connection is  $\tilde{\Gamma}^{\alpha}_{\phantom{\alpha}\mu\nu}=0$ and if a coordinate transformation is performed starting from a vanishing connection, as the curvature tensor continues to be zero, the transformed affine connection will remain to be flat.\\
The general linear affine connection coefficients, Eq. \eqref{eq: 3.10}, transform under a general coordinate map $\xi^{\lambda}\to x^{\lambda}$ as:
\begin{equation}
    \Gamma^{\rho}{}_{\mu\nu}(x^\lambda) = \frac{\partial x^{\rho}}{\partial \xi^{\gamma}} \frac{\partial \xi^{\alpha}}{\partial x^{\mu}} \frac{\partial \xi^\beta}{\partial x^{\nu}} \Gamma^{\gamma}{}_{\alpha \beta}(\xi^{\lambda}) + \frac{\partial^{2} \xi^{\alpha}}{\partial x^{\mu} \partial x^{\nu}} \frac{\partial x^{\rho}}{\partial \xi^{\alpha}}.\label{eq:transf coordinate CG}
\end{equation}
Thus, the transformed affine connection 
\begin{equation} \label{eq:SUBS_STEGR}
\Gamma^\alpha_{\ \mu\nu}=\frac{\partial x^\alpha}{\partial \xi^\lambda}\partial_\mu\partial_\nu \xi^\lambda
\end{equation}
is still flat.
The flatness condition restricts the connection to be purely inertial and Eq. \eqref{eq:SUBS_STEGR} can be parameterized by the general element $\Lambda^{\alpha}_{\phantom{\alpha}\beta}$ of $GL(4,\mathbb{R})$ as follows:
\begin{equation}\label{3.45}
	\Gamma^{\alpha}_{\phantom{\alpha}\mu\nu}=(\Lambda^{-1})^{\alpha}_{\phantom{\alpha}\lambda}\partial_{\mu}\Lambda^{\lambda}_{\phantom{\lambda}\nu}, \quad\textrm{with}\thinspace\Lambda\in GL(4,\mathbb{R})
\end{equation}
because of the independence of the details of the transformation \cite{JHK}, see Sec \ref{sec: Teleparallel framework}.
\\ Now we want to investigate the consequences of combining the above discussed constraint with either the metric compatibility or the torsionless condition.
In the first case, one proceeds through the substitution of Eq.$\thinspace$\eqref{3.45} in Eq.$\thinspace$\eqref{eq: 3.7} and imposes it to be null, obtaining
\begin{equation}\label{3.46}
	\partial_{\alpha}g_{\mu\nu}=(\Lambda^{-1})^{\rho}_{\phantom{\rho}\lambda}[g_{\rho\nu}\partial_{\alpha}\Lambda^{\lambda}_{\phantom{\lambda}\mu}+g_{\rho\mu}\partial_{\alpha}\Lambda^{\lambda}_{\phantom{\lambda}\nu}].
\end{equation}
Eq. \eqref{3.46} relates the metric tensor to the affine-connection and it gives a description in terms of the 16-component matrix $\Lambda$ instead of the 10 components $g_{\mu\nu}$. It turns out that this redundancy is linked to the possibility of performing local Lorentz transformations under which the theory is invariant \cite{fQ}.
\\ Finally, the following expression of the torsion tensor can be outlined
\begin{equation}\label{eq: t}
	T^{\alpha}_{\phantom{\alpha}\mu\nu}=(\Lambda^{-1})^{\alpha}_{\phantom{\alpha}\lambda}\partial_{\mu}\Lambda^{\lambda}_{\phantom{\lambda}\nu}-(\Lambda^{-1})^{\alpha}_{\phantom{\alpha}\lambda}\partial_{\nu}\Lambda^{\lambda}_{\phantom{\lambda}	\mu}
\end{equation}
where the relation \eqref{3.45} has been employed.
\\Still referring to the discussion in Sec. \ref{sec: Teleparallel framework}, it is particularly interesting to notice  from the comparison between Eq.$\thinspace$\eqref{eq: t} and Eq.$\thinspace$\eqref{eq: tttn} that the transformation of the vanishing connection (leading to the purely inertial affine connection \eqref{3.45}) generates the vierbein in the Weitzenb\"{o}ck gauge, namely, in the one such that the spin connection vanishes identically \cite{JHK}.
\\\\Instead, if the torsionless condition is implemented, the $\Lambda$ matrices, defining the flat connection of Eq. \eqref{3.45}, satisfy the following relation

\begin{equation}
	\partial_{\mu}\Lambda^{\lambda}_{\phantom{\lambda}\nu}-\partial_{\nu}\Lambda^{\lambda}_{\phantom{\lambda}\mu}=0.
\end{equation}

Such an equality is trivially realized introducing functions of the $x$ coordinates $\xi^{\mu}=\xi^{\mu}(x^{\nu})$ \cite{TSTformalism,DAmbrosio2022} and defining 
\begin{equation}\label{3.48}
	\Lambda^{\mu}_{\phantom{\mu}\nu}=\partial_{\nu}\xi^{\mu}.
\end{equation}
The flat and torsionless connections are thus obtained by plugging Eq.$\thinspace$\eqref{3.48} into Eq.$\thinspace$\eqref{3.45}
\begin{equation}\label{eq: coincident gauge}
	\Gamma^{\alpha}_{\phantom{\alpha}\mu\nu}=\frac{\partial x^{\alpha}}{\partial \xi^{\lambda}}\partial_{\mu}\frac{\partial \xi^{\lambda}}{\partial x^{\nu}}.
\end{equation}
It is then clear that the connection ascribed to the symmetric teleparallel theories can be removed by dint of a diffeomorphism. A null connection is, for example, obtained when setting
\begin{equation}
	\xi^{\mu}=x^{\mu}
\end{equation}
or the more general relation
\begin{equation}
	\xi^{\mu}=M^{\mu}_{\phantom{\mu}\nu}x^{\nu}+\xi_c^{\mu}
\end{equation}
where $M$ is an invertible matrix with constant components and $\xi_c^{\mu}$ are constants \cite{DAmbrosio2022}. 
\\Such a specific choice is referred to as \textit{coincident gauge} and its availability is always guaranteed when imposing the curvatureless and torsionless postulates.

\begin{figure*}[ht!]
\includegraphics[trim=0cm 4cm 0cm 5cm, scale=0.15]{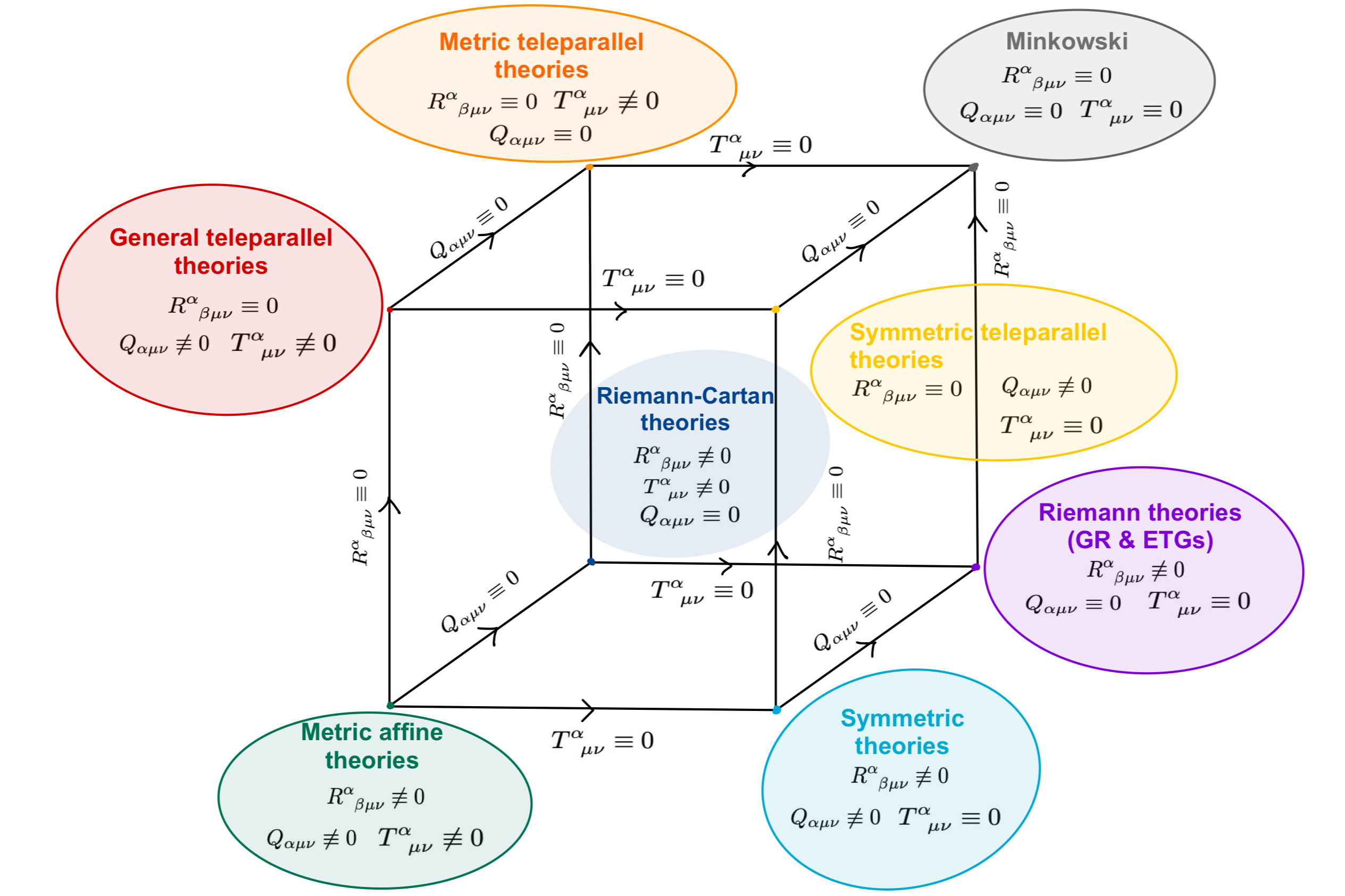}
\centering
\caption{A comprehensive picture of Affine Theories of Gravity, considering their geometric invariant and dynamical variables. See also \cite{Teleparallel, CdFF}.}
\label{Figure Cubo}
\end{figure*}

A summary of Affine Theories of Gravity with respect to their dynamical variables is reported in Fig.\ref{Figure Cubo}.

\section{The Extended Geometric Trinity of Gravity}\label{Section 3.3}

Recent discoveries, as the accelerated expansion of the Universe and the structure formation, have increased the interest of the scientific community in extending Einstein's theory of gravity. A very popular extension is the so-called $f(\accentset{\circ}{R})$ gravity, where the Lagrangian is a function of the Ricci scalar \cite{Capozziello:2002rd, Sotiriou2010, FC}. As discussed in Sec. \ref{sec: Trinity}, GR can be formulated in terms of other two dynamical variables, namely torsion (TEGR) and non-metricity (STEGR). Each formulation of the trinity gravity can be derived from a variational principle, Eqs \eqref{eq:GR action}, \eqref{eq:TEGR action}, and \eqref{STEGR action}, and even if the actions are not the same, they provide the equivalent field equations, with the same DoFs, and the same solutions \cite{JHK, ComparingEG}. This led to consider generalizations of the GTG, in the spirit of $f(\accentset{\circ}{R})$ gravity, replacing the scalars $\accentset{\circ}{R}$, $\hat{T}$, and $\stg{Q}$, with arbitrary smooth functions in the respective actions \cite{Ferraro06,Martin2019,Teleparallel,fQ,Carloni:2023egi}. \\ In this section, we discuss  the EGTG, considering the dynamical equivalent formulations of $f(\accentset{\circ}{R})$ with respect to torsion, $f(\hat{T}, \tilde{B})$, and  non-metricity, $f(\stg{Q}, B)$. In particular, we study the respective field equations, considering also the $f(\hat{T})$ and $f(\stg{Q})$ cases \cite{CdFF, Bahamonde15, Capozziello2019GW, fQ,fQB}.

\subsection{Extended Metric Gravity}\label{subsec: f(R)}

As said in the Introduction \ref{sec:Intro}, the straightforward   extension of  GR is  considering  $f(\accentset{\circ}{R})$ gravity, in which the Lagrangian density is an arbitrary function of the Ricci scalar $\accentset{\circ}{R}$. However, there are two approaches to derive the field equations using the variational method of $f(\accentset{\circ}{R})$ gravity, depending on the formalism we use \cite{SF, DeFelice10}. The first is the \textit{metric formalism} which is based on the initial assumption that the metric tensor and the connection are related, and thus the variation of the action is performed only with respect to $g_{\mu\nu}$.
Note that, in this case, we are considering only torsion-free theories. The other possibility consists in considering both metric and affine connection as independent quantities and varying the action with respect to both of them. Whereas in GR this process  leads to the same equations of motion as those obtained within the metric formalism, the application of different variational principles to $S_{f(\accentset{\circ}{R})}$  leads to different results and consequently one may distinguish  between distinct versions of $f(\accentset{\circ}{R})$ gravity: the \textit{metric} and the \textit{Palatini} $f(R)$ gravity, respectively obtained in the metric and Palatini formalism (i.e. metric affine formalism). See \cite{Allemandi:2004yx} for details.  However, in this section, we will take into account the metric formalism.\\
With these considerations in mind, one may write the $f(\mathring{R})$ action as
\begin{equation}\label{eq: A.1}
	S_{f(\mathring{R})}=\int d^4x [\sqrt{-g}f(\mathring{R})+2\chi \mathcal{L}_{m}]
\end{equation}
whose variation with respect to the metric leads to the desired field equations. Indeed
\begin{align}\label{eq: A.2}
	\delta_g S_{f(\mathring{R})}&=\underbrace{\int d^4x \delta_g (\sqrt{-g})f(\mathring{R})}_{(A)}+\underbrace{\int d^4x \sqrt{-g}\delta_g[f(\mathring{R})]}_{(B)}+\nonumber\\
    &+\underbrace{2\chi\int d^4x \delta_g\mathcal{L}_m}_{(C)}=0
\end{align}
where the term (A) can be immediately rewritten as follows
\begin{equation}\label{eq: A.3}
	\textrm{(A)}=-\frac{1}{2}\int d^4x f(\mathring{R})g_{\mu\nu}\sqrt{-g}\delta_g g^{\mu\nu}
\end{equation}
considering the validity of the following chains of equalities
\begin{align}
	\delta g&=gg^{\mu\nu}\delta_g g_{\mu\nu}=-gg_{\mu\nu}\delta_g g^{\mu\nu}\\
	\delta_g (\sqrt{-g})&=-\frac{1}{2\sqrt{-g}}\delta g=-\frac{1}{2}\sqrt{-g}g_{\mu\nu}\delta_gg^{\mu\nu}\label{eq: eg},
\end{align}
the term (B) can instead be momentarily recast as
\begin{equation}\label{eq: A.6}
	\textrm{(B)}
    =\int d^4x f_{\mathring{R}}(\mathring{R}) \sqrt{-g} [\mathring{R}_{\mu\nu}\delta_g g^{\mu\nu}+g^{\mu\nu}\delta_g \mathring{R}_{\mu\nu}]
\end{equation}
and (C), containing the matter part of the Lagrangian $\mathcal{L}_{m}$, can eventually be outlined in the following convenient form
\begin{equation}\label{Matter}
	(C)=-\chi \int d^4x  T_{\mu\nu}\sqrt{-g}\delta_g g^{\mu\nu}
\end{equation}
with 
\begin{equation}\label{eq: S-ET}
	T_{\mu\nu}=-\frac{2}{\sqrt{-g}}\frac{\delta_g\mathcal{L}_{m}}{\delta_g g^{\mu\nu}}.
\end{equation}
Here, $f_{\mathring{R}}(\mathring{R})$ is the total derivative with respect to 
$\mathring{R}$.

In the Appendix \ref{appedice f(R)} all the detailed calculation are exposed. Using Eqs.$\thinspace$\eqref{eq: A.2}, \eqref{eq: A.3}, \eqref{Matter} and \eqref{eq: A.14}, we achieve the $f(\mathring{R})$ field equations
\begin{equation}\label{eq: f(R) field equations}
	f_{\mathring{R}}(\mathring{R})\mathring{R}_{\mu\nu}-\frac{1}{2}g_{\mu\nu}f(\mathring{R})-\mathring{\nabla}_{\mu}\mathring{\nabla}_{\nu}f_{\mathring{R}}(\mathring{R})+\mathring{\Box}f_{\mathring{R}}(\mathring{R})g_{\mu\nu}=\chi T_{\mu\nu}.
\end{equation}
These equations can be written in the Einstein-like form by introducing the Einstein tensor $G_{\mu\nu}=\accentset{\circ}{R}_{\mu\nu}-\frac{1}{2}g_{\mu\nu}\accentset{\circ}{R}$, that is 
\begin{align}
    f_{\mathring{R}}(\accentset{\circ}{R}_{\mu\nu}-\frac{1}{2}g_{\mu\nu}\accentset{\circ}{R} + \frac{1}{2}g_{\mu\nu}\accentset{\circ}{R})-\frac{1}{2}g_{\mu\nu}f(\mathring{R})&\\ \notag -\mathring{\nabla}_{\mu}\mathring{\nabla}_{\nu}f_{\mathring{R}}(\mathring{R})+\mathring{\Box}f_{\mathring{R}}(\mathring{R})g_{\mu\nu}=\chi T_{\mu\nu}, 
\end{align}
obtaining 
\begin{align}\label{eq: f(R) in GR}
    G_{\mu\nu}=\frac{1}{f_{\mathring{R}}}\bigg[ \mathring{\nabla}_{\mu}\mathring{\nabla}_{\nu}f_{\mathring{R}}(\mathring{R})-\mathring{\Box}f_{\mathring{R}}(\mathring{R})g_{\mu\nu} &\\ \notag + g_{\mu\nu}\frac{f(\mathring{R})-f_{\mathring{R}}(\mathring{R})\mathring{R}}{2}\bigg] + 
    \chi T_{\mu\nu}
\end{align}
We can interpret the first terms of the RHS of Eq. \eqref{eq: f(R) in GR} as an effective stress-energy tensor, $T^{\textit{eff}}_{\mu\nu}$, sourcing the effective Einstein equations \cite{FC}.
Furthermore, the trace of Eq. \eqref{eq: f(R) in GR} is:
\begin{equation}\label{eq: traccia f(R)}
    3\mathring{\Box}f_{\mathring{R}}(\mathring{R})-2f(\mathring{R})+\mathring{R}f_{\mathring{R}}(\mathring{R})=\chi T\,,
\end{equation}
corresponding to a Klein-Gordon equation. It is worth noticing that, with respect to GR, here the trace is a dynamical equation.
The $f(\mathring{R})$ gravity is characterized by fourth order field equations, as consequence there are more different solutions than the GR case \cite{FC}. 
For instance, by choosing the special case $f(\mathring{R})=\mathring{R}$ and $f_{\mathring{R}}(\mathring{R})=1$ in  Eq. \eqref{eq: f(R) in GR}, it is possible to recover the GR field equations:

\begin{equation}
    G_{\mu\nu} = \chi T_{\mu\nu}.
\end{equation}

In this case, the trace of Eq. \eqref{eq: traccia f(R)} reduces to $\mathring{R}=\chi T$ and hence the Ricci scalar $\mathring{R}$ is directly determined by the matter. In $f(\mathring{R})$ gravity, there is a propagating scalar DoF, $\mathring{\Box}f_{\mathring{R}}$, represented by $\phi=f_{\mathring{R}}(\mathring{R})$, whose dynamics is determined by the trace equation \eqref{eq: traccia f(R)}; this scalar field $\phi$ is the so-called \qm{scalaron} \cite{STAROBINSKY80}.\\
In general, the standard procedure to derive the gravitational waves is  considering the linearized theory, perturbing around the Minkowski background, obtaining small metric perturbations \cite{Maggiore07}:
\begin{equation}
    g_{\mu\nu}=\eta_{\mu\nu} + h_{\mu\nu}.
\end{equation}
Thus, one has to verify which are the components of the Riemann tensor physically relevant. In the linearized theory, it is possible to show that there are six algebraically independent components. Considering the propagation of the gravitational waves in the $z$ direction, we can recognize six polarization modes: plus (+), cross (x), breathing (b), longitudinal (l), vector-x (x), and vector-y (y) modes. The plus and the cross modes are tensor-type (spin 2) gravitational waves, the vector-x and vector-y modes are vector-type (spin 1) gravitational waves, and the breathing and longitudinal modes are scalar-type (spin 0) gravitational waves \cite{Bogdanos:2009tn}.
In GR, the gravitational waves have only two polarization states,
the plus and cross modes. However, when we consider the case of extended theories of gravity, we can have more non-null components of the Riemann tensor and hence of polarization modes can be greater than two. 
The difference between gravitational waves in $f(\accentset{\circ}{R})$ and in GR is caused by the extra scalar degree of freedom contained in the non-vanishing trace of the field equation of $f(\accentset{\circ}{R})$ \eqref{eq: traccia f(R)}. In fact, Eq. \eqref{eq: traccia f(R)} can be rewritten with respect to the scalaron $\phi$:
\begin{equation}
    \phi\to f_{\mathring{R}} \mbox{    and    }\frac{dV_{eff}}{d\phi} \to \frac{2f(\mathring{R})-\mathring{R}f_{\mathring{R}}(\mathring{R}) - \chi \rho}{3},
\end{equation}
obtaining the Klein-Gordon equation:
\begin{equation}
    \mathring{\Box} \phi = \frac{dV_{eff}}{d\phi}.
\end{equation}

Using the weak field approximation on the field equations, several studies investigated the additional polarization modes in all versions of $f(\accentset{\circ}{R})$
theories \cite{DeLaurentis08,Vignolo09,Vignolo10,Corda07,Corda2007,Taishi19}, and it has been shown that exists only one massive longitudinal (scalar) mode besides the two massless tensor modes of  GR \cite{Yang11,Jetzer09}.

\subsection{Extended Teleparallel Gravity}\label{subsec: f(T,B)}

Turning now the attention to the derivation of the $f(\hat{T}, \tilde{B})$ metric field equations, it is worth noticing that the content of the calculations reported below is strictly dependent on the tetrad formulation, introduced in Sec. \ref{sec: Teleparallel framework}.
The $f(\hat{T}, \tilde{B})$ action can be written as
\begin{equation}\label{eq: f(T,B) azione}
	S_{f(\hat{T}, \tilde{B})}=\int d^4x [ef(\hat{T}, \tilde{B})+2\chi\mathcal{L}_{\text{m}}].
\end{equation}
In order to derive the $f(\hat{T}, \tilde{B})$ field equations, we have to introduce an equivalent expression of the torsion scalar defined in Eq.$\thinspace$\eqref{eq: TSFP}:
\begin{equation}
	\hat{T}=-\hat{S}_{a}^{\phantom{a}\mu\nu}\hat{T}^{a}_{\phantom{a}\mu\nu}
\end{equation}
which strictly depends on the so-dubbed superpotential:
\begin{equation}\label{eq: superpotential_a}
	\hat{S}_{a}^{\phantom{a}\mu\nu}=\frac{1}{2}[\hat{K}^{\mu\phantom{a}\nu}_{\phantom{\mu}a}+e_{a}^{\phantom{a}\mu}\hat{T}^{\nu}-e_{a}^{\phantom{a}\nu}\hat{T}^{\mu}].
\end{equation}
The detailed calculations have been reported in the Appendix \ref{appendice f(T,B)}, from which we can derive the $f(\hat{T}, \tilde{B})$ field equations:

\begin{align}\label{eq: equazioni di campo f(T,B)}
	&g_{\mu\nu}[-f(\hat{T}, \tilde{B})+(f_{\tilde{B}}+f_{\hat{T}})\tilde{B}]+2\accentset{\circ}{R}_{\mu\nu}f_{\hat{T}}-4[(\partial^{\rho}f_{\hat{T}})+\nonumber\\
 &+(\partial^{\rho}f_{\tilde{B}})]\hat{S}_{\mu\rho\nu}
	+2\mathring{\nabla}_{\mu}\mathring{\nabla}_{\nu}f_{\tilde{B}}-2g_{\mu\nu}\mathring{\Box}f_{\tilde{B}}=2\chi T_{\mu\nu}.
\end{align}

The $f(\mathring{R})$ gravity is recovered in this model if we consider Eq. \eqref{eq: TEGR}, which allows to our function $f$ to assume the particular form
\begin{equation}\label{eq: f(R)=f(T,B)}
    f(\hat{T}, \tilde{B}) = f(\hat{T}-\tilde{B})= f(\accentset{\circ}{R})
\end{equation}
This lead to the following  condition:

\begin{align}\label{eq: f(T-B)}
	f(\accentset{\circ}{R})&=f(\hat{T}-\tilde{B})\to \notag \\ f_{\accentset{\circ}{R}}(\accentset{\circ}{R})&=f_{\hat{T}}=-f_{\tilde{B}}.
\end{align}

Therefore, if we insert the conditions \eqref{eq: f(R)=f(T,B)} and \eqref{eq: f(T-B)} in the $f(\hat{T}, \tilde{B})$ field equations, Eq. \eqref{eq: equazioni di campo f(T,B)}, our model  reduces to the $f(\accentset{\circ}{R})$ gravity \eqref{eq: f(R) field equations}, proving the dynamical equivalence between the two theories.
It is possible to derive the gravitational waves also for $f(\hat{T}, \tilde{B})$ gravity. In the Weak Field Limit, they result to be equivalent to $f(\accentset{\circ}{R})$ gravity, showing three polarizations: the two standard of GR, plus and cross, which are purely transverse with two-helicity, massless tensor polarization modes, and an additional massive scalar mode with zero-helicity. The boundary term $\tilde{B}$ excites the extra scalar polarization and the mass of scalar field adds a new degree of freedom, namely a single mixed scalar polarization \cite{Capozziello2019GW}.

\subsection{$f(T)$ gravity}\label{subsec: f(T) gravity}

In analogy to $f(\accentset{\circ}{R})$, one can consider the $f(\hat{T})$ gravity:
\begin{equation}
     S_{f(\hat{T})}=\int d^4x [e f(\hat{T})] \label{eq:f(T) action},
\end{equation}
in which the boundary term $\tilde{B}$ has been neglected.\\
Since, torsion scalar is only built of first derivatives of the tetrads, we will have second-order field equations \cite{FC, Teleparallel, Cai2016}. Furthermore, considering an $f(\hat{T})$ gravity means to neglect Eq. \eqref{eq: TEGR}, since the torsion tensor does not differ from the Ricci tensor up to a divergence boundary term, and, as consequence $f(\hat{T})$ will not be different from $f(\accentset{\circ}{R})$ by a total derivative term.
Therefore, let us choose a function $f$ not dependent on the boundary term $B$, i.e.:
\begin{equation}
    f(\hat{T}, \tilde{B})= f(\hat{T})
\end{equation}
In this case, we have $f_{\tilde{B}}=0$ and the $f(\hat{T}, \tilde{B})$ field equations, given in Eq. \eqref{eq: equazioni di campo f(T,B)}, become:

\begin{equation}
    -g_{\mu\nu}f(\hat{T})+2\accentset{\circ}{R}_{\mu\nu}f_{\hat{T}}   
    -4(\partial^{\rho}f_{\hat{T}})\hat{S}_{\mu\rho\nu}	=2\chi T_{\mu\nu}.
\end{equation}
 or, in the standard form, using Eqs \eqref{3.55} and \eqref{3.56}:
 \begin{align}\label{eq: f(T,B) letteratura}
     -e\delta^{\lambda}_{\nu}f(\hat{T})+4ef_{\hat{T}}\hat{S}_{\rho}{}^{\lambda\mu}T^{\rho}{}_{\mu\nu}-4f_{\hat{T}}\partial_{\mu}(e S_{c}{}^{\mu\lambda})& \\-4e\partial_{\mu}f_{\hat{T}}S_{\nu}{}^{\mu\lambda}=2\chi e T_{\nu}{}^{\lambda}
 \end{align}
where the last term on the LHS can be written as $4e\partial_{\mu}f_{\hat{T}}S_{\nu}{}^{\mu\lambda}=4ef_{\hat{T}\hat{T}}(\partial_{\mu}T)S_{\nu}{}^{\mu\lambda}$. \\
If we choose $f(\hat{T})=\hat{T}$ Eq. \eqref{eq: f(T,B) letteratura} reduces to the TEGR case \cite{ComparingEG, Barrow11}. Therefore, for $f(\hat{T})=\hat{T}$, GR is recovered and the equivalence among GR and TEGR is restored. In this case, the field equations are covariant, being identically recovered from the second Bianchi identity and the theory is also local Lorentz invariant \cite{ComparingEG}. \\
On the other hand, in the general case $f(\hat{T})\neq \hat{T}$, the field equations are not invariant under a local Lorentz transformation \cite{Barrow11}, due to the imposition of a vanishing spin connection. This assumption has the good motivation to simplify the derivation of the solutions, since the independent components of the Riemann tensor and of the spin connections are reduced and hence the latter can be set to zero through a local Lorentz transformation \cite{Cai2016}. However, in $f(\hat{T})$ gravity, this assumption makes the theory frame-dependent since a solution of the field equations depends on the choice of the frame. As consequence, in order to avoid the dependence on the frame, $f(\hat{T})$ has to be covariant, which means that we need to find the appropriate spin connection to the tetrad \cite{Krššák_2016}.  

It is possible also to investigate gravitational waves in this theory. In particular, considering the Minkowskian limit and a class of analytic $f(\hat{T})$ functions in the Lagrangian, one has that the
gravitational wave modes in $f(\hat{T})$ gravity are the same
as those in GR \cite{BAMBA13}. Since in the perturbed equations the terms of the scalar field are of  zeroth order, it does not propagate at the
first order of perturbations. As consequence, the Einstein equation of GR
with a cosmological constant proportional to $f(\hat{T} = 0)$
can be recovered; the cosmological constant can exist only if $f(\hat{T} = 0) \neq 0$ \cite{Cai2016}.

\subsection{Extended Symmetric Teleparallel Gravity}\label{subsec: f(Q,B)}

In the previous section, Sec. \ref{subsec: f(T,B)}, we achieved the equivalence of $f(\mathring{R})$ to a particular case of the $f(\hat{T}, \tilde{B})$ gravity, since our function had to assume the form $f(\hat{T}, \tilde{B})=f(\hat{T}-\tilde{B})$.\\
In this section, we apply the same procedure of Sec. \ref{subsec: f(T,B)} to obtain the equivalence between $f(\accentset{\circ}{R})$ and extended symmetric teleparallel theories with boundary terms $f(\stg{Q},B)$, proving that they reduce to the field equations of $f(\accentset{\circ}{R})$ gravity, Eq. \eqref{eq: f(R) field equations}.
\\The action of a $f(\stg{Q},B)$ theory is 
\begin{equation}\label{eq: azione f(Q,B)}
	S_{f(\stg{Q}, B)}=\int d^4x [\sqrt{-g}f(\stg{Q}, B)+2\chi\mathcal{L}_{m}]
\end{equation}
In order to determine the variation of $\stg{Q}$, we can introduce an alternative way to define the non-metricity scalar \eqref{eq: non-metricity scalar} \cite{fQB}
\begin{equation}
	\stg{Q}=\stg{Q}_{\alpha\mu\nu}\stg{P}^{\alpha\mu\nu},
\end{equation}
considering the \textit{non-metricity conjugate} or \textit{superpotential}
\begin{align}
	\stg{P}^{\alpha}_{\phantom{\alpha}\mu\nu}&=\frac{1}{4}\Big[-\stg{Q}^{\alpha}_{\phantom{\alpha}\mu\nu}+\stg{Q}^{\phantom{\mu}\alpha}_{\mu\phantom{\alpha}\nu}+\stg{Q}^{\phantom{\nu}\alpha}_{\nu\phantom{\alpha}\mu}+(\stg{Q}^{\alpha}-\tilde{\stg{Q}}^{\alpha})g_{\mu\nu}+\nonumber\\
  &\phantom{=\frac{1}{4}}-\frac{1}{2}(\delta^{\alpha}_{\mu}\stg{Q}_{\nu}+\delta^{\alpha}_{\nu}\stg{Q}_{\mu})\Big],
\end{align}
and then proceed with its variation. 
Following the calculations described in  Appendix \ref{appendice f(Q,B)}, we obtain the  $f(\stg{Q},B)$ field equations.
\begin{align}\label{eq: 3.66}
	&f_{\stg{Q}}(\accentset{\circ}{R}_{\mu\nu}-\frac{1}{2}g_{\mu\nu}\accentset{\circ}{R})+2\stg{P}{^{\lambda}_{\phantom{\lambda}\mu\nu}}\accentset{\circ}{\nabla}_{\lambda}(f_{\stg{Q}}+f_B)
    +\accentset{\circ}{\nabla}_{\mu}\accentset{\circ}{\nabla}_{\nu}f_B+\nonumber\\
    &-g_{\mu\nu}\accentset{\circ}{\Box}f_B
	-\frac{1}{2}g_{\mu\nu}(f-f_BB-f_{\stg{Q}}\stg{Q})=\chi T_{\mu\nu}.
\end{align}
We can recover $f(\accentset{\circ}{R})$ gravity considering the equivalence given in Eq. \eqref{3.27},  choosing the function $f$ as
\begin{equation}\label{eq: f(R)=f(q-B)}
    f(\stg{Q},B) = f(\stg{Q}-B) = f(\accentset{\circ}{R}).
\end{equation}
As consequence 
\begin{align}\label{eq: f(Q-B)}
	f(\accentset{\circ}{R})&=f(\stg{Q}-B)\to \notag \\
 f_{\accentset{\circ}{R}}(\accentset{\circ}{R})&=f_{\stg{Q}}=-f_{B}.
\end{align}
It is once again evident from Eqs \eqref{eq: f(R)=f(q-B)} and $\thinspace$\eqref{eq: f(Q-B)} that Eq.$\thinspace$\eqref{eq: 3.66} reduces to Eq.$\thinspace$\eqref{eq: f(R) field equations}, proving the dynamical equivalence of $f(\accentset{\circ}{R})$ and $f(\stg{Q}-B)$.\\
It is worth noticing that the presence of the boundary term $B$ allows to
render the $f(\stg{Q},B)$ gravity to be characterized by fourth order field equations, as the $f(\hat{T}, \tilde{B})$ case \cite{Bahamonde15, Capozziello2019GW, bahamonde2017}.
Furthermore, it is possible to vary the Lagrangian \eqref{eq: azione f(Q,B)} with respect to the connection \eqref{3.18}, (i.e. $\delta_{\Gamma}S_{f(\stg{Q}, B)}=0$), in order to obtain the connection field equations in $f(\stg{Q},B)$ gravity, namely

\begin{equation}\label{eq: f(Q,B) connessione}
    \mathcal{C}_{\alpha}=\nabla_{\mu}\nabla_{\nu}[\stg{P}^{\mu\nu}{}_{\alpha}e (f_{\stg{Q}}(\stg{Q})+f_{B}(B))],
\end{equation}
which represent an additional constraint to be considered.

As in the $f(\hat{T}, \tilde{B})$ case, it is possible to consider gravitational waves generated in $f(\stg{Q}, B)$ gravity. This
theory exhibits gravitational waves regardless of the gauge adopted: they are the standard massless tensors plus a massive scalar gravitational wave like in the case of $\accentset{\circ}{R}$ gravity. It is precisely the boundary term B that generates the massive scalar mode with an effective
mass associated to a Klein-Gordon equation in the linearized boundary term. In the $f(\stg{Q}, B)$ gravity, considering approximate linear
equation, we find three polarization modes: two massless transverse
tensor radiation modes, with helicity equal to 2, reproducing the standard plus and cross modes, and an additional massive scalar wave mode with transverse polarization of zero helicity. These results can be achieve considering both the coincidence gauge and leaving the gauge free. Thus, both $f(\stg{Q}, B)$ and $f(\accentset{\circ}{R})$ gravity have the same massive transverse scalar perturbation \cite{Capriolo2024}.

\subsection{$f(Q)$ gravity}\label{subsec: f(Q) gravity}

Let us describe now the easiest extension of STEGR, based on the non-metricity scalar Eq. \eqref{3.26}, in analogy to $f(\accentset{\circ}{R})$ and $f(\hat{T})$ theories. The $f(\stg{Q})$ gravity is expressed by the following action:
\begin{equation}\label{eq: azione f(Q)}
    S_{f(\stg{Q})}=\int d^4x [\sqrt{-g}f(\stg{Q})+2\chi\mathcal{L}_{m}].
\end{equation}
As in $f(\hat{T})$ case, we will obtain second order field equations, since the non-metricity scalar $\stg{Q}$ is built up on the first derivative of the metric tensor Eq. \eqref{3.26} \cite{fQ, CdFF}. Moreover, if we consider an action only dependent on $\stg{Q}$, the equivalence with respect to GR is no-more valid, Eq. \eqref{3.27}, since the respective Lagragians do not differ up to a boundary term $B$. This means that also the extended symmetric teleparallel case is affected by the absence of $B$, since $f(\accentset{\circ}{R})$ and $f(\stg{Q})$ will not differ by a total derivative term, and thus these theories are no longer equivalent.\\
Let us begin with examining our field equations \eqref{eq: 3.66} when the function $f$ is independent of the boundary term. Thus, we simply set:
\begin{equation}
    f(\stg{Q}, B)=f(\stg{Q}),
\end{equation}
obtaining also $f_B=0$. Doing this, our $f(\stg{Q}, B)$ field equations \eqref{eq: 3.66} become:
\begin{equation}\label{eq: f(Q) field equation}
    f_{\stg{Q}}(\accentset{\circ}{R}_{\mu\nu}-\frac{1}{2}g_{\mu\nu}\accentset{\circ}{R})+2\stg{P}{^{\lambda}_{\phantom{\lambda}\mu\nu}}\partial_{\lambda}f_{\stg{Q}}-\frac{1}{2}g_{\mu\nu}(f-f_{\stg{Q}}\stg{Q})=\chi T_{\mu\nu},
\end{equation}
considering that the second term on RHS of \eqref{eq: f(Q) field equation}  can be written as $2\stg{P}{^{\lambda}_{\phantom{\lambda}\mu\nu}}\partial_{\lambda}f_{\stg{Q}}=2\stg{P}{^{\lambda}_{\phantom{\lambda}\mu\nu}}f_{\stg{Q}\stg{Q}}\partial_{\lambda}\stg{Q}$. 
\\ For $f(\stg{Q})=\stg{Q}$ and $f_{\stg{Q}}=1$, we recover STEGR and, as consequence, GR. As for the TEGR and $f(\hat{T})$ cases, in which we choose the Weitzenb\"ock connection inducing a vanishing spin connection, here, under the teleparallelism and torsion-free constraints, we can always choose the coincident gauge Eq. \eqref{eq: coincident gauge} in which the affine connection locally vanishes. In STEGR,  this is indeed a gauge condition because other choices only contribute a surface term in the action, but when we consider extended STEGR theories, such as $f(\stg{Q})$, the choice of the coincident gauge and of the metric as the only fundamental
variable will produce a different evolution of metric in different coordinate systems. Therefore, we could seek any coordinate
systems which are compatible with coincident gauge, or consider affine connections that are valid for any coordinate systems in extended STEGR theories \cite{Zhao21}. \\ In addition, we also have to take into account the field equations obtained by the variation of the action \eqref{eq: azione f(Q)} with respect to the linear affine connection \eqref{3.18} \cite{fQ,DAmbrosio2022,CdFF}; in this case Eq. \eqref{eq: f(Q,B) connessione} reduces to:
\begin{equation}
    \mathcal{C}_{\alpha}=\nabla_{\mu}\nabla_{\nu}(\stg{P}^{\mu\nu}{}_{\alpha}e f_{\stg{Q}}(\stg{Q})).
\end{equation}

Thus, in $f(\stg{Q})$ gravity, the field equations are no longer equivalent to the $f(\accentset{\circ}{R})$ case, because the connection cannot be absorbed into a boundary term and it encodes additional DoFs, different from the ones of the metric.

It is also possible to consider gravitational waves in  $f(\stg{Q})$ gravity. Here, there are only two massless and tensor modes, whose polarizations exactly reproduce the plus and cross tensor modes of GR. Thus, in the $f(\stg{Q})$ gravity, the gravitational waves behave as those in $f(\hat{T})$ gravity \cite{Capozziello2019GW} and it is not possible to distinguish them from those of GR only by wave polarization measurements. This shows that the situation is different with respect to the curvature-based $f(\accentset{\circ}{R})$ gravity where, as shown in Sec. \ref{subsec: f(R)}, there is always an additional scalar mode for $f(\accentset{\circ}{R}) \neq \accentset{\circ}{R}$ \cite{CAPOZZIELLO2024}.

\section{Geometric extensions as  effective stress-energy tensors}\label{sec: discussion}
An important approach to discuss extensions of gravity theories  is constructing effective stress-energy tensors containing the further geometric DoFs. In this picture, the more \qm{geometry} appear as more \qm{effective matter fields}. We want to conclude this paper considering also this perspective.

In Sec \ref{sec: Trinity}, we obtained the field equations of GTG, comparing their Lagragians and using the identities \eqref{eq: TEGR} and \eqref{3.27}. An alternative way to obtain the dynamical equivalence of GR, TEGR, and STEGR consists in considering the conservation laws. This approach was proposed by Einstein himself and it is based on: \\ \\
(1) the second Bianchi Identity
\begin{equation}\label{eq: BIanchi Identity}
     \accentset{\circ}{\nabla}_{\lambda}\accentset{\circ}{R}^{\alpha}_{\phantom{\alpha}\beta\mu\nu} + \accentset{\circ}{\nabla}_{\mu}\accentset{\circ}{R}^{\alpha}_{\phantom{\alpha}\beta\nu\lambda} + \accentset{\circ}{\nabla}_{\nu}\accentset{\circ}{R}^{\alpha}_{\phantom{\alpha}\beta\lambda\mu} = 0,
\end{equation}
with the identification of a divergenceless tensor $\accentset{\circ}{\nabla}_{\mu} G^{\mu\nu}=0$, the Einstein tensor; \\ \\
(2) the covariant conservation of the stress-energy tensor, $\accentset{\circ}{\nabla}_{\mu} T^{\mu\nu}=0$. 
This led to the Einstein field equations
\begin{equation}\label{eq: Einstein Field Eq}
    G_{\mu\nu}=\accentset{\circ}{R}_{\mu\nu}-\frac{1}{2}g_{\mu\nu}\accentset{\circ}{R}=\kappa T_{\mu\nu},
\end{equation}
Indeed, starting from the second Bianchi Identity for a general linear affine connection, Eq. \eqref{eq: 3.10}, it is possible to infer the field equations in TEGR and STEGR framework; see \cite{ComparingEG} for a detailed discussion.

\begin{figure*}[ht!]
\includegraphics[trim=0cm 5cm 0cm 6cm,scale=0.2]{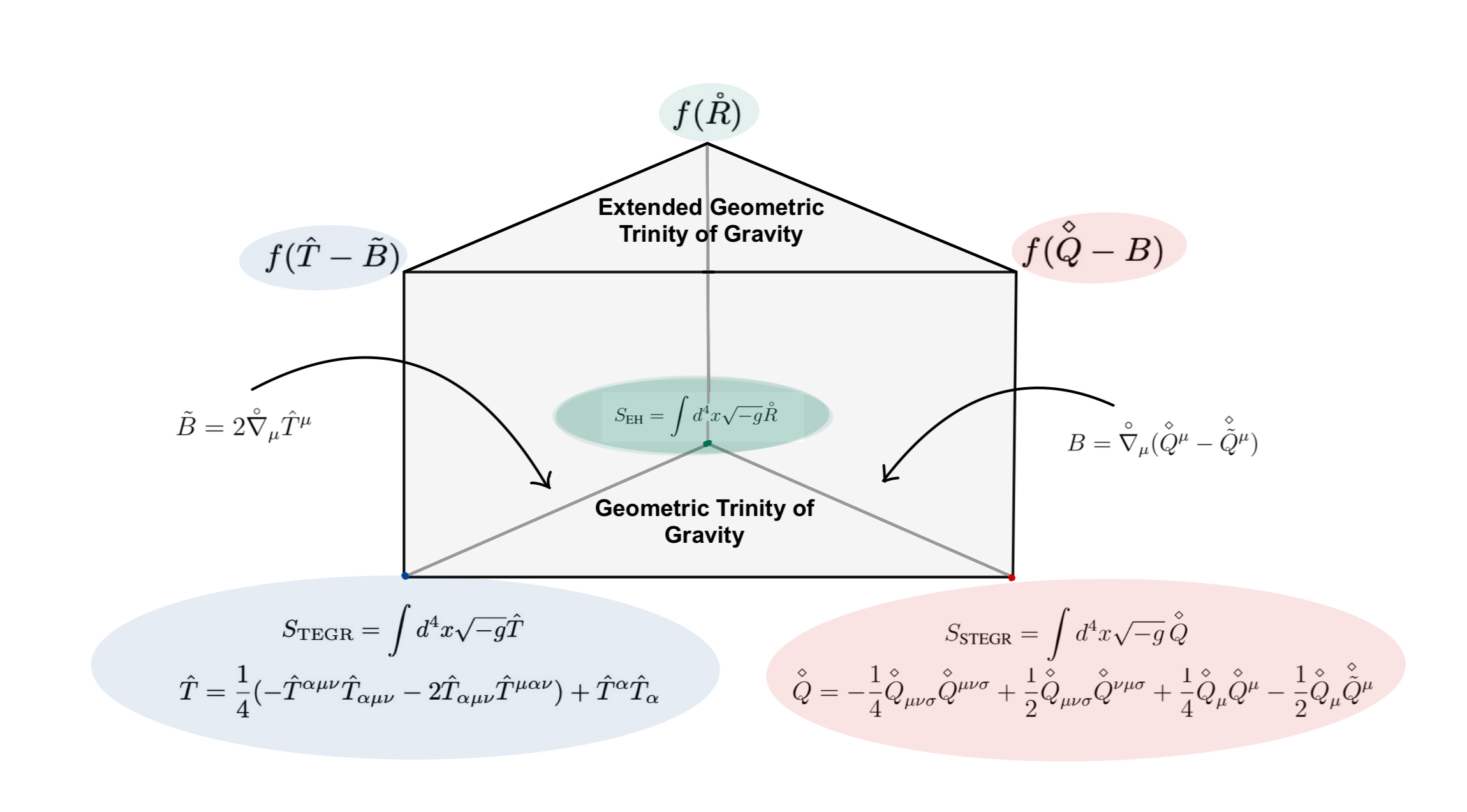}
\centering
\caption{Improving the Geometric Trinity of Gravity to the Extended Geometric Trinity of Gravity \cite{CdFF}.}
\label{Figure EGTG}
\end{figure*}

Thus, for a specific choice of the free parameters, TEGR is completely equivalent to GR. Because of the equivalence between GR and TEGR, curvature and torsion are able to provide equivalent descriptions of the gravitational interaction. \\
In GR, geometry replaces the concept of gravitational force, and the trajectories are determined by geodesics.
On the other hand, in Teleparallel Gravity, the gravitational interaction is an effect of  torsion of a zero-curvature Lorentz connection, see Sec. \ref{sec: Teleparallel framework}. This is the reason why there are no geodesics in Teleparallel Gravity, but force equations, as in Electrodynamics, or autoparallels \cite{ATG, Romano2019}.\\
STEGR shares many similar properties to TEGR. In this theory, we require that curvature and torsion are both zero and gravity generated by the non-metricity tensor. The metric and the affine connection are the fundamental objects, and, similarly to TEGR, under the teleparallelism constraint, we can always choose  the coincident gauge, in which the affine connection vanishes. \\ 
Thus, gravitation is characterized by three equivalent descriptions because of its most peculiar property of being a universal field \cite{ATG}. All objects, regardless of their structure and nature, feel  gravity and, just like the other fundamental interactions of Nature, it can be described in terms of a gauge theory. Additionally, universality of free fall makes it possible to realize a geometrized description, based on the Equivalence Principle, that is valid in GR, TEGR, and STEGR \cite{ComparingEG, Ferrara24}. However, it can be a fundamental or an emergent principle as discussed in \cite{Mancini:2025asp}. In this perspective, curvature, torsion, and non-metricity are simply alternative ways of representing the same gravitational field, accounting for the same DoFs. In this sense, GR, TEGR, and STEGR form the Geometric Trinity of Gravity (see Fig.$\thinspace$\ref{Figure EGTG}).
The conservation laws can also be applied to EGTG. In fact, as shown in Sec. \ref{subsec: f(R)}$, f(\accentset{\circ}{R})$ field equations can be recast in a GR-like form, i.e. Eq. \eqref{eq: f(R) in GR}, whose RHS can be rewritten as
\begin{equation}\label{eq: GR modificata}
    G_{\mu\nu} = \chi(T^{m}_{\mu\nu}+T^{eff}_{\mu\nu}),
\end{equation}
having separated the matter energy-momentum tensor, obtained by the variation \eqref{eq: S-ET}, and the the effective energy-momentum
tensor given by the geometrical contribution \cite{DeFelice10,Capozziello:2012uv}:
\begin{align}
    T^{eff}_{\mu\nu} =& \frac{1}{f_{\mathring{R}}} \bigg[ \mathring{\nabla}_{\mu}\mathring{\nabla}_{\nu}f_{\mathring{R}} (\mathring{R}) \\ -&\mathring{\Box}f_{\mathring{R}}(\mathring{R})g_{\mu\nu} + g_{\mu\nu}\frac{f(\mathring{R})-f_{\mathring{R}}(\mathring{R})\mathring{R}}{2}\bigg].
\end{align}

Therefore, since the Einstein tensor $G_{\mu\nu}$ respects the Bianchi identity \eqref{eq: BIanchi Identity}, $\nabla^{\mu}G_{\mu\nu}=0$, as well as the energy-momentum tensor is divergence free
\begin{equation}
    \nabla^{\mu}T^{m}_{\mu\nu}=0\,.
\end{equation}
It follows that:
\begin{equation}
    \nabla^{\mu}T^{eff}_{\mu\nu} = 0.
\end{equation}
Hence, the continuity equation holds not only for the Einstein tensor and $T^{m}_{\mu\nu}$, but also for the effective energy-momentum tensor $T^{eff}_{\mu\nu}$.
As a consequence, generalized conservation will be also  valid for the field equations of $f(\hat{T}-\tilde{B})$ and $f(\stg{Q}-B)$, Eqs. \eqref{eq: equazioni di campo f(T,B)} and \eqref{eq: 3.66}, according to their dynamical equivalence to $f(\accentset{\circ}{R})$ \cite{Bahamonde15, fQB}. Therefore, the teleparallel and the symmetric teleparallel equivalent of $f(\accentset{\circ}{R})$ gravity, i.e. $f(\hat{T}-\tilde{B})$ and $f(\stg{Q}-B)$, are  Lorentz invariant theories of gravity that can be formulated with respect to $\accentset{\circ}{R}$, $\hat{T}$, $\stg{Q}$ and the boundary terms. On the other hand, we have to take into account  the presence of higher order derivative terms.\\
In general,  it is possible to consider extensions recasting the field equations as \cite{Capozziello:2013, Capozziello:2014}
\begin{equation}
    g(\Psi^i)(G_{\mu\nu}+H_{\mu\nu})=\chi T_{\mu\nu},
\end{equation}
where the factor $g(\Psi^i)$ parametrizes the coupling with  matter fields of $T_{\mu\nu}$ and $\Psi^i$ generically represents either curvature invariants or other gravitational fields contributing to the dynamics, such as scalar fields \cite{Harko:2014, Bertolami:2007}. The tensor $H_{\mu\nu}$ represents  additional geometric quantities  expressing 
modifications introduced by extensions  with respect to GR, as the $T^{eff}_{\mu\nu}$ of Eq. \eqref{eq: GR modificata}. Obviously, the Einstein theory is recovered for $H_{\mu\nu}=0$ and $g(\Psi^i)=1$. This implies that the covariant conservation of the energy-momentum tensor, $\nabla^{\mu}T_{\mu\nu}$, produces a combination of $G_{\mu\nu}$ with $H_{\mu\nu}$:
\begin{equation}\label{eq:conservazione modificata}
    \nabla^{\mu}H_{\mu\nu}=-\frac{\chi}{g^2}T_{\mu\nu}\nabla^{\mu}g
\end{equation}
It is worth mentioning that in literature, $H_{\mu\nu}$ is treated as an effective energy-momentum tensor, by transporting it on the RHS of Eq. \eqref{eq: GR modificata}. However, it can be misleading to consider these impositions as conditions of the energy, since they emerge from Eq. \eqref{eq:conservazione modificata} \cite{Capozziello:2013, Capozziello:2014}.

However, if the ETG under consideration can be equivalently described throughout an appropriate conformal transformation, it then becomes correct to associate the transformed $H_{\mu\nu}$ to the new $T_{\mu\nu}$, in the conformally transformed Einstein frame, i.e. scalar-tensor gravity theories and in $f(\accentset{\circ}{R})$ gravity \cite{CDL}. Indeed, conformal transformations allow to consider the further DoFs coming from ETGs under the form of curvature invariants and scalar fields.

Thus, several generalized theories of gravity can be redefined as GR plus a number of appropriate fields coupled to matter by means of a conformal transformation in the so-called Einstein frame, see \cite{Capozziello:2013, Capozziello:2014} for a detailed discussion. 

In this perspective, $f(\accentset{\circ}{R})$, $f(\hat{T}-\tilde{B})$, and $f(\stg{Q}-B)$ can be considered the \qm{Extended Geometric Trinity of Gravity} (see Fig.$\thinspace$\ref{Figure EGTG}).


\section{Conclusions and  perspectives}
\label{sec:end}

In this paper, we aimed to give a comprehensive approach to  metric, teleparallel and symmetric teleparallel equivalent to GR in view of  comparing the possible  extensions starting from  $f(\accentset{\circ}{R})$.
Specifically, two triplets of dynamically equivalent theories have been examined in detail: the GTG and the EGTG.
Each of the above frameworks combines seemingly different theories describing gravity in terms of diverse geometric objects, namely curvature, torsion, and non-metricity. \\
In Sec. \ref{Section3.2}, we considered the Metric-Affine Theories of Gravity; in this framework,  the linear affine connection is treated as an independent geometric field with respect to the metric tensor  allowing to define curvature, torsion, and non-metricity tensors. These latter are the dynamical variables of the theories and are features of the linear affine connection. The restriction to a specific subclass of MAGs depends on the choice of the linear affine connection. \\
In Sec. \ref{sec: Teleparallel framework}, we provided the fundamental geometric tools to formulate the Teleparallel Gravity: tetrads and spin connection. Whilst the former solder the tangent space to the spacetime manifold, the latter depends on the tetrads \cite{Martin2019} and it represents the inertial effects, as a result of local Lorentz transformations \cite{ATG}. \\
We focused on Teleparallel Gravity in Sec. \ref{sec: Trinity}.
We gave an overview of the GTG, showing how its constituent theories (i.e. GR, TEGR, STEGR) are dynamically equivalent. Starting from a flat linear connection, defined with respect to the contorsion tensor and the Levi-Civita connection, we analyzed the metric-teleparallel case. We obtained the expression of the Ricci scalar in the TEGR framework Eq.$\thinspace$\eqref{eq: TEGR}, using the Riemann tensor of Eq.$\thinspace$\eqref{eq: Riemann TT}. The Ricci and the torsion scalars differ for a boundary term, due to the choice of the connection (Weitzenb\"{o}ck gauge). The same approach can be applied to the non-metricity case in STEGR, obtaining the Ricci scalar in terms of the non-metricity; again the two theories are equivalent up to a boundary term, caused by the coincident gauge Eq.$\thinspace$\eqref{3.27}.
Then, in Sec. \ref{Section 3.3}, we compared the actions of the non-linear extended versions of the GTG theories, i.e. $f(\accentset{\circ}{R})$, $f(\hat{T})$, and $f(\stg{Q})$. In this case, the equivalence is lost but it can be restored when a specific functional dependence is considered. In particular, to recover $f(\accentset{\circ}{R})$ gravity, we have to include the respective boundary terms as variables of $f$. The proof of this statement is based on deriving the field equations of $f(\tg{T}, \tilde{B})$ and $f(\stg{Q}, B)$ gravity and on noticing their reduction to the $f(\accentset{\circ}{R})$ case, when $f(\tg{T},\tilde{B})=f(\tg{T}-\tilde{B})$ and $f(\stg{Q}, B)=f(\stg{Q}-B)$. \\
In Sec. \ref{sec: discussion}, we discuss about the equivalence of the Geometric Trinity and the Extended Geometric Trinity using  conservation laws and the possibility to recast  extra DoFs in terms of effective stress-energy tensors.  In these cases, the compatibility of theories is guaranteed  by the second Bianchi identity.\\
As previously mentioned, curvature, torsion, and non-metricity represent the same gravitational field, even if they give rise to conceptually different theories. The equivalence of GR, TEGR, and STEGR, as well as of $f(\accentset{\circ}{R})$, $f(\tg{T}-\tilde{B})$, and $f(\stg{Q}-B)$, depends on a gauge choice of the affine-connection and the subsequent constraints on  dynamical variables.
Thus, we may ask whether there could be any other gauge to fix the affine connection, so that the equivalence among the theories is still valid. Moreover, additional questions could be related to the definition of the fundamental dynamical variables and the role of both  boundary terms and  Equivalence Principle  \cite{ComparingEG, Ferrara24, Mancini:2025asp}.
\\As a concluding remark,  several investigations  may be conducted in the framework of  GTG and EGTG. 
An example could be the  cosmological investigation of  EGTG. It could give hints towards addressing  tensions in cosmic parameters, defining the nature of dark energy  as well as the large scale structure formation \cite{Caruana23, Colgain:2024xqj, Odintsov:2024woi}. 
Finally, given the ever-growing interest  in  quantum information and black hole physics \cite{Hashimoto2016, Capozziello:2024mxh}, another research line  can  be to examine the Maldacena-Shenker-Stanford Conjecture \cite{MSS} in the context of metric and symmetric teleparallel theories (as well as in their extensions) in order to understand to what extent its validity holds according to the various representations of gravity \cite{AC,AC1}.

\section*{Acknowledgements}
The authors acknowledge the Istituto Italiano di Fisica Nucleare (INFN) iniziative specifiche QGSKY and MOONLIGHT2.  SC thanks the Gruppo Nazionale di Fisica Matematica (GNFM)  of Istituto Nazionale di Alta Matematica (INDAM) for the support. 
This paper is based upon work from COST Action CA21136
Addressing observational tensions in cosmology with systematics and fundamental physics (CosmoVerse) supported by COST (European Cooperation in Science and Technology). 

\appendix
\section{Field equations of Extended  Gravity} \label{Appendix}

In these Appendices, we give  details about the derivation of the field equations in $f(\mathring{R})$, $f(\hat{T}, \tilde{B})$, and $f(\stg{Q}, B)$ gravity, starting from the variations of the respectives Lagrangians, Eqs \eqref{eq: A1}, \eqref{eq: f(T,B) azione}, and \eqref{eq: azione f(Q,B)}. 

\subsection{Derivation of $f(R)$ field equations}\label{appedice f(R)}

Let us derive Eq. \eqref{eq: equazioni di campo f(T,B)} from the variation of the Lagrangian given in Eq. \eqref{eq: A.1}.
In Sec \ref{subsec: f(R)}, we wrote the variation of the $f(\mathring{R})$ Lagrangian, obtaining Eqs. \eqref{eq: A.3}, \eqref{eq: A.6}, and \eqref{Matter}.
At this point one is interested in expressing the variation $\delta_g \mathring{R}_{\mu\nu}$, constituting part of the last term in the RHS of Eq.$\thinspace$\eqref{eq: A.6}, in terms of the variation of the metric tensor with contravariant indices $\delta_g g^{\mu\nu}$. 
The calculations may be performed in a local inertial frame (LIF), so that they can be carried out in a simpler way. In fact, the Riemann tensor reduces to the sum of solely two terms, as the Levi-Civita connection coefficient are zero in such a frame
\begin{equation}\label{eq: A.7}
	\mathring{R}^{\alpha}_{\phantom{\alpha}\beta\mu\nu}\overset{\textrm{LIF}}{=}\mathring{\Gamma}^{\alpha}_{\phantom{\alpha}\beta\nu,\mu}-\mathring{\Gamma}^{\alpha}_{\phantom{\alpha}\beta\mu,\nu}
\end{equation}

and the expression of the Ricci tensor in a LIF can be obtained as usual from Eq.$\thinspace$\eqref{eq: A.7} contracting the contravariant index with the second covariant one
\begin{equation}\label{eq: A.8}
	\mathring{R}_{\mu\nu}\overset{\textrm{LIF}}{=}\mathring{\Gamma}^{\alpha}_{\phantom{\alpha}\mu\nu,\alpha}-\mathring{\Gamma}^{\alpha}_{\phantom{\alpha}\mu\alpha,\nu}.
\end{equation}
The variation of Eq.$\thinspace$\eqref{eq: A.8} is clearly dependent on the variation of the general-relativistic connection coefficients \eqref{eq: 3.6} that, following the introduction of the expressions
\begin{equation}
	\delta_g g_{\mu\nu}=h_{\mu\nu}\quad\textrm{and}\quad\delta_g g^{\mu\nu}=-h^{\mu\nu},
\end{equation}
can be outlined as {}
\begin{align}\label{eq: A.10}
	\delta_g\mathring{\Gamma}^{\alpha}_{\phantom{\alpha}\mu\nu}&=-\frac{1}{2}h^{\alpha\sigma}(g_{\mu\sigma,\nu}+g_{\sigma\nu,\mu}-g_{\mu\nu,\sigma})+\nonumber\\
    &+\frac{1}{2}g^{\alpha\sigma}(h_{\mu\sigma,\nu}+h_{\sigma\nu,\mu}-h_{\nu\mu,\sigma})=\nonumber\\
	&=\frac{1}{2}g^{\alpha\sigma}(-2h_{\sigma\lambda}\mathring{\Gamma}^{\lambda}_{\phantom{\lambda}\mu\nu}+h_{\mu\sigma,\nu}+h_{\sigma\nu,\mu}-h_{\nu\mu,\sigma})=\nonumber\\
	&=\frac{1}{2}g^{\alpha\sigma}(\mathring{\nabla}_{\mu}h_{\sigma\nu}+\mathring{\nabla}_{\nu}h_{\mu\sigma}-\mathring{\nabla}_{\sigma}h_{\mu\nu})
\end{align}
where the second equality is justified considering that 
\begin{align}
&\mathring{\Gamma}^{\lambda}_{\phantom{\lambda}\mu\nu}g_{\lambda\sigma}=\frac{1}{2}(g_{\mu\sigma,\nu}+g_{\sigma\nu,\mu}-g_{\mu\nu,\sigma})\nonumber\\
	&h^{\alpha\sigma}g_{\lambda\sigma}=-\delta_g(g^{\alpha\sigma})g_{\lambda\sigma}=\delta_g(g_{\lambda\sigma})g^{\alpha\sigma}=h_{\lambda\sigma}g^{\alpha\sigma}\nonumber
\end{align}
and the third one follows from noticing that
\begin{align}
	&\mathring{\nabla}_{\mu}h_{\sigma\nu}+\mathring{\nabla}_{\nu}h_{\mu\sigma}-\mathring{\nabla}_{\sigma}h_{\mu\nu}=\nonumber\\
	&=h_{\mu\sigma,\nu}+h_{\sigma\nu,\mu}-h_{\nu\mu,\sigma}-\mathring{\Gamma}^{\lambda}_{\phantom{\lambda}\mu\nu}h_{\lambda\sigma}
	-\mathring{\Gamma}^{\lambda}_{\phantom{\lambda}\sigma\nu}h_{\lambda\mu}+\nonumber\\
    &-\mathring{\Gamma}^{\lambda}_{\phantom{\lambda}\mu\sigma}h_{\lambda\nu}-\mathring{\Gamma}^{\lambda}_{\phantom{\lambda}\mu\nu}h_{\lambda\sigma}+\mathring{\Gamma}^{\lambda}_{\phantom{\lambda}\sigma\mu}h_{\lambda\nu}+\mathring{\Gamma}^{\lambda}_{\phantom{\lambda}\sigma\nu}h_{\lambda\mu}=\nonumber\\
    &=h_{\mu\sigma,\nu}+h_{\sigma\nu,\mu}-h_{\nu\mu,\sigma}-2\mathring{\Gamma}^{\lambda}_{\phantom{\lambda}\mu\nu}h_{\lambda\sigma}.\nonumber
\end{align}
Additionally, the choice of operating in a LIF guarantees to write the variation of Eq.$\thinspace$\eqref{eq: A.8} in terms of the covariant derivatives instead of the ordinary ones and thus to obtain{}
\begin{equation}\label{eq: A.11}
	\delta_g \mathring{R}_{\mu\nu}=\mathring{\nabla}_{\alpha}\delta_g\mathring{\Gamma}^{\alpha}_{\phantom{\alpha}\mu\nu}-\mathring{\nabla}_{\nu}\delta_g\mathring{\Gamma}^{\alpha}_{\phantom{\alpha}\mu\alpha}.
\end{equation}
An important remark has to be made on Eq.$\thinspace$\eqref{eq: A.11} because we have not reported 'LIF' over the equality sign. Whereas the Christoffel symbols do not follow the tensorial transformation law, their variation is instead a tensor. This implies that the RHS of Eq.$\thinspace$\eqref{eq: A.11} is a tensor and that the equation itself is a tensorial equation, given that the variation of a tensor (in the LHS) is still a tensor. Then, the above cited equation is invariant when changing the reference frame and thus not only can it be considered valid in a LIF but also in any other frame. \\
Now one can proceed to perform the substitution of Eq.$\thinspace$\eqref{eq: A.10} in Eq.$\thinspace$\eqref{eq: A.11} and, keeping the metric-compatibility condition in mind, the following relation is straightforwardly derived
\begin{align}\label{eq: A.12}
	\delta_g \mathring{R}_{\mu\nu}=\frac{1}{2}g^{\alpha\sigma}&[(\mathring{\nabla}_{\alpha}\mathring{\nabla}_{\mu}h_{\sigma\nu}+\mathring{\nabla}_{\alpha}\mathring{\nabla}_{\nu}h_{\mu\sigma}-\mathring{\nabla}_{\alpha}\mathring{\nabla}_{\sigma}h_{\mu\nu})+\nonumber\\
	&-(\mathring{\nabla}_{\nu}\mathring{\nabla}_{\mu}h_{\sigma\alpha}+\mathring{\nabla}_{\nu}\mathring{\nabla}_{\alpha}h_{\mu\sigma}-\mathring{\nabla}_{\nu}\mathring{\nabla}_{\sigma}h_{\mu\alpha})].
\end{align}
Multiplying both sides of Eq.$\thinspace$\eqref{eq: A.12} by $g^{\mu\nu}$ one gets the equality{}
\begin{align}\label{eq: A.13}
	g^{\mu\nu}\delta_g \mathring{R}_{\mu\nu}&=\mathring{\nabla}^{\sigma}\mathring{\nabla}^{\mu}h_{\mu\sigma}-\mathring{\Box} h=\nonumber\\
    &=\mathring{\nabla}^{\sigma}\mathring{\nabla}^{\mu}\delta_gg_{\mu\sigma}-\mathring{\Box} g^{\mu\nu}\delta_g g_{\mu\nu}
\end{align}
that can finally be inserted in Eq.$\thinspace$\eqref{eq: A.6} obtaining{}
\begin{align}\label{eq: A.14}
	\textrm{(B)}&=\int d^4x f_{\mathring{R}}(\mathring{R}) \sqrt{-g} [\mathring{R}_{\mu\nu}\delta_g g^{\mu\nu}+\mathring{\nabla}^{\sigma}\mathring{\nabla}^{\mu}\delta_gg_{\mu\sigma}+\nonumber\\
    &-\mathring{\Box} g^{\mu\nu}\delta_g g_{\mu\nu}]
	=\int d^4x\sqrt{-g}[f_{\mathring{R}}(\mathring{R}) \mathring{R}_{\mu\nu}+\nonumber\\
    &-\mathring{\nabla}_{\mu}\mathring{\nabla}_{\nu}f_{\mathring{R}}(\mathring{R})
    +\mathring{\Box}f_{\mathring{R}}(\mathring{R})g_{\mu\nu}]\delta_g g^{\mu\nu}
\end{align}
where the integration by parts has been used twice together with the metric-compatibility. Finally we get
\begin{equation}\label{eq: f(R) field equationsA}
	f_{\mathring{R}}(\mathring{R})\mathring{R}_{\mu\nu}-\frac{1}{2}g_{\mu\nu}f(\mathring{R})-\mathring{\nabla}_{\mu}\mathring{\nabla}_{\nu}f_{\mathring{R}}(\mathring{R})+\mathring{\Box}f_{\mathring{R}}(\mathring{R})g_{\mu\nu}=\chi T_{\mu\nu}.
\end{equation}

\subsection{Derivation of $f(T,B)$ field equations}\label{appendice f(T,B)}

Let us derive now the field equations of $f(\hat{T},\tilde{B})$ gravity. In order to avoid a worthless burdening of the equations, a different notation will be used throughout this derivation, removing the over-hat on quantities dependent on the Weitzenb\"{o}ck connection and indicating the general-relativistic objects (derived from the Levi-Civita connection) inserting an over-circle as usual. The standard notation will then be restored at the end of the calculations.
In particular, we start from Eq \eqref{eq: f(T,B) azione}, considering its variation with respect to the tetrad field $e^{c}_{\phantom{c}\lambda}$:
\begin{align}\label{eq: f(T, B) action}
	\delta_e S_{f(T, \tilde{B})}&=\underbrace{\int d^4x (\delta_e e) f(T, \tilde{B})}_{(A')}+\underbrace{\int d^4x e\delta_e f(T, \tilde{B})}_{(B')}+\nonumber\\
    &+2\chi\underbrace{\int d^4x\delta_e\mathcal{L}_{\text{m}}}_{(C')}.
\end{align}
The $(A')$ term may be rewritten as
\begin{equation}
	(A')=\int d^4x e f(T, \tilde{B}) \thinspace e_{c}^{\phantom{c}\lambda}\delta_ee^{c}_{\phantom{c}\lambda}.
\end{equation}
Indeed, referring to Eqs$\thinspace$\eqref{eq: gcontravariant}, \eqref{eq: e}, \eqref{eq: eg} , one easily obtains{}
	\begin{align}\label{eq: de}
		\delta_e(e)&=\delta_e(\sqrt{-g})=-\frac{1}{2}e\thinspace g_{\mu\nu}\delta_eg^{\mu\nu}=-\frac{1}{2}e\thinspace g_{\mu\nu}\frac{\partial g^{\mu\nu}}{\partial{e^{c}_{\phantom{c}\lambda}}}\delta_e e^{c}_{\phantom{c}\lambda}=\nonumber\\
		&=\frac{1}{2}e\thinspace g_{\mu\nu}[e_{c}^{\phantom{c}\mu}e_{a}^{\phantom{a}\lambda}e_{b}^{\phantom{b}\nu}\eta^{ab}+e_{a}^{\phantom{a}\mu}e_{c}^{\phantom{c}\nu}e_{b}^{\phantom{b}\lambda}\eta^{ab}]\delta_e e^{c}_{\phantom{c}\lambda}=\nonumber\\
		&=\frac{1}{2}e\thinspace g_{\mu\nu}[g^{\lambda\nu}e_{c}^{\phantom{c}\mu}+g^{\mu\lambda}e_{c}^{\phantom{c}\nu}]\delta_e e^{c}_{\phantom{c}\lambda}= e\thinspace e_{c}^{\phantom{c}\lambda}\delta_ee^{c}_{\phantom{c}\lambda}
	\end{align}
where the fourth equality follows from deriving Eq.$\thinspace$\eqref{eq: orthogonality relation} with respect to the tetrad field and performing additional manipulations, i.e.{}
\begin{align}
	&\frac{\partial e_{a}^{\phantom{a}\mu}}{\partial e^{c}_{\phantom{c}\lambda}}e^{a}_{\phantom{a}\nu}=-e_{c}^{\phantom{c}\mu}g^{\lambda}_{\nu};\nonumber\\
	&\frac{\partial e_{a}^{\phantom{a}\mu}}{\partial e^{c}_{\phantom{c}\lambda}}e^{a}_{\phantom{a}\nu}e_{b}^{\phantom{b}\nu}=-e_{c}^{\phantom{c}\mu}g^{\lambda}_{\nu}e_{b}^{\phantom{b}\nu};\nonumber\\
	&\frac{\partial e_{a}^{\phantom{a}\mu}}{\partial e^{c}_{\phantom{c}\lambda}}=-e_{c}^{\phantom{c}\mu}e_{a}^{\phantom{a}\lambda}.
\end{align}
Analogously to the $(C)$ term of the previous section, the $(C')$ term can be recast as{}
\begin{equation}
	(C')=2\chi \int d^4x  e\thinspace T_{c}^{\phantom{c}\lambda}\delta_e e^{c}_{\phantom{c}\lambda}
\end{equation}
with 
\begin{equation}
	T_{c}^{\phantom{c}\lambda}=\frac{1}{e}\frac{\delta_e\mathcal{L}_{m}}{\delta_e e^{c}_{\phantom{c}\lambda}}.
\end{equation}
Now one wants to outline an expression of the $(B')$ term 
\begin{equation}\label{eq: (B')}
	(B')= \int d^4x \thinspace[\underbrace{e\thinspace f_T(T, \tilde{B})\delta_e T}_{(D')}+\underbrace{e\thinspace f_{\tilde{B}}(T, \tilde{B})\delta_e \tilde{B}}_{(E')}]
\end{equation}
in terms of the variation of the tetrad field, as it has already been done for the $(A')$ and $(C')$ term. 
Starting from the first integrand term $(D')$ in Eq.$\thinspace$\eqref{eq: (B')}, one may write{}
\begin{align}\label{eq: D'equiv}
	(D')&=-ef_T\delta_e(S_{a}^{\phantom{a}\mu\nu}T^{a}_{\phantom{a}\mu\nu})=\nonumber\\
    &=-ef_T[\delta_e (S_{a}^{\phantom{a}\mu\nu})T^{a}_{\phantom{a}\mu\nu}+ S_{a}^{\phantom{a}\mu\nu}\delta_e (T^{a}_{\phantom{a}\mu\nu})]
\end{align}
\begin{itemize}
\item The first term in the RHS of Eq.$\thinspace$\eqref{eq: D'equiv} can be straightforwardly rewritten using Eq.$\thinspace$\eqref{eq: superpotential_a} as follows
\begin{align}
	-ef_T\delta_e(S_{a}^{\phantom{a}\mu\nu})T^{a}_{\phantom{a}\mu\nu}&=-\frac{1}{2}ef_TT^{a}_{\phantom{a}\mu\nu}[\underbrace{\delta_e(K^{\mu\phantom{a}\nu}_{\phantom{\mu}a})}_{(D'_1)}+\nonumber\\
    &+\underbrace{e_{a}^{\phantom{a}\mu}\delta_e(T^{\nu})}_{(D'_2)}+
    -\underbrace{e_{a}^{\phantom{a}\nu}\delta_e(T^{\mu})}_{(D'_3)}+\nonumber\\
	&+\underbrace{\delta_e(e_{a}^{\phantom{a}\mu})T^{\nu}-\delta_e(e_{a}^{\phantom{a}\nu})T^{\mu}}_{(D'_4)}].\label{eq: B29}
\end{align}
One can now perform the calculations needed to achieve the above-mentioned aim of expressing each term appearing in Eq.$\thinspace$\eqref{eq: B29} in terms of the variation of the tetrad field. Starting from the term involving $(D'_1)$, one has:
\begin{align}
	&-ef_TT^{a}_{\phantom{a}\mu\nu}(D'_1)=-\frac{1}{2}ef_TT^{a}_{\phantom{a}\mu\nu}(\delta_e T^{\mu\phantom{a}\nu}_{\phantom{\mu}a}+\delta_e T_{a}^{\phantom{a}\mu\nu}+\nonumber\\
    &+\delta_eT^{\nu\mu}_{\phantom{\nu\mu}a})=-\frac{1}{2}ef_TT^{a}_{\phantom{a}\mu\nu}(\delta_e T_{a}^{\phantom{a}\mu\nu}-2\delta_eT^{\mu\nu}_{\phantom{\mu\nu}a})=\nonumber\\
    &=e[2f_T(T_{\rho}^{\phantom{\rho}\mu\lambda}
    +T^{\lambda\mu}_{\phantom{\lambda\mu}\rho}+T^{\mu\phantom{\rho}\lambda}_{\phantom{\mu}\rho})T^{\rho}_{\phantom{\rho}\mu c}+		\nonumber\\
	&+(T_{c}^{\phantom{c}\mu\lambda}+T^{\lambda\mu}_{\phantom{\lambda\mu}c}+T^{\mu\phantom{c}\lambda}_{\phantom{\mu}c})(f_TW^{\rho}_{\phantom{\rho}\mu\rho}+\partial_{\mu}f_{T})+\nonumber\\
    &+f_T\partial_{\mu}(T_{c}^{\phantom{c}\mu\lambda}+T^{\lambda\mu}_{\phantom{\lambda\mu}c}
	+T^{\mu\phantom{c}\lambda}_{\phantom{\mu}c})]\delta_e(e^{c}_{\phantom{c}\lambda})=\nonumber\\
    &=2e[f_T(4S_{\rho}^{\phantom{\rho}\mu\lambda}T^{\rho}_{\phantom{\rho}\mu c}+2T^{\lambda}T^{c}+2T^{\mu}T^{\lambda}_{\phantom{\lambda}\mu c})+\nonumber\\
	&+(f_TW^{\rho}_{\phantom{\rho}\mu\rho}+\partial_{\mu}f_{T})K^{\mu\phantom{c}\lambda}_{\phantom{\mu}c}+f_T\partial_{\mu}(K^{\mu\phantom{c}\lambda}_{\phantom{\mu}c})]\delta_e(e^{c}_{\phantom{c}\lambda}).
\end{align}
From Eq.$\thinspace$\eqref{eq: tttn}, one derives
\begin{align}\label{eq: deltaT}
	&\delta_e T^{\mu}=\delta_e(g^{\mu\rho}T^{\alpha}_{\phantom{\alpha}\rho\alpha})=\delta_e(g^{\mu\rho})[e_{a}^{\phantom{a}\alpha}(\partial_{\rho}e^{a}_{\phantom{a}\alpha}-\partial_{\alpha}e^{a}_{\phantom{a}\rho})]+\nonumber\\
	&+g^{\mu\rho}[\delta_e(e_{a}^{\phantom{a}\alpha})(\partial_{\rho}e^{a}_{\phantom{a}\alpha}-\partial_{\alpha}e^{a}_{\phantom{a}\rho})]+g^{\mu\rho}e_{a}^{\phantom{a}\alpha}[\partial_{\rho}\delta_e(e^{a}_{\phantom{a}\alpha})+\nonumber\\
    &-\partial_{\alpha}\delta_e(e^{a}_{\phantom{a}\rho})]=-[(g^{\lambda\rho}e_{c}^{\phantom{c}\mu}+g^{\lambda\mu}e_{c}^{\phantom{c}\rho})T^{\alpha}_{\phantom{\alpha}\rho\alpha}+\nonumber\\
    &-T^{a\mu}_{\phantom{a\mu}\alpha}e_{c}^{\phantom{c}\alpha}e_{a}^{\phantom{a}\lambda}]\delta_e(e^{c}_{\phantom{c}\lambda})+g^{\mu\rho}e_{a}^{\phantom{a}\alpha}[\partial_{\rho}\delta_e(e^{a}_{\phantom{a}\alpha})+\nonumber\\
    &-\partial_{\alpha}\delta_{e}(e^{a}_{\phantom{a}\rho})]=-(T^{\lambda}e_{c}^{\phantom{c}\mu}
	+g^{\lambda\mu}T_{c}-T^{\lambda\phantom{c}\mu}_{\phantom{\lambda}c})\delta_e(e^{c}_{\phantom{c}\lambda})+\nonumber\\
 &+g^{\mu\rho}e_{a}^{\phantom{a}\alpha}[\partial_{\rho}\delta_e(e^{a}_{\phantom{a}\alpha})-\partial_{\alpha}\delta_{e}(e^{a}_{\phantom{a}\rho})]
\end{align}
then, $-ef_TT^{a}_{\phantom{a}\mu\nu}(D'_2)$ reads
\begin{align}
	&-ef_TT^{a}_{\phantom{a}\mu\nu}(D'_2)=-ef_T(2T_{c}T^{\lambda}+g^{\lambda\nu}T_{c}+\nonumber\\
    &-T^{\lambda\phantom{c}\nu}_{\phantom{\lambda}c}T_{\nu})\delta_e(e^{c}_{\phantom{c}\lambda})-\partial_{\rho}(ef_TT^{\rho}e_{c}^{\phantom{c}\alpha})\delta_e(e^{a}_{\phantom{a}\alpha})+\nonumber\\
    &+\partial_{\alpha}(ef_TT^{\rho}e_{c}^{\phantom{c}\lambda})\delta_{e}(e^{a}_{\phantom{a}\rho})=e\{-f_T(2T_{c}T^{\lambda}+\nonumber\\
	&+g^{\lambda\nu}T_{c}-T^{\lambda\phantom{c}\nu}_{\phantom{\lambda}c}T_{\nu})-f_TW^{\mu}_{\phantom{\mu}\rho\mu}(e_{c}^{\phantom{c}\lambda}T^{\rho}-e_{c}^{\phantom{c}\rho}T^{\lambda})+\nonumber\\
    &-\partial_{\rho}(f_T)(e_{c}^{\phantom{c}\lambda}T^{\rho}-e_{c}^{\phantom{c}\rho}T^{\lambda})+f_T(W^{\lambda}_{\phantom{\lambda}\rho c}T^{\rho}+W_cT^{\lambda})+\nonumber\\
    &-f_T[e_c^{\phantom{c}\lambda}(\partial_{\rho}T^{\rho})-e_{c}^{\phantom{c}\rho}(\partial_{\rho}T^{\lambda})]\}\delta_e(e^{c}_{\phantom{c}\lambda})
\end{align}
where total divergence terms have been neglected. The term $-ef_TT^{a}_{\phantom{a}\mu\nu}(D'_3)$ can be obtained by exchanging the $\mu$ and $\nu$ indices in $(D'_2)$. Renaming those summed indices (each in turn as $\nu$ and $\mu$) and using the antisymmetry property of the torsion tensor, one has:
\begin{align}
	-ef_TT^{a}_{\phantom{a}\mu\nu}(D'_3)=ef_TT^{a}_{\phantom{a}\mu\nu}(D'_2).
\end{align}
Finally, the last term in Eq.$\thinspace$\eqref{eq: B29} can be rewritten as 
\begin{align}
	&-ef_TT^{a}_{\phantom{a}\mu\nu}(D'_4)=-ef_TT^{a}_{\phantom{a}\mu\nu}[\delta_e(e_{a}^{\phantom{a}\mu})T^{\nu}-\delta_e(e_{a}^{\phantom{a}\nu})T^{\mu}]=\nonumber\\
	&=-2ef_TT^{a}_{\phantom{a}\mu\nu}\delta_ee_{a}^{\phantom{a}\mu}T^{\nu}=2ef_TT^{a}_{\phantom{a}\mu\nu} T^{\nu} e_{a}^{\phantom{a}\lambda}e_{c}^{\phantom{c}\mu}\delta_e e^{c}_{\phantom{c}\lambda}=\nonumber\\
    &=2ef_TT^{\lambda}_{\phantom{\lambda}c\nu} 	T^{\nu}\delta_e e^{c}_{\phantom{c}\lambda}.
\end{align}
\item The second term in the RHS of Eq.$\thinspace$\eqref{eq: D'equiv} is recast in the equivalent form
\begin{align}\label{eq: B35}
	&-ef_TS_{a}^{\phantom{a}\mu\nu}\delta_e T^{a}_{\phantom{a}\mu\nu}=-ef_TS_{a}^{\phantom{a}\mu\nu}[\partial_{\mu}\delta_e(e^{a}_{\phantom{a}\nu})-\partial_{\nu}(\delta_ee^{a}_{\phantom{a}\mu})]=\nonumber\\
	&=2e[f_TW^{\rho}_{\phantom{\rho}\mu\rho}S_{c}^{\phantom{c}\mu\lambda}+(\partial_{\mu}f_T)S_{c}^{\phantom{c}\mu\lambda}+f_T(\partial_{\mu}S_{c}^{\phantom{c}\mu\lambda})]\delta_e(e^{c}_{\phantom{c}\lambda}).
\end{align}
having considered that the superpotential is antisymmetric under the interchange of the last two indices and having followed the usual strategy of integration by parts neglecting the boundary terms.
\end{itemize}
Collecting all the results in Eqs$\thinspace$\eqref{eq: B29}-\eqref{eq: B35} and plugging them in $(D')$ \eqref{eq: D'equiv} lead to
\begin{align}
	(D')=4e&[f_T(S_{\rho}^{\phantom{\rho}\mu\lambda}T^{\rho}_{\phantom{\rho}\mu c}+W^{\rho}_{\phantom{\rho}\mu\rho}S_{c}^{\phantom{c}\mu\lambda})+f_T(\partial_{\mu}S_{c}^{\phantom{c}\mu\lambda})+\nonumber\\
    &+(\partial_{\mu}f_T)S_{c}^{\phantom{c}\mu\lambda}]\delta_e(e^{c}_{\phantom{c}\lambda}).
\end{align}
\\\\The last term to consider is the one labeled as $(E')$, which depends on the boundary term $\tilde{B}$ defined in Eq.$\thinspace$\eqref{eq: TEGR boundary term}. Ignoring once again the boundary terms, one finds
\begin{align}\label{eq: B36}
	(E')=ef_{\tilde{B}}\delta_{e}\Big[\frac{2}{e}\partial_{\mu}(eT^{\mu})\Big]=&-[f_{\tilde{B}}\tilde{B}+2\partial_{\mu}(f_{\tilde{B}})T^{\mu}]\delta_{e}e+\nonumber\\
     &-2e\partial_{\mu}(f_{\tilde{B}})\delta_eT^{\mu}
\end{align}
where both $\delta_ee$ and $\delta_e T^{\mu}$ have already been rewritten in terms of $\delta_ee_{c}^{\phantom{c}\lambda}$ as shown in Eqs$\thinspace$\eqref{eq: de} and \eqref{eq: deltaT}.
Whilst one may immediately write 
\begin{equation}\label{eq: B37}
	-[f_{\tilde{B}}\tilde{B}+2\partial_{\mu}(f_{\tilde{B}})T^{\mu}]\delta_{e}e=-e\thinspace e_{c}^{\phantom{c}\lambda}[f_{\tilde{B}}\tilde{B}+2\partial_{\mu}(f_{\tilde{B}})T^{\mu}]\delta_ee^{c}_{\phantom{c}\lambda},
\end{equation}
additional computations are needed when reworking the remaining term in Eq.$\thinspace$\eqref{eq: B36}. After having substituted the aforementioned expression of $\delta_e T^{\mu}$ (which contains, as it has been noticed before, derivatives of the variation of the tetrad fields), one integrates by parts and neglects the boundary terms as usual, obtaining
\begin{align}\label{eq: B38}
	-2e\partial_{\mu}f_{\tilde{B}}\delta_eT^{\mu}&=-2\{\partial_{\mu}[e\partial^{\lambda}(f_{\tilde{B}})e_{c}^{\phantom{c}\mu}]-\partial_{\mu}[e\partial^{\mu}(f_{\tilde{B}})e_{c}^{\phantom{c}\lambda}]+\nonumber\\
	&-\partial_{\mu}(f_{\tilde{B}})e(T^{\lambda}e_{c}^{\phantom{c}\mu}+g^{\lambda\mu}T_{c}-T^{\lambda\phantom{c}\mu}_{\phantom{\lambda}c})\}\delta_e(e^{c}_{\phantom{c}\lambda}).
\end{align}
For future purposes it may be convenient to rewrite the first two terms in Eq.$\thinspace$\eqref{eq: B38} replacing the ordinary partial derivative with the general-relativistic covariant derivative introducing suitable additional terms that depend on the Levi-Civita connection
\begin{equation}\label{eq: B39}	\mathring{\Gamma}^{\alpha}_{\phantom{\alpha}\mu\nu}=W^{\alpha}_{\phantom{\alpha}\mu\nu}-K^{\alpha}_{\phantom{\alpha}\mu\nu}
\end{equation}
written in the notation that is being used in the current section.
\\Of course the just mentioned substitution is of immediate realization when one is dealing with the ordinary derivative of a scalar (as $f_{\tilde{B}}$), due to it being equal to its covariant derivative (i.e. $\nabla_{\mu}f=\partial_{\mu}f$).
\\Through the additional use of the symmetry properties of Eq.$\thinspace$\eqref{eq: B39} under the interchange of its last two indices, Eq.$\thinspace$\eqref{eq: B38} becomes
\begin{align}\label{eq: B40}
-2e\partial_{\mu}f_{\tilde{B}}\delta_eT^{\mu}=&-2e[e_{c}^{\phantom{c}\mu}\mathring{\nabla}_{\mu}\mathring{\nabla}^{\lambda}f_{\tilde{B}}-e_{c}^{\phantom{c}\lambda}\mathring{\Box}f_{\tilde{B}}+\nonumber\\
 &-(\partial_{\mu}f_{\tilde{B}})K^{\mu\phantom{c}\lambda}_{\phantom{\mu}c}+T^{\lambda}e_{c}^{\phantom{c}\mu}]		\delta_e(e^{c}_{\lambda}).
\end{align}
 Inserting Eqs$\thinspace$\eqref{eq: B37} and \eqref{eq: B40} into Eq.$\thinspace$\eqref{eq: B37}, the following final expression of the $(E')$ term is obtained
 \begin{align}
 	(E')&=-e\{e_{c}^{\phantom{c}\lambda}[f_{\tilde{B}}\tilde{B}+2\partial_{\mu}(f_{\tilde{B}})T^{\mu}]+2[e_{c}^{\phantom{c}\mu}\mathring{\nabla}_{\mu}\mathring{\nabla}^{\lambda}f_{\tilde{B}}+\nonumber\\
  &-e_{c}^{\phantom{c}\lambda}\mathring{\Box}f_{\tilde{B}}-(\partial_{\mu}f_{\tilde{B}})(K^{\mu\phantom{c}\lambda}_{\phantom{\mu}c}+T^{\lambda}e_{c}^{\phantom{c}\mu})]\}\delta_e(e^{c}_{\phantom{c}\lambda})=\nonumber\\
  &-e[e_{c}^{\phantom{c}\lambda}f_{\tilde{B}}\tilde{B}+2e_{c}^{\phantom{c}\mu}\mathring{\nabla}_{\mu}\mathring{\nabla}^{\lambda}f_{\tilde{B}}-2e_{c}^{\phantom{c}\lambda}\mathring{\Box}f_{\tilde{B}}+\nonumber\\
  &-4S_{c}^{\phantom{c}\mu\lambda}(\partial_{\mu}f_{\tilde{B}})]\delta_e(e^{c}_{\phantom{c}\lambda}).
 \end{align}
The $f(T, \tilde{B})$ metric field equations are finally determined from Eq.$\thinspace$\eqref{eq: f(T, B) action} and from the above-calculated $(A')$, $(B')$ and $(C')$ terms and read as
\begin{align}\label{eq: B42}
	&-f e_{c}^{\phantom{c}\lambda}-4[f_{\hat{T}}(S_{\rho}^{\phantom{\rho}\mu\lambda}\hat{T}^{\rho}_{\phantom{\rho}\mu c}+W^{\rho}_{\phantom{\rho}\mu\rho}\hat{S}_{c}^{\phantom{c}\mu\lambda})+f_{\hat{T}}(\partial_{\mu}\hat{S}_{c}^{\phantom{c}\mu\lambda})+\nonumber\\
    &+(\partial_{\mu}f_{\hat{T}})\hat{S}_{c}^{\phantom{c}\mu\lambda}]+[e_{c}^{\phantom{c}\lambda}f_{\tilde{B}}\tilde{B}+2e_{c}^{\phantom{c}\mu}\mathring{\nabla}_{\mu}\mathring{\nabla}^{\lambda}f_{\tilde{B}}-2e_{c}^{\phantom{c}\lambda}\mathring{\Box}f_{\tilde{B}}+\nonumber\\
    &-4\hat{S}_{c}^{\phantom{c}\mu\lambda}(\partial_{\mu}f_{\tilde{B}})]=2\chi T_{c}^{\phantom{c}\lambda}
\end{align}
where the usual notation employed in metric teleparallel theories of gravity has been restored.
The multiplication of both sides by the tetrad field $e^{c}_{\phantom{c}\nu}$ leads to the following equivalent expression in the manifold indices 
\begin{align}\label{3.55}
	&-\delta^{\lambda}_{\nu}f(\hat{T}, \tilde{B})-4f_{\hat{T}}\hat{S}_{\rho}^{\phantom{\rho}\mu\lambda}\hat{T}^{\rho}_{\phantom{\rho}\mu \nu}-4f_{\hat{T}}W^{\rho}_{\phantom{\rho}\mu\rho}\hat{S}_{\nu}^{\phantom{\nu}\mu\lambda}+\nonumber\\
 &-4e^{c}_{\phantom{c}\nu}f_{\hat{T}}(\partial_{\mu}\hat{S}_{c}^{\phantom{c}\mu\lambda})-4\partial_{\mu}f_{\hat{T}}\hat{S}_{\nu}^{\phantom{\nu}\mu\lambda}+\delta^{\lambda}_{\nu}f_{\tilde{B}}\tilde{B}+\nonumber\\
&+2\mathring{\nabla}_{\nu}\mathring{\nabla}^{\lambda}f_{\tilde{B}}-2\delta^{\lambda}_{\nu}\mathring{\Box}f_{\tilde{B}}-4\hat{S}_{\nu}^{\phantom{\nu}\mu\lambda}(\partial_{\mu}f_{\tilde{B}})=2\chi T_{\nu}^{\phantom{\nu}\lambda}.
\end{align}
The third and fourth terms in the LHS of Eq.$\thinspace$\eqref{3.55} can be conveniently rewritten as
\begin{align}\label{3.56}
	&4f_{\hat{T}}W^{\rho}_{\phantom{\rho}\mu\rho}\hat{S}_{\nu}^{\phantom{\nu}\mu\lambda}+4e^{c}_{\phantom{c}\nu}f_{\hat{T}}(\partial_{\mu}\hat{S}_{c}^{\phantom{c}\mu\lambda})=\frac{4}{e}e^{c}_{\phantom{c}\nu}f_{\hat{T}}\partial_{\mu} (e\hat{S}_{c}^{\phantom{c}\mu\lambda})=\nonumber\\
	&=\frac{2}{e}f_{\hat{T}}[\partial_{\mu}(e\hat{K}^{\mu\phantom{\nu}\lambda}_{\phantom{\mu}\nu})+\delta^{\mu}_{\nu}\partial_{\mu}(e\hat{T}^{\lambda})-\delta^{\lambda}_{\nu}\partial_{\mu}(e\hat{T}^{\mu})]+\nonumber\\
 &-4f_{\hat{T}}\hat{S}_{\rho}^{\phantom{\rho}\mu\lambda}W^{\rho}_{\phantom{\rho}\mu\nu}
	=\frac{2}{e}f_{\hat{T}}[\partial_{\mu}(e\hat{K}^{\mu\phantom{\nu}\lambda}_{\phantom{\mu}\nu})+\delta^{\mu}_{\nu}\partial_{\mu}(e\hat{T}^{\lambda})]+\nonumber\\
 &-\delta^{\lambda}_{\nu}f_{\hat{T}}\tilde{B}-4f_{\hat{T}}\hat{S}_{\rho}^{\phantom{\rho}\mu\lambda}W^{\rho}_{\phantom{\rho}\mu\nu}.
\end{align}
The convenience lies in the possibility of tracing part of the previous expression back to the Ricci tensor. Indeed, one has
\begin{align}\label{3.57}
	\accentset{\circ}{R}^{\lambda}_{\phantom{\lambda}\nu}&=-\accentset{\circ}{\nabla}_{\mu}(\hat{K}^{\mu\phantom{\nu}\lambda}_{\phantom{\mu}\nu})+\accentset{\circ}{\nabla}_{\nu}(\hat{K}^{\mu\phantom{\mu}\lambda}_{\phantom{\mu}\mu})-\hat{K}^{\sigma\phantom{\nu}\lambda}_{\phantom{\sigma}\nu}
	\hat{K}^{\mu}_{\phantom{\mu}\mu\sigma}+\nonumber\\
 &+\hat{K}^{\sigma\phantom{\mu}\lambda}_{\phantom{\sigma}\mu}\hat{K}^{\mu}_{\phantom{\mu}\nu\sigma}=-\frac{1}{e}[\partial_{\mu}(e\hat{K}^{\mu\phantom{\nu}\lambda}_{\phantom{\mu}\nu})+\partial_{\nu}(e\hat{T}^{\lambda})]+\nonumber\\
 &+T_{\mu}\hat{K}^{\mu\phantom{\nu}\lambda}_{\phantom{\mu}\nu}+T^{\lambda}W^{\mu}_{\phantom{\mu}\nu\mu}+(W^{\rho}_{\phantom{\rho}\mu\nu}-\hat{K}^{\rho}_{\phantom{\rho}\mu\nu})\hat{K}			^{\mu\phantom{\rho}\lambda}_{\phantom{\mu}\rho}+\nonumber\\
	&-W^{\lambda}_{\phantom{\lambda}\mu\rho}\hat{K}^{\mu\phantom{\nu}\rho}_{\phantom{\mu}\nu}-(W^{\lambda}_{\phantom{\lambda}\nu\rho}-\hat{K}^{\lambda}_{\phantom{\lambda}\nu\rho})T^{\rho}
\end{align}
where 
\begin{align}
\accentset{\circ}{\nabla}_{\mu}(\hat{K}^{\mu\phantom{\nu}\lambda}_{\phantom{\mu}\nu})&=\frac{1}{e}\partial_{\mu}(e\hat{K}^{\mu\phantom{\nu}\lambda}_{\phantom{\mu}\nu})-W^{\rho}_{\phantom{\rho}\mu\rho}\hat{K}^{\mu\phantom{\nu}\lambda}_{\phantom{\mu}\nu}
	+(W^{\mu}_{\phantom{\mu}\mu\rho}+\nonumber\\
 &-\hat{K}^{\mu}_{\phantom{\mu}\mu\rho})\hat{K}^{\rho\phantom{\nu}\lambda}_{\phantom{\rho}\nu}-(W^{\rho}_{\phantom{\rho}\mu\nu}-\hat{K}^{\rho}_{\phantom{\rho}\mu\nu})\hat{K}^{\mu\phantom{\rho}\lambda}_{\phantom{\mu}\rho}+\nonumber\\
 &+(W^{\lambda}_{\phantom{\lambda}\mu\rho}-\hat{K}^{\lambda}_{\phantom{\lambda}\mu\rho})\hat{K}^{\mu\phantom{\nu}\rho}_{\phantom{\mu}\nu};\\
	\accentset{\circ}{\nabla}_{\nu}(\hat{K}^{\mu\phantom{\mu}\lambda}_{\phantom{\mu}\mu})&=-\accentset{\circ}{\nabla}_{\nu}(\hat{T}^{\lambda})=-\frac{1}{e}\partial_{\nu}(e\hat{T}^{\lambda})-(W^{\lambda}_{\phantom{\lambda}\nu\rho}+\nonumber\\
 &-\hat{K}^{\lambda}_{\phantom{\lambda}\nu\rho})T^{\rho}+W^{\mu}_{\phantom{\mu}\nu\mu}T^{\lambda};
 \end{align}
 \begin{align}\label{3.58}
	&(W^{\rho}_{\phantom{\rho}\mu\nu}-\hat{K}^{\rho}_{\phantom{\rho}\mu\nu})\hat{K}^{\mu\phantom{\rho}\lambda}_{\phantom{\mu}\rho}-W^{\lambda}_{\phantom{\lambda}\mu\rho}\hat{K}^{\mu\phantom{\nu}\rho}_{\phantom{\mu}\nu}=W^{\rho}_{\phantom{\rho}\nu\mu}\hat{K}^{\mu\phantom{\rho}\lambda}_{\phantom{\mu}\rho}+\nonumber\\&+\hat{T}^{\rho}_{\phantom{\rho}\mu\nu}\hat{K}^{\mu\phantom{\rho}\lambda}_{\phantom{\mu}\rho}-\hat{K}^{\rho}_{\phantom{\rho}\mu\nu}\hat{K}^{\mu\phantom{\rho}\lambda}_{\phantom{\mu}\rho}-W^{\lambda}_{\phantom{\lambda}\mu\rho}\hat{K}^{\mu\phantom{\nu}\rho}_{\phantom{\mu}\nu}=W^{\rho}_{\phantom{\rho}\nu\mu}\hat{K}^{\mu\phantom{\rho}\lambda}_{\phantom{\mu}\rho}+\nonumber\\
    &-(\hat{T}^{\rho}_{\phantom{\rho}\mu\nu}+\hat{K}_{\nu\mu}^{\phantom{\nu\mu}\rho})\hat{K}^{\lambda\phantom{\rho}\mu}_{\phantom{\lambda}\rho}-W^{\lambda}_{\phantom{\lambda}\mu\rho}\hat{K}^{\mu\phantom{\nu}\rho}_{\phantom{\mu}\nu}=W^{\rho}_{\phantom{\rho}\nu\mu}\hat{K}^{\mu\phantom{\rho}\lambda}_{\phantom{\mu}\rho}+\nonumber\\
 &+\hat{K}^{\mu\phantom{\nu}\rho}_{\phantom{\mu}\nu}(W^{\lambda}_{\phantom{\lambda}\mu\rho}-\accentset{\circ}{\Gamma}^{\lambda}_{\phantom{\lambda}\mu\rho})-W^{\lambda}_{\phantom{\lambda}\mu\rho}\hat{K}				^{\mu\phantom{\nu}\rho}_{\phantom{\mu}\nu}=W^{\rho}_{\phantom{\rho}\nu\mu}\hat{K}^{\mu\phantom{\rho}\lambda}_{\phantom{\mu}\rho}
\end{align}
\begin{align}\label{3.59}
\hat{T}_{\mu}\hat{K}^{\mu\phantom{\nu}\lambda}_{\phantom{\mu}\nu}+\hat{T}^{\lambda}W^{\mu}_{\phantom{\mu}\nu\mu}&-(W^{\lambda}_{\phantom{\lambda}\nu\rho}-\hat{K}^{\lambda}_{\phantom{\lambda}\nu\rho})\hat{T}^{\rho}=\nonumber\\
 &=W^{\mu}_{\phantom{\mu}\nu\mu}\hat{T}^{\lambda}-W^{\lambda}_{\phantom{\lambda}\nu\mu}\hat{T}^{\mu}.
\end{align}
Plugging Eqs$\thinspace$\eqref{3.58} and \eqref{3.59} into Eq.$\thinspace$\eqref{3.57}, the previously mentioned relation is obtained
\begin{equation}\label{3.60}
	\accentset{\circ}{R}^{\lambda}_{\phantom{\lambda}\nu}=-\frac{1}{e}[\partial_{\mu}(e\hat{K}^{\mu\phantom{\nu}\lambda}_{\phantom{\mu}\nu})+\delta^{\mu}_{\nu}\partial_{\mu}(e\hat{T}^{\lambda})]+2\hat{S}_{\rho}^{\phantom{\rho}\mu\lambda}W^{\rho}_{\phantom{\rho}\nu\mu}.
\end{equation}
Finally, one uses Eq.$\thinspace$\eqref{3.56} and Eq.$\thinspace$\eqref{3.60} in order to rewrite Eqs$\thinspace$\eqref{3.55} in the standard form \eqref{eq: equazioni di campo f(T,B)} that we report here
\begin{align}\label{eq: equazioni di campo f(T,B)A}
	&g_{\mu\nu}[-f(\hat{T}, \tilde{B})+(f_{\tilde{B}}+f_{\hat{T}})\tilde{B}]+2\accentset{\circ}{R}_{\mu\nu}f_{\hat{T}}-4[(\partial^{\rho}f_{\hat{T}})+\nonumber\\
 &+(\partial^{\rho}f_{\tilde{B}})]\hat{S}_{\mu\rho\nu}
+2\mathring{\nabla}_{\mu}\mathring{\nabla}_{\nu}f_{\tilde{B}}-2g_{\mu\nu}\mathring{\Box}f_{\tilde{B}}=2\chi T_{\mu\nu}.
\end{align}

\subsection{Derivation of $f(Q,B)$ field equations}\label{appendice f(Q,B)}

Finally we derive  the $f(\stg{Q},B)$ field equations. For the reasons  already  outlined in the previous Appendices, \ref{appedice f(R)} and \ref{appendice f(T,B)}, a different notation will be momentarily employed: the over-diamonds on symbols of the symmetric teleparallel framework are removed and the over-circles will remain in use to indicate the dependence from the Levi-Civita connection. 
In order to do this, we have to consider the variation of the Lagrangian \eqref{eq: azione f(Q,B)} with respect to the metric tensor $g_{\mu\nu}$:
\begin{align}\label{eq: fQBav}
	\delta_g S_{f(Q, B)}&=\underbrace{\int d^4x \delta_g(\sqrt{-g})f(Q, B)}_{(A'')}+\underbrace{\int d^4x \sqrt{-g}\delta_g f(Q, B)}_{(B'')} \nonumber\\
    &+\underbrace{2\chi\int d^4x \delta_g \mathcal{L}_{m}}_{(C'')}.
\end{align}
Whereas $(A'')$ and $(C'')$ can be treated identically as the $(A)$ and $(C)$ terms in Eq.$\thinspace$\eqref{eq: A.2}, leading to the expressions \eqref{eq: A.3}, \eqref{Matter} and \eqref{eq: S-ET} with the obvious replacement in Eq.$\thinspace$\eqref{eq: A.3} of $f(\mathring{R})$ with $f(Q, B)$, further calculations are needed to deal with the $(B'')$ term. The latter can be outlined as
\begin{equation}\label{eq: B''}
	(B'')=\int d^4x \sqrt{-g} [f_Q(Q, B)\delta_gQ+ f_B(Q, B)\delta_g B],
\end{equation}
so the problem is reduced to calculate the variations with respect to the metric of the non-metricity scalar $Q$ and of the boundary term $B$.
One easily obtains{}
\begin{align}\label{eq: Q1}
\delta_gQ&=\delta_g(Q_{\alpha\mu\nu})P^{\alpha\mu\nu}+Q_{\alpha\mu\nu}\delta_g P^{\alpha\mu\nu}=\nonumber\\ &=\nabla_{\alpha}(\delta_g g_{\mu\nu})P^{\alpha\mu\nu}+Q_{\alpha\mu\nu}\delta_g(g^{\mu\rho}g^{\nu\sigma}P^{\alpha}_{\phantom{\alpha}\rho\sigma})=\nonumber\\
	&=-\nabla_{\alpha}(g_{\rho\mu}g_{\sigma\nu}\delta_g g^{\rho\sigma})P^{\alpha\mu\nu}+Q_{\alpha\mu\nu}P^{\alpha\nu}_{\phantom{\alpha\nu}\rho}\delta_g g^{\mu\rho}+\nonumber\\ &+Q_{\alpha\mu\nu}P^{\alpha\mu}_{\phantom{\alpha\mu}\sigma}\delta_gg^{\nu\sigma}+Q^{\phantom{\alpha}\rho\sigma}_{\alpha}\delta_g P^{\alpha}_{\phantom{\alpha}\rho\sigma}=\nonumber\\
&=-2Q_{\alpha\lambda\mu}P^{\alpha\lambda}_{\phantom{\alpha\lambda}\nu}+P^{\lambda}_{\phantom{\lambda}\mu\nu}\nabla_{\lambda}-Q_{\alpha\mu\lambda}P^{\alpha\lambda}_{\phantom{\alpha\lambda}\nu}+\nonumber\\  &-Q_{\alpha\lambda\nu}P^{\alpha\lambda}		_{\phantom{\alpha\lambda}\mu})\delta_g g^{\mu\nu}+Q^{\phantom{\alpha}\rho\sigma}_{\alpha}\delta_g P^{\alpha}_{\phantom{\alpha}\rho\sigma}=\nonumber\\ &=-P^{\lambda}_{\phantom{\lambda}\mu\nu}\nabla_{\lambda}\delta_g g^{\mu\nu}+Q^{\phantom{\alpha}\rho\sigma}_{\alpha}\delta_g P^{\alpha}_{\phantom{\alpha}\rho\sigma}.
\end{align}
The last term in the RHS of Eq.$\thinspace$\eqref{eq: Q1} may be re-expressed in terms of $\delta_g g^{\mu\nu}$ as follows:
\begin{align}\label{eq: QB1}
	Q^{\phantom{\alpha}\rho\sigma}_{\alpha}&\delta_g P^{\alpha}_{\phantom{\alpha}\rho\sigma}=\frac{1}{4}Q^{\phantom{\alpha}\rho\sigma}_{\alpha}\Big[\underbrace{-\delta_g Q^{\alpha}_{\phantom{\alpha}\rho\sigma}}_{(B''_1)}+\underbrace{\delta_g Q^{\phantom{\rho}\alpha}_{\rho\phantom{\alpha}\sigma}}_{(B''_2)}+\underbrace{\delta_g Q^{\phantom{\sigma}\alpha}_{\sigma\phantom{\alpha}\rho}}_{(B''_3)}+\nonumber\\
    &+(\underbrace{\delta_g Q^{\alpha}}_{(B''_4)}\underbrace{-\delta_g \tilde{Q}^{\alpha}}_{(B''_5)})g_{\sigma\rho}
+\underbrace{(Q^{\alpha}-\tilde{Q}^{\alpha})\delta_g g_{\sigma\rho}}_{(B''_6)}\underbrace{-\frac{1}{2}\delta^{\alpha}_{\rho}\delta_g Q_{\sigma}}_{(B''_7)}\nonumber\\
    &\underbrace{-\frac{1}{2}\delta^{\alpha}_{\sigma}\delta_g Q_{\rho}}_{(B''_8)}\Big].
\end{align}
where
\begin{align}
	&(B''_1)=-\delta_g(g^{\alpha\lambda}Q_{\lambda\rho\sigma})=-\delta_g (g^{\alpha\lambda})Q_{\lambda\rho\sigma}-g^{\alpha\lambda}\delta_g (Q_{\lambda\rho\sigma})=\nonumber\\
    &\phantom{(B''_1)}-\delta_g (g^{\alpha\lambda})Q_{\lambda\rho\sigma}-g^{\alpha\lambda}\nabla_{\lambda}\delta_g g_{\rho\sigma}=-Q_{\lambda\rho\sigma}\delta_g g^{\alpha\lambda}+\nonumber\\
    &\phantom{(B''_1)}+g^{\alpha\lambda}\nabla_{\lambda}(g_{\mu\rho}g_{\nu\sigma}\delta_g g^{\mu\nu})=-Q_{\lambda\rho\sigma}\delta_g g^{\alpha\lambda}+\nonumber\\
	&\phantom{(B''_1)}+(Q^{\alpha}_{\phantom{\alpha}\mu\rho}g_{\nu\sigma}+Q^{\alpha}_{\phantom{\alpha}\nu\sigma}g_{\mu\rho}+g^{\alpha\lambda}g_{\mu\rho}g_{\nu\sigma}\nabla_{\lambda})\delta_g g^{\mu\nu};	\label{eq: fQB1}\\\nonumber\\
	&(B''_2)=-Q_{\rho\mu}^{\phantom{\rho\mu}\alpha}g_{\sigma\nu}\delta_g g^{\mu\nu}-g_{\sigma\nu}\nabla_{\rho}\delta_g g^{\alpha\nu};\\\nonumber\\
	&(B''_3)=-Q_{\sigma\mu}^{\phantom{\sigma\mu}\alpha}g_{\rho\nu}\delta_gg^{\mu\nu}-g_{\rho\nu} \nabla_{\sigma}\delta_g g^{\alpha\nu};\\\nonumber\\
	&(B''_4)=\delta_g(Q^{\alpha\mu}_{\phantom{\alpha\mu}\mu})=Q_{\lambda}\delta_g g^{\alpha\lambda}-Q^{\alpha}_{\phantom{\alpha}\beta\rho}\delta_g g^{\beta\rho}+\nonumber\\
    &\phantom{(B''_4)}-g^{\alpha\lambda}g_{\beta\nu}\nabla_{\lambda}\delta_g g^{\beta\nu};\\\nonumber\\
	&(B''_5)=-\delta_g(Q^{\mu\alpha}_{\phantom{\mu\alpha}\mu})=\nabla_{\nu}\delta_g g^{\alpha\nu};\\\nonumber\\
	&(B''_6)=-(Q^{\alpha}-\tilde{Q}^{\alpha})g_{\mu\sigma}g_{\nu\rho}\delta_gg^{\mu\nu};\\\nonumber\\
	&(B''_7)=-\frac{1}{2}\delta^{\alpha}_{\rho}\delta_g (Q_{\sigma\lambda}^{\phantom{\sigma\lambda}\lambda})=\frac{1}{2}\delta^{\alpha}_{\rho}(Q_{\sigma\mu\nu}+g_{\mu\nu}\nabla_{\sigma})\delta_g g^{\mu\nu};\\\nonumber\\
	&(B''_8)=\frac{1}{2}\delta^{\alpha}_{\sigma}(Q_{\rho\mu\nu}+g_{\mu\nu}\nabla_{\rho})\delta_g g^{\mu\nu}.\label{eq: fQB8}
\end{align}
Through the substitution of the equalities \eqref{eq: fQB1}-\eqref{eq: fQB8} into Eq.$\thinspace$\eqref{eq: QB1} and conveniently renaming the repeated indices, one gets {}
\begin{align}\label{eq: A.32}
	Q^{\phantom{\alpha}\rho\sigma}_{\alpha}\delta_g P^{\alpha}_{\phantom{\alpha}\rho\sigma}&=\frac{1}{4}\Big[Q^{\lambda}_{\phantom{\lambda}\mu\nu}-2Q_{\mu\nu}^{\phantom{\mu\nu}\lambda}-Q^{\lambda}g_{\mu\nu}+\delta^{\lambda}_{\nu}Q_{\mu}+\nonumber\\
    &+\tilde{Q}^{\lambda}g_{\mu\nu}\Big]\nabla_{\lambda}\delta_g g^{\mu\nu}+\frac{1}{4}\Big[-Q_{\mu}^{\phantom{\mu}\rho\sigma}Q_{\nu\rho\sigma}+\nonumber\\
    &+Q^{\alpha\rho}_{\phantom{\alpha\rho}\mu}Q_{\alpha\rho\nu}+
    Q_{\alpha\mu\sigma}Q^{\alpha\sigma}_{\phantom{\alpha\sigma}\nu}-Q_{\mu}^{\phantom{\mu}	\rho\alpha}
	Q_{\alpha\rho\nu}+\nonumber\\
	&Q^{\alpha\sigma}_{\phantom{\alpha\sigma}\nu}Q_{\sigma\mu\alpha}+Q_{\mu}Q_{\nu}-2Q_{\alpha}Q^{\alpha}_{\phantom{\alpha}\mu\nu}+\nonumber\\
    &+2\tilde{Q}_{\alpha}Q^{\alpha}_{\phantom{\alpha}\mu\nu}\Big]\delta_g g^{\mu\nu}=-P_{\lambda\mu\nu}\nabla^{\lambda}\delta_g g^{\mu\nu}+\nonumber\\
    &+\frac{1}{4}Q^{\alpha\rho}_{\phantom{\alpha\rho}\mu}[2Q_{\alpha\rho\nu}
    -2Q_{\rho\alpha\nu}-2Q_{\nu\alpha\rho}+\nonumber\\
	&-2(Q_{\alpha}-\tilde{Q}_{\alpha})g_{\rho\nu}+g_{\alpha\rho}Q_{\nu}+g_{\alpha\nu}Q_{\rho}]\delta_g g^{\mu\nu}+\nonumber\\
    &+\frac{1}{4}Q_{\mu}^{\phantom{\mu}\rho\sigma}[-Q_{\nu\rho\sigma}
	-Q_{\sigma\rho\nu}+g_{\rho\sigma}Q_{\nu}]\delta_g g^{\mu\nu}+\nonumber\\
    &+\frac{1}{4}Q^{\alpha\rho}_{\phantom{\alpha\rho}\mu}[Q_{\rho\alpha\nu}+2Q_{\nu\alpha\rho}-g_{\alpha\rho}Q_{\nu}+\nonumber\\
	&-g_{\alpha\nu}Q_{\rho}]\delta_g g^{\mu\nu}=(Q_{\mu}^{\phantom{\mu}\rho\sigma}P_{\nu\rho\sigma}+\nonumber\\
    &-P_{\lambda\mu\nu}\nabla^{\lambda})\delta_g g^{\mu\nu}-2Q^{\alpha\rho}_{\phantom{\alpha\rho}\mu}P_{\alpha\rho\nu}\delta_gg^{\mu\nu}
\end{align}
where some terms have been added and subtracted so to obtain a shorter expression in terms of the $Q_{\alpha\mu\nu}$ and $P_{\alpha\mu\nu}$ tensors.
\\\\Plugging the relation \eqref{eq: A.32} into Eq.$\thinspace$\eqref{eq: Q1}, one gets the following equality
\begin{align}
	\sqrt{-g}f_Q(Q, B)&\delta_g Q=\sqrt{-g}f_Q(Q,B)(Q_{\mu}^{\phantom{\mu}\rho\sigma}P_{\nu\rho\sigma}+\nonumber\\
    &-2P^{\lambda}_{\phantom{\lambda}\mu\nu}\nabla_{\lambda}-2Q^{\alpha\rho}_{\phantom{\alpha\rho}\mu}P_{\alpha\rho\nu})\delta_g g^{\mu\nu},
\end{align}
which can be rewritten as
\begin{align}\label{eq: B.61}
	&\sqrt{-g}f_Q(Q, B)\delta_g Q=\sqrt{-g}[f_Q(Q_{\mu}^{\phantom{\mu}\rho\sigma}P_{\nu\rho\sigma}+Q_{\lambda}P^{\lambda}_{\phantom{\lambda}\mu\nu}+\nonumber\\
    &+2\nabla_{\lambda}P^{\lambda}_{\phantom{\lambda}\mu\nu}-2Q^{\alpha\rho}_{\phantom{\alpha\rho}\mu}P_{\alpha\rho\nu})+2P^{\lambda}_{\phantom{\lambda}\mu\nu}\nabla_{\lambda}f_Q]\delta_g g^{\mu\nu},
\end{align}
having omitted boundary terms that do not contribute to the field equations and having noticed that 
\begin{align}\label{eq: B.62}
	\nabla_{\lambda}\sqrt{-g}&=(\nabla_{\lambda}-\accentset{\circ}{\nabla}_{\lambda})\sqrt{-g}=-(\Gamma^{\sigma}_{\phantom{\sigma}\lambda\sigma}-\accentset{\circ}{\Gamma}^{\sigma}_{\phantom{\sigma}\lambda\sigma})\sqrt{-g}=\nonumber\\
    &=-L^{\sigma}_{\phantom{\sigma}\lambda\sigma}\sqrt{-g}=\frac{1}{2}Q_{\lambda}\sqrt{-g}.
\end{align}
Finally, the remaining integrand term in $(B'')$ may be computed recalling the definition of the boundary term $B$ and denoting the difference $Q^{\mu}-\tilde{Q}^{\mu}$ as $\bar{Q}^{\mu}$ for compactness purposes. One easily obtains:
\begin{align}\label{eq: A.21}
	&\sqrt{-g}f_B\delta_g B =\sqrt{-g} f_B \delta_g [\mathring{\nabla}_{\alpha}\bar{Q}^{\alpha}]=\sqrt{-g}f_B\delta_g [\partial_{\alpha}\bar{Q}^{\alpha}+\nonumber\\
    &+\mathring{\Gamma}^{\alpha}_{\phantom{\alpha}\alpha\mu}\bar{Q}^{\mu}]
	=\sqrt{-g}f_B\delta_g\Big[\frac{1}{\sqrt{-g}}\partial_{\mu}(\sqrt{-g}\bar{Q}^{\mu})\Big]=\nonumber\\
    &=\sqrt{-g}f_B\delta_g\Big(\frac{1}{\sqrt{-g}}\Big)\partial_{\mu}(\sqrt{-g}\bar{Q}^{\mu})+\nonumber\\
	&+f_B \partial_{\mu}[\delta_g(\sqrt{-g})\bar{Q}^{\mu}+\sqrt{-g}\delta_g (\bar{Q}^{\mu})]=\nonumber\\
    &=\frac{1}{2}\sqrt{-g}g_{\mu\nu}f_B B\delta_g g^{\mu\nu}+\nonumber\\
    &+\frac{1}{2}\sqrt{-g}g_{\mu\nu}\partial_{\alpha}(f_B)\bar{Q}^{\alpha}\delta_g g^{\mu\nu}
    -\sqrt{-g}\partial_{\alpha}(f_B)\delta_g (\bar{Q}^{\alpha})
\end{align}
where once again total divergence terms that can be traced back to boundary terms are neglected.
\\The aim of gaining an expression in terms of $\delta g^{\mu\nu}$ implies the necessity of rewriting the third term in the last equality of Eq.$\thinspace$\eqref{eq: A.21} as follows{}
\begin{align}
	&-\sqrt{-g}\partial_{\alpha}(f_B)\delta_g\bar{Q}^{\alpha}=-\sqrt{-g}\partial_{\alpha}(f_B)\delta_g[g^{\lambda\alpha}g^{\mu\nu}(Q_{\lambda\mu\nu}+\nonumber\\
    &-Q_{\mu\nu\lambda})]
	=-\sqrt{-g}\partial_{\alpha}(f_B)\{[-2Q^{\alpha}_{\phantom{\alpha}\mu\nu}-g_{\mu\nu}\nabla^{\alpha}+Q^{\phantom{\mu\nu}\alpha}_{\mu\nu}+\nonumber\\
    &+\delta^{\alpha}_{\mu}(\tilde{Q}_{\nu}
	+\nabla_{\nu})]\delta_g g^{\mu\nu}+(Q_{\lambda}-\tilde{Q}_{\lambda})\delta_g g^{\lambda\alpha}+(Q^{\alpha}_{\phantom{\alpha}\mu\nu}+\nonumber\\
    &-Q^{\phantom{\mu\nu}\alpha}_{\mu\nu})\delta_g g^{\mu\nu}\}=-\sqrt{-g}\partial_{\alpha}(f_B)[-Q^{\alpha}_{\phantom{\alpha}\mu\nu}-g_{\mu\nu}\nabla^{\alpha}+\nonumber\\
    &+\delta^{\alpha}_{\mu}(Q_{\nu}+\nabla_{\nu})]\delta_g g^{\mu\nu}.
\end{align}
This result may now be inserted in Eq.$\thinspace$\eqref{eq: A.21} and, consequently, the expression below is obtained {}
\begin{align}\label{eq: B.65}
	&\sqrt{-g}f_B\delta_g B=\sqrt{-g}\Big[\frac{1}{2}g_{\mu\nu}f_B B+\frac{1}{2}g_{\mu\nu}\partial_{\alpha}(f_B)\bar{Q}^{\alpha}+\nonumber\\
    &+\partial_{\alpha}(f_B)(Q^{\alpha}_{\phantom{\alpha}\mu\nu}
	+g_{\mu\nu}\nabla^{\alpha})-\partial_{\mu}(f_B)(Q_{\nu}+\nabla_{\nu})\Big]\delta_g g^{\mu\nu}.
\end{align}
Using Eq.$\thinspace$\eqref{eq: B.62} and being
\begin{equation}
	\nabla_{\alpha}(\accentset{\circ}{\nabla}^{\alpha}f_B)=\accentset{\circ}{\nabla}_{\alpha}\accentset{\circ}{\nabla}^{\alpha}f_B+L^{\alpha}_{\phantom{\alpha}\alpha\lambda}\accentset{\circ}{\nabla}^{\lambda}f_B,
\end{equation}
the relation \eqref{eq: B.65} is then straightforwardly recast as{}
\begin{align}\label{eq: 3.59}
	&\sqrt{-g}f_B\delta_g B=\sqrt{-g}\Big[\frac{1}{2}g_{\mu\nu}f_B B-g_{\mu\nu}\accentset{\circ}{\nabla}_{\alpha}\accentset{\circ}{\nabla}^{\alpha}f_B+\nonumber\\
    &+\accentset{\circ}{\nabla}_{\nu}\accentset{\circ}{\nabla}_{\mu}f_B
   +2\accentset{\circ}{\nabla}_{\alpha}f_BP^{\alpha}_{\phantom{\alpha}\mu\nu}\Big]\delta_g g^{\mu\nu}.
\end{align}
The $f(\stg{Q}, B)$ metric field equations are thus determined by substituting Eqs$\thinspace$\eqref{eq: B.61} and \eqref{eq: 3.59} into Eq.$\thinspace$\eqref{eq: B''} and by inserting the latter jointly with $(A'')$ and $(C'')$ in \eqref{eq: fQBav} equated to zero. In the standard notation they read as Eq. \eqref{eq: 3.59}
\begin{align}\label{eq: B.68}
	&f_{\stg{Q}}(\stg{Q}{_{\mu}^{\phantom{\mu}\rho\sigma}}\stg{P}{_{\nu\rho\sigma}}+\stg{Q}{_{\lambda}}\stg{P}{^{\lambda}_{\phantom{\lambda}\mu\nu}}+2\stg{\nabla}{_{\lambda}}\stg{P}{^{\lambda}_{\phantom{\lambda}\mu\nu}}-2\stg{Q}{^{\alpha\rho}{}_{\mu}}\stg{P}{_{\alpha\rho\nu}})+\nonumber\\
 &+2\stg{P}{^{\lambda}_{\phantom{\lambda}\mu\nu}}\accentset{\circ}{\nabla}_{\lambda}(f_{\stg{Q}}+f_B)+\accentset{\circ}{\nabla}_{\mu}\accentset{\circ}{\nabla}_{\nu}f_B-g_{\mu\nu}\accentset{\circ}{\Box}f_B+\nonumber\\
 &-\frac{1}{2}g_{\mu\nu}(f-f_BB)=\chi T_{\mu\nu}.
\end{align}
It turns out that the first term in parenthesis in Eq. \eqref{eq: 3.59} can be equated to the Einstein tensor
\begin{equation}
	\accentset{\circ}{G}_{\mu\nu}=\accentset{\circ}{R}_{\mu\nu}-\frac{1}{2}g_{\mu\nu}\accentset{\circ}{R}
\end{equation}
plus an additional term depending on the non-metricity scalar.
\\Indeed, from Eq.$\thinspace$\eqref{eq: STT Riemann} and Eq.$\thinspace$\eqref{3.27}, one has
\begin{align}\label{eq: additional}
	&\accentset{\circ}{R}_{\mu\nu}-\frac{1}{2}g_{\mu\nu}\accentset{\circ}{R}=-\accentset{\circ}{\nabla}_{\rho}\stg{L}{^{\rho}_{\phantom{\rho}\mu\nu}}+\accentset{\circ}{\nabla}_{\nu}\stg{\tilde{L}}{_{\mu}}-\stg{\tilde{L}}{_{\sigma}}\stg{L}{^{\sigma}_{\phantom{\sigma}\mu\nu}}+\nonumber\\
    &+\stg{L}{^{\rho}_{\phantom{\rho}\nu\sigma}}\stg{L}{^{\sigma}_{\phantom{\sigma}\mu\nu}}-\frac{1}{2}g_{\mu\nu}(\stg{Q}-B)
	=\stg{Q}{_{\mu}^{\phantom{\mu}\rho\sigma}}\stg{P}{_{\nu\rho\sigma}}+\stg{Q}{_{\lambda}}\stg{P}{^{\lambda}_{\phantom{\lambda}\mu\nu}}+\nonumber\\
    &+2\stg{\nabla}{_{\lambda}}\stg{P}{^{\lambda}_{\phantom{\lambda}\mu\nu}}-2\stg{Q}{^{\alpha\rho}_{\phantom{\alpha\rho}\mu}}\stg{P}{_{\alpha\rho\nu}}-\frac{1}{2}g_{\mu\nu}\stg{Q}
\end{align}
being
\begin{align}
    &\accentset{\circ}{\nabla}_{\rho}\stg{L}{^{\rho}_{\mu\nu}}=\stg{\nabla}{_{\rho}}\stg{L}{^{\rho}_{\phantom{\rho}\mu\nu}}-\stg{\tilde{L}}{_{\sigma}}\stg{L}
	{^{\sigma}_{\phantom{\sigma}\mu\nu}}+\stg{L}{^{\rho}_{\phantom{\rho}\nu\sigma}}\stg{L}{^{\sigma}_{\phantom{\sigma}\mu\nu}}+\stg{L}{^{\sigma}_{\phantom{\sigma}\rho\nu}}\stg{L}{^{\rho}_{\phantom{\rho}\mu\sigma}};\nonumber\\
	&\accentset{\circ}{\nabla}_{\nu}\stg{\tilde{L}}{_{\mu}}=-\accentset{\circ}{\nabla}_{\nu}\stg{Q}{_{\mu}}=-\frac{1}{2}(\stg{\nabla}{_{\nu}}\stg{Q}{_{\mu}}+
	\stg{L}{^{\rho}_{\phantom{\rho}\mu\nu}}\stg{Q}{_{\rho}});\nonumber
 \end{align}
 \begin{align}
	\frac{1}{2}g_{\mu\nu}(\stg{Q}-B)=\frac{1}{2}g_{\mu\nu}&[\stg{Q}-\stg{\nabla}{_{\rho}}(\stg{Q}{^{\rho}}-\stg{\tilde{Q}}{^{\rho}})+\nonumber\\
    &-\stg{Q}{_{\rho\phantom{\sigma}\sigma}^{\phantom{\rho}\sigma}}(\stg{Q}{^{\rho}}-\stg{\tilde{Q}}{^{\rho}})].\nonumber
\end{align}
Then, Eq. \eqref{eq: additional} can thus be inserted in Eqs$\thinspace$\eqref{eq: 3.59} leading to the desired field Eqs. \eqref{eq: 3.66} that we report here for the sake of completeness
\begin{align}\label{eq: 3.66A}
	&f_{\stg{Q}}(\accentset{\circ}{R}_{\mu\nu}-\frac{1}{2}g_{\mu\nu}\accentset{\circ}{R})+2\stg{P}{^{\lambda}_{\phantom{\lambda}\mu\nu}}\accentset{\circ}{\nabla}_{\lambda}(f_{\stg{Q}}+f_B)
    +\accentset{\circ}{\nabla}_{\mu}\accentset{\circ}{\nabla}_{\nu}f_B+\nonumber\\
    &-g_{\mu\nu}\accentset{\circ}{\Box}f_B
	-\frac{1}{2}g_{\mu\nu}(f-f_BB-f_{\stg{Q}}\stg{Q})=\chi T_{\mu\nu}.
\end{align}

\medskip
\bibliography{ref_EGTG}

\begin{thebibliography}{98}%
\makeatletter
\providecommand \@ifxundefined [1]{%
 \@ifx{#1\undefined}
}%
\providecommand \@ifnum [1]{%
 \ifnum #1\expandafter \@firstoftwo
 \else \expandafter \@secondoftwo
 \fi
}%
\providecommand \@ifx [1]{%
 \ifx #1\expandafter \@firstoftwo
 \else \expandafter \@secondoftwo
 \fi
}%
\providecommand \natexlab [1]{#1}%
\providecommand \enquote  [1]{``#1''}%
\providecommand \bibnamefont  [1]{#1}%
\providecommand \bibfnamefont [1]{#1}%
\providecommand \citenamefont [1]{#1}%
\providecommand \href@noop [0]{\@secondoftwo}%
\providecommand \href [0]{\begingroup \@sanitize@url \@href}%
\providecommand \@href[1]{\@@startlink{#1}\@@href}%
\providecommand \@@href[1]{\endgroup#1\@@endlink}%
\providecommand \@sanitize@url [0]{\catcode `\\12\catcode `\$12\catcode
  `\&12\catcode `\#12\catcode `\^12\catcode `\_12\catcode `\%12\relax}%
\providecommand \@@startlink[1]{}%
\providecommand \@@endlink[0]{}%
\providecommand \url  [0]{\begingroup\@sanitize@url \@url }%
\providecommand \@url [1]{\endgroup\@href {#1}{\urlprefix }}%
\providecommand \urlprefix  [0]{URL }%
\providecommand \Eprint [0]{\href }%
\providecommand \doibase [0]{http://dx.doi.org/}%
\providecommand \selectlanguage [0]{\@gobble}%
\providecommand \bibinfo  [0]{\@secondoftwo}%
\providecommand \bibfield  [0]{\@secondoftwo}%
\providecommand \translation [1]{[#1]}%
\providecommand \BibitemOpen [0]{}%
\providecommand \bibitemStop [0]{}%
\providecommand \bibitemNoStop [0]{.\EOS\space}%
\providecommand \EOS [0]{\spacefactor3000\relax}%
\providecommand \BibitemShut  [1]{\csname bibitem#1\endcsname}%
\let\auto@bib@innerbib\@empty
\bibitem [{\citenamefont {Wald}(1984)}]{Wald}%
  \BibitemOpen
  \bibfield  {author} {\bibinfo {author} {\bibfnamefont {R.~M.}\ \bibnamefont
  {Wald}},\ }\href {\doibase 10.7208/chicago/9780226870373.001.0001} {\emph
  {\bibinfo {title} {{General Relativity}}}}\ (\bibinfo  {publisher} {Chicago
  Univ. Pr.},\ \bibinfo {address} {Chicago, USA},\ \bibinfo {year}
  {1984})\BibitemShut {NoStop}%
\bibitem [{\citenamefont {Straumann}(2013)}]{Straumann}%
  \BibitemOpen
  \bibfield  {author} {\bibinfo {author} {\bibfnamefont {N.}~\bibnamefont
  {Straumann}},\ }\href {\doibase 10.1007/978-94-007-5410-2} {\emph {\bibinfo
  {title} {{General Relativity}}}},\ Graduate Texts in Physics\ (\bibinfo
  {publisher} {Springer},\ \bibinfo {address} {Dordrecht},\ \bibinfo {year}
  {2013})\BibitemShut {NoStop}%
\bibitem [{\citenamefont {Capozziello}\ and\ \citenamefont
  {De~Laurentis}(2011)}]{CDL}%
  \BibitemOpen
  \bibfield  {author} {\bibinfo {author} {\bibfnamefont {S.}~\bibnamefont
  {Capozziello}}\ and\ \bibinfo {author} {\bibfnamefont {M.}~\bibnamefont
  {De~Laurentis}},\ }\href {\doibase 10.1016/j.physrep.2011.09.003} {\bibfield
  {journal} {\bibinfo  {journal} {Phys. Rept.}\ } (\bibinfo {year} {2011}),\
  10.1016/j.physrep.2011.09.003},\ \Eprint {http://arxiv.org/abs/1108.6266}
  {1108.6266} \BibitemShut {NoStop}%
\bibitem [{\citenamefont {Capozziello}\ and\ \citenamefont
  {Faraoni}(2011)}]{FC}%
  \BibitemOpen
  \bibfield  {author} {\bibinfo {author} {\bibfnamefont {S.}~\bibnamefont
  {Capozziello}}\ and\ \bibinfo {author} {\bibfnamefont {V.}~\bibnamefont
  {Faraoni}},\ }\href {https://doi.org/10.1007%2F978-94-007-0165-6} {\emph
  {\bibinfo {title} {Beyond Einstein Gravity}}}\ (\bibinfo  {publisher}
  {Springer Netherlands},\ \bibinfo {year} {2011})\BibitemShut {NoStop}%
\bibitem [{\citenamefont {{Will}}(1993)}]{Will1993}%
  \BibitemOpen
  \bibfield  {author} {\bibinfo {author} {\bibfnamefont {C.~M.}\ \bibnamefont
  {{Will}}},\ }\href {\doibase 10.1017/9781316338612} {\emph {\bibinfo {title}
  {{Theory and Experiment in Gravitational Physics}}}}\ (\bibinfo  {publisher}
  {Cambridge University Press},\ \bibinfo {year} {1993})\BibitemShut {NoStop}%
\bibitem [{\citenamefont {{Ni}}(2016)}]{Ni2016}%
  \BibitemOpen
  \bibfield  {author} {\bibinfo {author} {\bibfnamefont {W.-T.}\ \bibnamefont
  {{Ni}}},\ }\href {\doibase 10.1142/S0218271816300032} {\bibfield  {journal}
  {\bibinfo  {journal} {International Journal of Modern Physics D}\ }\textbf
  {\bibinfo {volume} {25}},\ \bibinfo {eid} {1630003-786} (\bibinfo {year}
  {2016})},\ \Eprint {http://arxiv.org/abs/1611.06025} {arXiv:1611.06025
  [gr-qc]} \BibitemShut {NoStop}%
\bibitem [{\citenamefont {{De Marchi}}\ and\ \citenamefont
  {{Cascioli}}(2020)}]{DeMarchi2020}%
  \BibitemOpen
  \bibfield  {author} {\bibinfo {author} {\bibfnamefont {F.}~\bibnamefont {{De
  Marchi}}}\ and\ \bibinfo {author} {\bibfnamefont {G.}~\bibnamefont
  {{Cascioli}}},\ }\href {\doibase 10.1088/1361-6382/ab6ae0} {\bibfield
  {journal} {\bibinfo  {journal} {Classical and Quantum Gravity}\ }\textbf
  {\bibinfo {volume} {37}},\ \bibinfo {eid} {095007} (\bibinfo {year}
  {2020})},\ \Eprint {http://arxiv.org/abs/1911.05561} {arXiv:1911.05561
  [gr-qc]} \BibitemShut {NoStop}%
\bibitem [{\citenamefont {Abbott}\ \emph {et~al.}(2016)\citenamefont {Abbott}
  \emph {et~al.}}]{Abbott2016}%
  \BibitemOpen
  \bibfield  {author} {\bibinfo {author} {\bibfnamefont {B.~P.}\ \bibnamefont
  {Abbott}} \emph {et~al.} (\bibinfo {collaboration} {LIGO Scientific
  Collaboration and Virgo Collaboration}),\ }\href {\doibase
  10.1103/PhysRevLett.116.061102} {\bibfield  {journal} {\bibinfo  {journal}
  {Phys. Rev. Lett.}\ }\textbf {\bibinfo {volume} {116}},\ \bibinfo {pages}
  {061102} (\bibinfo {year} {2016})}\BibitemShut {NoStop}%
\bibitem [{\citenamefont {Abbott}\ \emph
  {et~al.}(2021{\natexlab{a}})\citenamefont {Abbott} \emph
  {et~al.}}]{Abbott2022}%
  \BibitemOpen
  \bibfield  {author} {\bibinfo {author} {\bibfnamefont {R.}~\bibnamefont
  {Abbott}} \emph {et~al.} (\bibinfo {collaboration} {LIGO Scientific, VIRGO,
  KAGRA}),\ }\href@noop {} {\bibfield  {journal} {\bibinfo  {journal} {arXiv
  e-prints}\ } (\bibinfo {year} {2021}{\natexlab{a}})},\ \Eprint
  {http://arxiv.org/abs/2111.03606} {arXiv:2111.03606 [gr-qc]} \BibitemShut
  {NoStop}%
\bibitem [{\citenamefont {{Sathyaprakash}}\ and\ \citenamefont
  {{Schutz}}(2009)}]{Sathyaprakash2009}%
  \BibitemOpen
  \bibfield  {author} {\bibinfo {author} {\bibfnamefont {B.~S.}\ \bibnamefont
  {{Sathyaprakash}}}\ and\ \bibinfo {author} {\bibfnamefont {B.~F.}\
  \bibnamefont {{Schutz}}},\ }\href {\doibase 10.12942/lrr-2009-2} {\bibfield
  {journal} {\bibinfo  {journal} {Living Reviews in Relativity}\ }\textbf
  {\bibinfo {volume} {12}},\ \bibinfo {eid} {2} (\bibinfo {year} {2009})},\
  \Eprint {http://arxiv.org/abs/0903.0338} {arXiv:0903.0338 [gr-qc]}
  \BibitemShut {NoStop}%
\bibitem [{\citenamefont {Bailes}\ \emph {et~al.}(2021)\citenamefont {Bailes}
  \emph {et~al.}}]{Bailes2021}%
  \BibitemOpen
  \bibfield  {author} {\bibinfo {author} {\bibfnamefont {M.}~\bibnamefont
  {Bailes}} \emph {et~al.},\ }\href@noop {} {\bibfield  {journal} {\bibinfo
  {journal} {Nature Reviews Physics}\ } (\bibinfo {year} {2021})}\BibitemShut
  {NoStop}%
\bibitem [{\citenamefont {Abbott}\ \emph
  {et~al.}(2021{\natexlab{b}})\citenamefont {Abbott} \emph
  {et~al.}}]{LIGOScientific2021psn}%
  \BibitemOpen
  \bibfield  {author} {\bibinfo {author} {\bibfnamefont {R.}~\bibnamefont
  {Abbott}} \emph {et~al.} (\bibinfo {collaboration} {LIGO Scientific, VIRGO,
  KAGRA}),\ }\href@noop {} {\  (\bibinfo {year} {2021}{\natexlab{b}})},\
  \Eprint {http://arxiv.org/abs/2111.03634} {arXiv:2111.03634 [astro-ph.HE]}
  \BibitemShut {NoStop}%
\bibitem [{\citenamefont {{Event Horizon Telescope Collaboration}}\ \emph
  {et~al.}(2019{\natexlab{a}})\citenamefont {{Event Horizon Telescope
  Collaboration}}, \citenamefont {{Akiyama}} \emph {et~al.}}]{EHC20191}%
  \BibitemOpen
  \bibfield  {author} {\bibinfo {author} {\bibnamefont {{Event Horizon
  Telescope Collaboration}}}, \bibinfo {author} {\bibfnamefont
  {K.}~\bibnamefont {{Akiyama}}},  \emph {et~al.},\ }\href {\doibase
  10.3847/2041-8213/ab0ec7} {\bibfield  {journal} {\bibinfo  {journal} {ApJ}\
  }\textbf {\bibinfo {volume} {875}},\ \bibinfo {eid} {L1} (\bibinfo {year}
  {2019}{\natexlab{a}})}\BibitemShut {NoStop}%
\bibitem [{\citenamefont {{Event Horizon Telescope Collaboration}}\ \emph
  {et~al.}(2019{\natexlab{b}})\citenamefont {{Event Horizon Telescope
  Collaboration}}, \citenamefont {{Akiyama}} \emph {et~al.}}]{EHC20192}%
  \BibitemOpen
  \bibfield  {author} {\bibinfo {author} {\bibnamefont {{Event Horizon
  Telescope Collaboration}}}, \bibinfo {author} {\bibfnamefont
  {K.}~\bibnamefont {{Akiyama}}},  \emph {et~al.},\ }\href {\doibase
  10.3847/2041-8213/ab0c96} {\bibfield  {journal} {\bibinfo  {journal} {ApJ}\
  }\textbf {\bibinfo {volume} {875}},\ \bibinfo {eid} {L2} (\bibinfo {year}
  {2019}{\natexlab{b}})}\BibitemShut {NoStop}%
\bibitem [{\citenamefont {{Event Horizon Telescope Collaboration}}\ \emph
  {et~al.}(2019{\natexlab{c}})\citenamefont {{Event Horizon Telescope
  Collaboration}}, \citenamefont {{Akiyama}} \emph {et~al.}}]{EHC20193}%
  \BibitemOpen
  \bibfield  {author} {\bibinfo {author} {\bibnamefont {{Event Horizon
  Telescope Collaboration}}}, \bibinfo {author} {\bibfnamefont
  {K.}~\bibnamefont {{Akiyama}}},  \emph {et~al.},\ }\href {\doibase
  10.3847/2041-8213/ab0c57} {\bibfield  {journal} {\bibinfo  {journal} {ApJ}\
  }\textbf {\bibinfo {volume} {875}},\ \bibinfo {eid} {L3} (\bibinfo {year}
  {2019}{\natexlab{c}})}\BibitemShut {NoStop}%
\bibitem [{\citenamefont {{Event Horizon Telescope Collaboration}}\ \emph
  {et~al.}(2019{\natexlab{d}})\citenamefont {{Event Horizon Telescope
  Collaboration}}, \citenamefont {{Akiyama}} \emph {et~al.}}]{EHC20194}%
  \BibitemOpen
  \bibfield  {author} {\bibinfo {author} {\bibnamefont {{Event Horizon
  Telescope Collaboration}}}, \bibinfo {author} {\bibfnamefont
  {K.}~\bibnamefont {{Akiyama}}},  \emph {et~al.},\ }\href {\doibase
  10.3847/2041-8213/ab0e85} {\bibfield  {journal} {\bibinfo  {journal} {ApJ}\
  }\textbf {\bibinfo {volume} {875}},\ \bibinfo {eid} {L4} (\bibinfo {year}
  {2019}{\natexlab{d}})}\BibitemShut {NoStop}%
\bibitem [{\citenamefont {{Event Horizon Telescope Collaboration}}\ \emph
  {et~al.}(2019{\natexlab{e}})\citenamefont {{Event Horizon Telescope
  Collaboration}}, \citenamefont {{Akiyama}} \emph {et~al.}}]{EHC20195}%
  \BibitemOpen
  \bibfield  {author} {\bibinfo {author} {\bibnamefont {{Event Horizon
  Telescope Collaboration}}}, \bibinfo {author} {\bibfnamefont
  {K.}~\bibnamefont {{Akiyama}}},  \emph {et~al.},\ }\href {\doibase
  10.3847/2041-8213/ab0f43} {\bibfield  {journal} {\bibinfo  {journal} {ApJ}\
  }\textbf {\bibinfo {volume} {875}},\ \bibinfo {eid} {L5} (\bibinfo {year}
  {2019}{\natexlab{e}})}\BibitemShut {NoStop}%
\bibitem [{\citenamefont {{Event Horizon Telescope Collaboration}}\ \emph
  {et~al.}(2019{\natexlab{f}})\citenamefont {{Event Horizon Telescope
  Collaboration}}, \citenamefont {{Akiyama}} \emph {et~al.}}]{EHC20196}%
  \BibitemOpen
  \bibfield  {author} {\bibinfo {author} {\bibnamefont {{Event Horizon
  Telescope Collaboration}}}, \bibinfo {author} {\bibfnamefont
  {K.}~\bibnamefont {{Akiyama}}},  \emph {et~al.},\ }\href {\doibase
  10.3847/2041-8213/ab1141} {\bibfield  {journal} {\bibinfo  {journal} {ApJ}\
  }\textbf {\bibinfo {volume} {875}},\ \bibinfo {eid} {L6} (\bibinfo {year}
  {2019}{\natexlab{f}})}\BibitemShut {NoStop}%
\bibitem [{\citenamefont {Lichtenegger}\ and\ \citenamefont
  {Mashhoon}(2008)}]{Mach}%
  \BibitemOpen
  \bibfield  {author} {\bibinfo {author} {\bibfnamefont {H.}~\bibnamefont
  {Lichtenegger}}\ and\ \bibinfo {author} {\bibfnamefont {B.}~\bibnamefont
  {Mashhoon}},\ }\href@noop {} {\enquote {\bibinfo {title} {Mach's
  principle},}\ } (\bibinfo {year} {2008}),\ \Eprint
  {http://arxiv.org/abs/physics/0407078} {arXiv:physics/0407078} \BibitemShut
  {NoStop}%
\bibitem [{\citenamefont {Capozziello}\ \emph {et~al.}(2011)\citenamefont
  {Capozziello}, \citenamefont {Basini},\ and\ \citenamefont
  {De~Laurentis}}]{Capozziello:2011}%
  \BibitemOpen
  \bibfield  {author} {\bibinfo {author} {\bibfnamefont {S.}~\bibnamefont
  {Capozziello}}, \bibinfo {author} {\bibfnamefont {G.}~\bibnamefont {Basini}},
  \ and\ \bibinfo {author} {\bibfnamefont {M.}~\bibnamefont {De~Laurentis}},\
  }\href {\doibase 10.1140/epjc/s10052-011-1679-1} {\bibfield  {journal}
  {\bibinfo  {journal} {Eur. Phys. J. C}\ }\textbf {\bibinfo {volume} {71}},\
  \bibinfo {pages} {1679} (\bibinfo {year} {2011})},\ \Eprint
  {http://arxiv.org/abs/1105.6193} {arXiv:1105.6193 [gr-qc]} \BibitemShut
  {NoStop}%
\bibitem [{\citenamefont {{'t Hooft}}\ and\ \citenamefont
  {{Veltman}}(1974)}]{oneloop1974}%
  \BibitemOpen
  \bibfield  {author} {\bibinfo {author} {\bibfnamefont {G.}~\bibnamefont {{'t
  Hooft}}}\ and\ \bibinfo {author} {\bibfnamefont {M.}~\bibnamefont
  {{Veltman}}},\ }\href@noop {} {\bibfield  {journal} {\bibinfo  {journal}
  {Annales de L'Institut Henri Poincare Section (A) Physique Theorique}\
  }\textbf {\bibinfo {volume} {20}},\ \bibinfo {pages} {69} (\bibinfo {year}
  {1974})}\BibitemShut {NoStop}%
\bibitem [{\citenamefont {{Obukhov}}(2017)}]{Obukhov2017}%
  \BibitemOpen
  \bibfield  {author} {\bibinfo {author} {\bibfnamefont {Y.~N.}\ \bibnamefont
  {{Obukhov}}},\ }\href {\doibase 10.1103/PhysRevD.95.084028} {\bibfield
  {journal} {\bibinfo  {journal} {Phys. Rev. D}\ }\textbf {\bibinfo {volume}
  {95}},\ \bibinfo {eid} {084028} (\bibinfo {year} {2017})},\ \Eprint
  {http://arxiv.org/abs/1702.05185} {arXiv:1702.05185 [gr-qc]} \BibitemShut
  {NoStop}%
\bibitem [{\citenamefont {Kibble}(1961)}]{Kible1961}%
  \BibitemOpen
  \bibfield  {author} {\bibinfo {author} {\bibfnamefont {T.~W.~B.}\
  \bibnamefont {Kibble}},\ }\href {\doibase 10.1063/1.1703702} {\bibfield
  {journal} {\bibinfo  {journal} {Journal of Mathematical Physics}\ }\textbf
  {\bibinfo {volume} {2}},\ \bibinfo {pages} {212} (\bibinfo {year} {1961})},\
  \Eprint {http://arxiv.org/abs/https://doi.org/10.1063/1.1703702}
  {https://doi.org/10.1063/1.1703702} \BibitemShut {NoStop}%
\bibitem [{\citenamefont {{Sciama}}(1962)}]{Sciama1962}%
  \BibitemOpen
  \bibfield  {author} {\bibinfo {author} {\bibfnamefont {D.~W.}\ \bibnamefont
  {{Sciama}}},\ }in\ \href@noop {} {\emph {\bibinfo {booktitle} {Recent
  Developments in General Relativity}}}\ (\bibinfo  {publisher} {Warsaw: Polish
  Scientific Publishers},\ \bibinfo {year} {1962})\ p.\ \bibinfo {pages}
  {415}\BibitemShut {NoStop}%
\bibitem [{\citenamefont {{Hehl}}\ \emph {et~al.}(1995)\citenamefont {{Hehl}},
  \citenamefont {{McCrea}}, \citenamefont {{Mielke}},\ and\ \citenamefont
  {{Ne'eman}}}]{Hehl1995}%
  \BibitemOpen
  \bibfield  {author} {\bibinfo {author} {\bibfnamefont {F.~W.}\ \bibnamefont
  {{Hehl}}}, \bibinfo {author} {\bibfnamefont {J.~D.}\ \bibnamefont
  {{McCrea}}}, \bibinfo {author} {\bibfnamefont {E.~W.}\ \bibnamefont
  {{Mielke}}}, \ and\ \bibinfo {author} {\bibfnamefont {Y.}~\bibnamefont
  {{Ne'eman}}},\ }\href {\doibase 10.1016/0370-1573(94)00111-F} {\bibfield
  {journal} {\bibinfo  {journal} {Phys. Rev.}\ }\textbf {\bibinfo {volume}
  {258}},\ \bibinfo {pages} {1} (\bibinfo {year} {1995})},\ \Eprint
  {http://arxiv.org/abs/gr-qc/9402012} {arXiv:gr-qc/9402012 [gr-qc]}
  \BibitemShut {NoStop}%
\bibitem [{\citenamefont {Beltr\'an~Jim\'enez}\ \emph
  {et~al.}(2019)\citenamefont {Beltr\'an~Jim\'enez}, \citenamefont
  {Heisenberg},\ and\ \citenamefont {Koivisto}}]{JHK}%
  \BibitemOpen
  \bibfield  {author} {\bibinfo {author} {\bibfnamefont {J.}~\bibnamefont
  {Beltr\'an~Jim\'enez}}, \bibinfo {author} {\bibfnamefont {L.}~\bibnamefont
  {Heisenberg}}, \ and\ \bibinfo {author} {\bibfnamefont {T.~S.}\ \bibnamefont
  {Koivisto}},\ }\href {\doibase 10.3390/universe5070173} {\bibfield  {journal}
  {\bibinfo  {journal} {Universe}\ }\textbf {\bibinfo {volume} {5}},\ \bibinfo
  {pages} {173} (\bibinfo {year} {2019})},\ \Eprint
  {http://arxiv.org/abs/1903.06830} {arXiv:1903.06830 [hep-th]} \BibitemShut
  {NoStop}%
\bibitem [{\citenamefont {Capozziello}\ \emph {et~al.}(2022)\citenamefont
  {Capozziello}, \citenamefont {Falco},\ and\ \citenamefont
  {Ferrara}}]{ComparingEG}%
  \BibitemOpen
  \bibfield  {author} {\bibinfo {author} {\bibfnamefont {S.}~\bibnamefont
  {Capozziello}}, \bibinfo {author} {\bibfnamefont {V.~D.}\ \bibnamefont
  {Falco}}, \ and\ \bibinfo {author} {\bibfnamefont {C.}~\bibnamefont
  {Ferrara}},\ }\href {\doibase
  https://doi.org/10.1140/epjc/s10052-022-10823-x} {\bibfield  {journal}
  {\bibinfo  {journal} {Eur. Phys. J. C}\ } (\bibinfo {year} {2022}),\
  https://doi.org/10.1140/epjc/s10052-022-10823-x},\ \Eprint
  {http://arxiv.org/abs/2208.03011} {2208.03011} \BibitemShut {NoStop}%
\bibitem [{\citenamefont {{Clifton}}(2006)}]{Clifton2006}%
  \BibitemOpen
  \bibfield  {author} {\bibinfo {author} {\bibfnamefont {T.}~\bibnamefont
  {{Clifton}}},\ }\emph {\bibinfo {title} {{Alternative Theories of
  Gravity}}},\ \href@noop {} {Ph.D. thesis},\ \bibinfo  {school} {-} (\bibinfo
  {year} {2006})\BibitemShut {NoStop}%
\bibitem [{\citenamefont {{Capozziello}}\ and\ \citenamefont
  {{Francaviglia}}(2008)}]{Capozziello2008}%
  \BibitemOpen
  \bibfield  {author} {\bibinfo {author} {\bibfnamefont {S.}~\bibnamefont
  {{Capozziello}}}\ and\ \bibinfo {author} {\bibfnamefont {M.}~\bibnamefont
  {{Francaviglia}}},\ }\href {\doibase 10.1007/s10714-007-0551-y} {\bibfield
  {journal} {\bibinfo  {journal} {General Relativity and Gravitation}\ }\textbf
  {\bibinfo {volume} {40}},\ \bibinfo {pages} {357} (\bibinfo {year} {2008})},\
  \Eprint {http://arxiv.org/abs/0706.1146} {arXiv:0706.1146 [astro-ph]}
  \BibitemShut {NoStop}%
\bibitem [{\citenamefont {Nojiri}\ and\ \citenamefont
  {Odintsov}(2011)}]{Sergei}%
  \BibitemOpen
  \bibfield  {author} {\bibinfo {author} {\bibfnamefont {S.}~\bibnamefont
  {Nojiri}}\ and\ \bibinfo {author} {\bibfnamefont {S.~D.}\ \bibnamefont
  {Odintsov}},\ }\href {\doibase 10.1016/j.physrep.2011.04.001} {\bibfield
  {journal} {\bibinfo  {journal} {Phys. Rept.}\ }\textbf {\bibinfo {volume}
  {505}},\ \bibinfo {pages} {59} (\bibinfo {year} {2011})},\ \Eprint
  {http://arxiv.org/abs/1011.0544} {arXiv:1011.0544 [gr-qc]} \BibitemShut
  {NoStop}%
\bibitem [{\citenamefont {Nojiri}\ \emph {et~al.}(2017)\citenamefont {Nojiri},
  \citenamefont {Odintsov},\ and\ \citenamefont {Oikonomou}}]{Vasilis}%
  \BibitemOpen
  \bibfield  {author} {\bibinfo {author} {\bibfnamefont {S.}~\bibnamefont
  {Nojiri}}, \bibinfo {author} {\bibfnamefont {S.~D.}\ \bibnamefont
  {Odintsov}}, \ and\ \bibinfo {author} {\bibfnamefont {V.~K.}\ \bibnamefont
  {Oikonomou}},\ }\href {\doibase 10.1016/j.physrep.2017.06.001} {\bibfield
  {journal} {\bibinfo  {journal} {Phys. Rept.}\ }\textbf {\bibinfo {volume}
  {692}},\ \bibinfo {pages} {1} (\bibinfo {year} {2017})},\ \Eprint
  {http://arxiv.org/abs/1705.11098} {arXiv:1705.11098 [gr-qc]} \BibitemShut
  {NoStop}%
\bibitem [{\citenamefont {Nojiri}\ and\ \citenamefont
  {Odintsov}(2006)}]{Nojiri}%
  \BibitemOpen
  \bibfield  {author} {\bibinfo {author} {\bibfnamefont {S.}~\bibnamefont
  {Nojiri}}\ and\ \bibinfo {author} {\bibfnamefont {S.~D.}\ \bibnamefont
  {Odintsov}},\ }\href {\doibase 10.1142/S0219887807001928} {\bibfield
  {journal} {\bibinfo  {journal} {eConf}\ }\textbf {\bibinfo {volume}
  {C0602061}},\ \bibinfo {pages} {06} (\bibinfo {year} {2006})},\ \Eprint
  {http://arxiv.org/abs/hep-th/0601213} {arXiv:hep-th/0601213} \BibitemShut
  {NoStop}%
\bibitem [{\citenamefont {{Sotiriou}}\ and\ \citenamefont
  {{Faraoni}}(2010)}]{Sotiriou2010}%
  \BibitemOpen
  \bibfield  {author} {\bibinfo {author} {\bibfnamefont {T.~P.}\ \bibnamefont
  {{Sotiriou}}}\ and\ \bibinfo {author} {\bibfnamefont {V.}~\bibnamefont
  {{Faraoni}}},\ }\href {\doibase 10.1103/RevModPhys.82.451} {\bibfield
  {journal} {\bibinfo  {journal} {Reviews of Modern Physics}\ }\textbf
  {\bibinfo {volume} {82}},\ \bibinfo {pages} {451} (\bibinfo {year} {2010})},\
  \Eprint {http://arxiv.org/abs/0805.1726} {arXiv:0805.1726 [gr-qc]}
  \BibitemShut {NoStop}%
\bibitem [{\citenamefont {{Cai}}\ \emph {et~al.}(2016)\citenamefont {{Cai}},
  \citenamefont {{Capozziello}}, \citenamefont {{De Laurentis}},\ and\
  \citenamefont {{Saridakis}}}]{Cai2016}%
  \BibitemOpen
  \bibfield  {author} {\bibinfo {author} {\bibfnamefont {Y.-F.}\ \bibnamefont
  {{Cai}}}, \bibinfo {author} {\bibfnamefont {S.}~\bibnamefont
  {{Capozziello}}}, \bibinfo {author} {\bibfnamefont {M.}~\bibnamefont {{De
  Laurentis}}}, \ and\ \bibinfo {author} {\bibfnamefont {E.~N.}\ \bibnamefont
  {{Saridakis}}},\ }\href {\doibase 10.1088/0034-4885/79/10/106901} {\bibfield
  {journal} {\bibinfo  {journal} {Reports on Progress in Physics}\ }\textbf
  {\bibinfo {volume} {79}},\ \bibinfo {eid} {106901} (\bibinfo {year}
  {2016})},\ \Eprint {http://arxiv.org/abs/1511.07586} {arXiv:1511.07586
  [gr-qc]} \BibitemShut {NoStop}%
\bibitem [{\citenamefont {Mancini}\ \emph {et~al.}(2025)\citenamefont
  {Mancini}, \citenamefont {Tino},\ and\ \citenamefont
  {Capozziello}}]{Mancini:2025asp}%
  \BibitemOpen
  \bibfield  {author} {\bibinfo {author} {\bibfnamefont {C.}~\bibnamefont
  {Mancini}}, \bibinfo {author} {\bibfnamefont {G.~M.}\ \bibnamefont {Tino}}, \
  and\ \bibinfo {author} {\bibfnamefont {S.}~\bibnamefont {Capozziello}},\
  }\href@noop {} {\  (\bibinfo {year} {2025})},\ \Eprint
  {http://arxiv.org/abs/2501.06487} {arXiv:2501.06487 [gr-qc]} \BibitemShut
  {NoStop}%
\bibitem [{\citenamefont {Misner}\ \emph {et~al.}(1973)\citenamefont {Misner},
  \citenamefont {Thorne},\ and\ \citenamefont {Wheeler}}]{Misner1973}%
  \BibitemOpen
  \bibfield  {author} {\bibinfo {author} {\bibfnamefont {C.~W.}\ \bibnamefont
  {Misner}}, \bibinfo {author} {\bibfnamefont {K.}~\bibnamefont {Thorne}}, \
  and\ \bibinfo {author} {\bibfnamefont {J.}~\bibnamefont {Wheeler}},\
  }\href@noop {} {\emph {\bibinfo {title} {{Gravitation}}}}\ (\bibinfo
  {publisher} {W. H. Freeman},\ \bibinfo {address} {San Francisco},\ \bibinfo
  {year} {1973})\BibitemShut {NoStop}%
\bibitem [{\citenamefont {Romano}\ and\ \citenamefont
  {Furnari}(2019)}]{Romano2019}%
  \BibitemOpen
  \bibfield  {author} {\bibinfo {author} {\bibfnamefont {A.}~\bibnamefont
  {Romano}}\ and\ \bibinfo {author} {\bibfnamefont {M.}~\bibnamefont
  {Furnari}},\ }\href {https://books.google.it/books?id=y7yxDwAAQBAJ} {\emph
  {\bibinfo {title} {The Physical and Mathematical Foundations of the Theory of
  Relativity: A Critical Analysis}}}\ (\bibinfo  {publisher} {Springer
  International Publishing},\ \bibinfo {year} {2019})\BibitemShut {NoStop}%
\bibitem [{\citenamefont {Aldrovandi}\ and\ \citenamefont
  {Pereira}(2013)}]{ATG}%
  \BibitemOpen
  \bibfield  {author} {\bibinfo {author} {\bibfnamefont {R.}~\bibnamefont
  {Aldrovandi}}\ and\ \bibinfo {author} {\bibfnamefont {J.~G.}\ \bibnamefont
  {Pereira}},\ }\href {\doibase 10.1007/978-94-007-5143-9} {\emph {\bibinfo
  {title} {Teleparallel Gravity: An Introduction}}}\ (\bibinfo  {publisher}
  {Springer},\ \bibinfo {year} {2013})\BibitemShut {NoStop}%
\bibitem [{\citenamefont {{Kr{\v{s}}{\v{s}}{\'a}k}}\ \emph
  {et~al.}(2019)\citenamefont {{Kr{\v{s}}{\v{s}}{\'a}k}}, \citenamefont {{van
  den Hoogen}}, \citenamefont {{Pereira}}, \citenamefont {{B{\"o}hmer}},\ and\
  \citenamefont {{Coley}}}]{Martin2019}%
  \BibitemOpen
  \bibfield  {author} {\bibinfo {author} {\bibfnamefont {M.}~\bibnamefont
  {{Kr{\v{s}}{\v{s}}{\'a}k}}}, \bibinfo {author} {\bibfnamefont {R.~J.}\
  \bibnamefont {{van den Hoogen}}}, \bibinfo {author} {\bibfnamefont {J.~G.}\
  \bibnamefont {{Pereira}}}, \bibinfo {author} {\bibfnamefont {C.~G.}\
  \bibnamefont {{B{\"o}hmer}}}, \ and\ \bibinfo {author} {\bibfnamefont
  {A.~A.}\ \bibnamefont {{Coley}}},\ }\href {\doibase 10.1088/1361-6382/ab2e1f}
  {\bibfield  {journal} {\bibinfo  {journal} {Classical and Quantum Gravity}\
  }\textbf {\bibinfo {volume} {36}},\ \bibinfo {eid} {183001} (\bibinfo {year}
  {2019})},\ \Eprint {http://arxiv.org/abs/1810.12932} {arXiv:1810.12932
  [gr-qc]} \BibitemShut {NoStop}%
\bibitem [{\citenamefont {Hehl}(1971)}]{Hehl1971}%
  \BibitemOpen
  \bibfield  {author} {\bibinfo {author} {\bibfnamefont {F.}~\bibnamefont
  {Hehl}},\ }\href {\doibase https://doi.org/10.1016/0375-9601(71)90433-6}
  {\bibfield  {journal} {\bibinfo  {journal} {Physics Letters A}\ }\textbf
  {\bibinfo {volume} {36}},\ \bibinfo {pages} {225 } (\bibinfo {year}
  {1971})}\BibitemShut {NoStop}%
\bibitem [{\citenamefont {Hehl}(1973)}]{Hehl1973b}%
  \BibitemOpen
  \bibfield  {author} {\bibinfo {author} {\bibfnamefont {F.}~\bibnamefont
  {Hehl}},\ }\href@noop {} {\bibfield  {journal} {\bibinfo  {journal} {General
  Relativity and Gravitation}\ }\textbf {\bibinfo {volume} {4}},\ \bibinfo
  {pages} {333 } (\bibinfo {year} {1973})}\BibitemShut {NoStop}%
\bibitem [{\citenamefont {Hehl}(1974)}]{Hehl1974b}%
  \BibitemOpen
  \bibfield  {author} {\bibinfo {author} {\bibfnamefont {F.}~\bibnamefont
  {Hehl}},\ }\href@noop {} {\bibfield  {journal} {\bibinfo  {journal} {General
  Relativity and Gravitation}\ }\textbf {\bibinfo {volume} {5}},\ \bibinfo
  {pages} {491 } (\bibinfo {year} {1974})}\BibitemShut {NoStop}%
\bibitem [{\citenamefont {Hehl}\ \emph {et~al.}(1976)\citenamefont {Hehl},
  \citenamefont {von~der Heyde}, \citenamefont {Kerlick},\ and\ \citenamefont
  {Nester}}]{Hehl1976}%
  \BibitemOpen
  \bibfield  {author} {\bibinfo {author} {\bibfnamefont {F.~W.}\ \bibnamefont
  {Hehl}}, \bibinfo {author} {\bibfnamefont {P.}~\bibnamefont {von~der Heyde}},
  \bibinfo {author} {\bibfnamefont {G.~D.}\ \bibnamefont {Kerlick}}, \ and\
  \bibinfo {author} {\bibfnamefont {J.~M.}\ \bibnamefont {Nester}},\ }\href
  {\doibase 10.1103/RevModPhys.48.393} {\bibfield  {journal} {\bibinfo
  {journal} {Rev. Mod. Phys.}\ }\textbf {\bibinfo {volume} {48}},\ \bibinfo
  {pages} {393} (\bibinfo {year} {1976})}\BibitemShut {NoStop}%
\bibitem [{\citenamefont {{Wheeler}}(2018)}]{Wheeler2018}%
  \BibitemOpen
  \bibfield  {author} {\bibinfo {author} {\bibfnamefont {J.~T.}\ \bibnamefont
  {{Wheeler}}},\ }\href {\doibase 10.1007/s10714-018-2401-5} {\bibfield
  {journal} {\bibinfo  {journal} {General Relativity and Gravitation}\ }\textbf
  {\bibinfo {volume} {50}},\ \bibinfo {eid} {80} (\bibinfo {year} {2018})},\
  \Eprint {http://arxiv.org/abs/1801.03178} {arXiv:1801.03178 [gr-qc]}
  \BibitemShut {NoStop}%
\bibitem [{\citenamefont {Capozziello}\ \emph {et~al.}(2012)\citenamefont
  {Capozziello}, \citenamefont {De~Laurentis},\ and\ \citenamefont
  {Lambiase}}]{Capozziello:2012uv}%
  \BibitemOpen
  \bibfield  {author} {\bibinfo {author} {\bibfnamefont {S.}~\bibnamefont
  {Capozziello}}, \bibinfo {author} {\bibfnamefont {M.}~\bibnamefont
  {De~Laurentis}}, \ and\ \bibinfo {author} {\bibfnamefont {G.}~\bibnamefont
  {Lambiase}},\ }\href {\doibase 10.1016/j.physletb.2012.07.007} {\bibfield
  {journal} {\bibinfo  {journal} {Phys. Lett. B}\ }\textbf {\bibinfo {volume}
  {715}},\ \bibinfo {pages} {1} (\bibinfo {year} {2012})},\ \Eprint
  {http://arxiv.org/abs/1201.2071} {arXiv:1201.2071 [gr-qc]} \BibitemShut
  {NoStop}%
\bibitem [{\citenamefont {Carroll}(2004)}]{Carroll}%
  \BibitemOpen
  \bibfield  {author} {\bibinfo {author} {\bibfnamefont {S.}~\bibnamefont
  {Carroll}},\ }\href@noop {} {\emph {\bibinfo {title} {"Spacetime and
  Geometry: An Introduction to General Relativity"}}}\ (\bibinfo  {publisher}
  {A. Wesley},\ \bibinfo {year} {2004})\BibitemShut {NoStop}%
\bibitem [{\citenamefont {Tino}\ \emph {et~al.}(2020)\citenamefont {Tino},
  \citenamefont {Cacciapuoti}, \citenamefont {Capozziello}, \citenamefont
  {Lambiase},\ and\ \citenamefont {Sorrentino}}]{Tino:2020nla}%
  \BibitemOpen
  \bibfield  {author} {\bibinfo {author} {\bibfnamefont {G.~M.}\ \bibnamefont
  {Tino}}, \bibinfo {author} {\bibfnamefont {L.}~\bibnamefont {Cacciapuoti}},
  \bibinfo {author} {\bibfnamefont {S.}~\bibnamefont {Capozziello}}, \bibinfo
  {author} {\bibfnamefont {G.}~\bibnamefont {Lambiase}}, \ and\ \bibinfo
  {author} {\bibfnamefont {F.}~\bibnamefont {Sorrentino}},\ }\href {\doibase
  10.1016/j.ppnp.2020.103772} {\bibfield  {journal} {\bibinfo  {journal} {Prog.
  Part. Nucl. Phys.}\ }\textbf {\bibinfo {volume} {112}},\ \bibinfo {pages}
  {103772} (\bibinfo {year} {2020})},\ \Eprint
  {http://arxiv.org/abs/2002.02907} {arXiv:2002.02907 [gr-qc]} \BibitemShut
  {NoStop}%
\bibitem [{\citenamefont {Bahamonde}\ \emph {et~al.}(2021)\citenamefont
  {Bahamonde}, \citenamefont {Dialektopoulos}, \citenamefont
  {Escamilla-Rivera}, \citenamefont {Farrugia}, \citenamefont {Gakis},
  \citenamefont {Hendry}, \citenamefont {Hohmann}, \citenamefont {Said},
  \citenamefont {Mifsud},\ and\ \citenamefont {Valentino}}]{Teleparallel}%
  \BibitemOpen
  \bibfield  {author} {\bibinfo {author} {\bibfnamefont {S.}~\bibnamefont
  {Bahamonde}}, \bibinfo {author} {\bibfnamefont {K.~F.}\ \bibnamefont
  {Dialektopoulos}}, \bibinfo {author} {\bibfnamefont {C.}~\bibnamefont
  {Escamilla-Rivera}}, \bibinfo {author} {\bibfnamefont {G.}~\bibnamefont
  {Farrugia}}, \bibinfo {author} {\bibfnamefont {V.}~\bibnamefont {Gakis}},
  \bibinfo {author} {\bibfnamefont {M.}~\bibnamefont {Hendry}}, \bibinfo
  {author} {\bibfnamefont {M.}~\bibnamefont {Hohmann}}, \bibinfo {author}
  {\bibfnamefont {J.~L.}\ \bibnamefont {Said}}, \bibinfo {author}
  {\bibfnamefont {J.}~\bibnamefont {Mifsud}}, \ and\ \bibinfo {author}
  {\bibfnamefont {E.~D.}\ \bibnamefont {Valentino}},\ }\href {\doibase
  https://doi.org/10.1088/1361-6633/ac9cef} {\bibfield  {journal} {\bibinfo
  {journal} {Rep. Prog. Phys. 86 026901 (2023)}\ } (\bibinfo {year} {2021}),\
  https://doi.org/10.1088/1361-6633/ac9cef},\ \Eprint
  {http://arxiv.org/abs/2106.13793} {2106.13793} \BibitemShut {NoStop}%
\bibitem [{\citenamefont {Capozziello}\ \emph {et~al.}(2001)\citenamefont
  {Capozziello}, \citenamefont {Lambiase},\ and\ \citenamefont
  {Stornaiolo}}]{Capozziello:2001mq}%
  \BibitemOpen
  \bibfield  {author} {\bibinfo {author} {\bibfnamefont {S.}~\bibnamefont
  {Capozziello}}, \bibinfo {author} {\bibfnamefont {G.}~\bibnamefont
  {Lambiase}}, \ and\ \bibinfo {author} {\bibfnamefont {C.}~\bibnamefont
  {Stornaiolo}},\ }\href {\doibase
  10.1002/1521-3889(200108)10:8<713::AID-ANDP713>3.0.CO;2-2} {\bibfield
  {journal} {\bibinfo  {journal} {Annalen Phys.}\ }\textbf {\bibinfo {volume}
  {10}},\ \bibinfo {pages} {713} (\bibinfo {year} {2001})},\ \Eprint
  {http://arxiv.org/abs/gr-qc/0101038} {arXiv:gr-qc/0101038} \BibitemShut
  {NoStop}%
\bibitem [{\citenamefont {Nakahara}(2003)}]{Nakahara}%
  \BibitemOpen
  \bibfield  {author} {\bibinfo {author} {\bibfnamefont {M.}~\bibnamefont
  {Nakahara}},\ }\href@noop {} {\emph {\bibinfo {title} {Geometry, Topology and
  Physics}}}\ (\bibinfo  {publisher} {Institute of Physics Publishing},\
  \bibinfo {year} {2003})\BibitemShut {NoStop}%
\bibitem [{\citenamefont {{Pereira}}(2012)}]{Pereira2012}%
  \BibitemOpen
  \bibfield  {author} {\bibinfo {author} {\bibfnamefont {J.~G.}\ \bibnamefont
  {{Pereira}}},\ }\bibfield  {booktitle} {\emph {\bibinfo {booktitle} {The
  Sixth International School on Field Theory and Gravitation-2012}},\ }\href
  {\doibase 10.1063/1.4756972} {\bibfield  {journal} {\bibinfo  {journal} {AIP
  Conference Proceedings}\ }\bibinfo {series} {American Institute of Physics
  Conference Series},\ \textbf {\bibinfo {volume} {1483}},\ \bibinfo {pages}
  {1} (\bibinfo {year} {2012})},\ \Eprint {http://arxiv.org/abs/1210.0379}
  {arXiv:1210.0379 [gr-qc]} \BibitemShut {NoStop}%
\bibitem [{\citenamefont {Aldrovandi}\ \emph {et~al.}(2003)\citenamefont
  {Aldrovandi}, \citenamefont {Barros},\ and\ \citenamefont
  {Pereira}}]{Aldrovandi03}%
  \BibitemOpen
  \bibfield  {author} {\bibinfo {author} {\bibfnamefont {R.}~\bibnamefont
  {Aldrovandi}}, \bibinfo {author} {\bibfnamefont {P.~B.}\ \bibnamefont
  {Barros}}, \ and\ \bibinfo {author} {\bibfnamefont {J.~G.}\ \bibnamefont
  {Pereira}},\ }\href {\doibase 10.1023/A:1024060732690} {\bibfield  {journal}
  {\bibinfo  {journal} {Gen. Rel. Grav.}\ }\textbf {\bibinfo {volume} {35}},\
  \bibinfo {pages} {991} (\bibinfo {year} {2003})},\ \Eprint
  {http://arxiv.org/abs/gr-qc/0301077} {arXiv:gr-qc/0301077} \BibitemShut
  {NoStop}%
\bibitem [{\citenamefont {Capozziello}\ and\ \citenamefont
  {De~Laurentis}(2009)}]{Capozziello:2009zza}%
  \BibitemOpen
  \bibfield  {author} {\bibinfo {author} {\bibfnamefont {S.}~\bibnamefont
  {Capozziello}}\ and\ \bibinfo {author} {\bibfnamefont {M.}~\bibnamefont
  {De~Laurentis}},\ }\href {\doibase 10.1142/S0219887809003400} {\bibfield
  {journal} {\bibinfo  {journal} {Int. J. Geom. Meth. Mod. Phys.}\ }\textbf
  {\bibinfo {volume} {6}},\ \bibinfo {pages} {1} (\bibinfo {year}
  {2009})}\BibitemShut {NoStop}%
\bibitem [{\citenamefont {Esposito}(2023)}]{GE}%
  \BibitemOpen
  \bibfield  {author} {\bibinfo {author} {\bibfnamefont {G.}~\bibnamefont
  {Esposito}},\ }\href@noop {} {\emph {\bibinfo {title} {Fundamentals of
  Classical Field Theory}}}\ (\bibinfo  {publisher} {Indipendently published},\
  \bibinfo {year} {2023})\BibitemShut {NoStop}%
\bibitem [{\citenamefont {Heisenberg}(2024)}]{fQ}%
  \BibitemOpen
  \bibfield  {author} {\bibinfo {author} {\bibfnamefont {L.}~\bibnamefont
  {Heisenberg}},\ }\href {\doibase 10.1016/j.physrep.2024.02.001} {\bibfield
  {journal} {\bibinfo  {journal} {Phys. Rept.}\ }\textbf {\bibinfo {volume}
  {1066}},\ \bibinfo {pages} {1} (\bibinfo {year} {2024})},\ \Eprint
  {http://arxiv.org/abs/2309.15958} {arXiv:2309.15958 [gr-qc]} \BibitemShut
  {NoStop}%
\bibitem [{\citenamefont {Capozziello}\ \emph {et~al.}(2021)\citenamefont
  {Capozziello}, \citenamefont {Finch}, \citenamefont {Said},\ and\
  \citenamefont {Magro}}]{TSTformalism}%
  \BibitemOpen
  \bibfield  {author} {\bibinfo {author} {\bibfnamefont {S.}~\bibnamefont
  {Capozziello}}, \bibinfo {author} {\bibfnamefont {A.}~\bibnamefont {Finch}},
  \bibinfo {author} {\bibfnamefont {J.~L.}\ \bibnamefont {Said}}, \ and\
  \bibinfo {author} {\bibfnamefont {A.}~\bibnamefont {Magro}},\ }\href
  {\doibase 10.1140/epjc/s10052-021-09944-6} {\bibfield  {journal} {\bibinfo
  {journal} {The European Physical Journal C}\ } (\bibinfo {year} {2021}),\
  10.1140/epjc/s10052-021-09944-6}\BibitemShut {NoStop}%
\bibitem [{\citenamefont {{D'Ambrosio}}\ \emph {et~al.}(2022)\citenamefont
  {{D'Ambrosio}}, \citenamefont {{Fell}}, \citenamefont {{Heisenberg}},\ and\
  \citenamefont {{Kuhn}}}]{DAmbrosio2022}%
  \BibitemOpen
  \bibfield  {author} {\bibinfo {author} {\bibfnamefont {F.}~\bibnamefont
  {{D'Ambrosio}}}, \bibinfo {author} {\bibfnamefont {S.~D.~B.}\ \bibnamefont
  {{Fell}}}, \bibinfo {author} {\bibfnamefont {L.}~\bibnamefont
  {{Heisenberg}}}, \ and\ \bibinfo {author} {\bibfnamefont {S.}~\bibnamefont
  {{Kuhn}}},\ }\href {\doibase 10.1103/PhysRevD.105.024042} {\bibfield
  {journal} {\bibinfo  {journal} {Phys. Rev. D}\ }\textbf {\bibinfo {volume}
  {105}},\ \bibinfo {eid} {024042} (\bibinfo {year} {2022})},\ \Eprint
  {http://arxiv.org/abs/2109.03174} {arXiv:2109.03174 [gr-qc]} \BibitemShut
  {NoStop}%
\bibitem [{\citenamefont {Capozziello}\ \emph {et~al.}(2023)\citenamefont
  {Capozziello}, \citenamefont {Falco},\ and\ \citenamefont {Ferrara}}]{CdFF}%
  \BibitemOpen
  \bibfield  {author} {\bibinfo {author} {\bibfnamefont {S.}~\bibnamefont
  {Capozziello}}, \bibinfo {author} {\bibfnamefont {V.~D.}\ \bibnamefont
  {Falco}}, \ and\ \bibinfo {author} {\bibfnamefont {C.}~\bibnamefont
  {Ferrara}},\ }\href {\doibase
  https://doi.org/10.1140/epjc/s10052-023-12072-y} {\bibfield  {journal}
  {\bibinfo  {journal} {Eur. Phys. J. C}\ } (\bibinfo {year} {2023}),\
  https://doi.org/10.1140/epjc/s10052-023-12072-y},\ \Eprint
  {http://arxiv.org/abs/2307.13280} {2307.13280} \BibitemShut {NoStop}%
\bibitem [{\citenamefont {Capozziello}(2002)}]{Capozziello:2002rd}%
  \BibitemOpen
  \bibfield  {author} {\bibinfo {author} {\bibfnamefont {S.}~\bibnamefont
  {Capozziello}},\ }\href {\doibase 10.1142/S0218271802002025} {\bibfield
  {journal} {\bibinfo  {journal} {Int. J. Mod. Phys. D}\ }\textbf {\bibinfo
  {volume} {11}},\ \bibinfo {pages} {483} (\bibinfo {year} {2002})},\ \Eprint
  {http://arxiv.org/abs/gr-qc/0201033} {arXiv:gr-qc/0201033} \BibitemShut
  {NoStop}%
\bibitem [{\citenamefont {Ferraro}\ and\ \citenamefont
  {Fiorini}(2007)}]{Ferraro06}%
  \BibitemOpen
  \bibfield  {author} {\bibinfo {author} {\bibfnamefont {R.}~\bibnamefont
  {Ferraro}}\ and\ \bibinfo {author} {\bibfnamefont {F.}~\bibnamefont
  {Fiorini}},\ }\href {\doibase 10.1103/PhysRevD.75.084031} {\bibfield
  {journal} {\bibinfo  {journal} {Phys. Rev. D}\ }\textbf {\bibinfo {volume}
  {75}},\ \bibinfo {pages} {084031} (\bibinfo {year} {2007})},\ \Eprint
  {http://arxiv.org/abs/gr-qc/0610067} {arXiv:gr-qc/0610067} \BibitemShut
  {NoStop}%
\bibitem [{\citenamefont {Carloni}\ and\ \citenamefont
  {Luongo}(2024)}]{Carloni:2023egi}%
  \BibitemOpen
  \bibfield  {author} {\bibinfo {author} {\bibfnamefont {Y.}~\bibnamefont
  {Carloni}}\ and\ \bibinfo {author} {\bibfnamefont {O.}~\bibnamefont
  {Luongo}},\ }\href {\doibase 10.1140/epjc/s10052-024-12878-4} {\bibfield
  {journal} {\bibinfo  {journal} {Eur. Phys. J. C}\ }\textbf {\bibinfo {volume}
  {84}},\ \bibinfo {pages} {519} (\bibinfo {year} {2024})},\ \Eprint
  {http://arxiv.org/abs/2312.16088} {arXiv:2312.16088 [gr-qc]} \BibitemShut
  {NoStop}%
\bibitem [{\citenamefont {Bahamonde}\ \emph {et~al.}(2015)\citenamefont
  {Bahamonde}, \citenamefont {B\"ohmer},\ and\ \citenamefont
  {Wright}}]{Bahamonde15}%
  \BibitemOpen
  \bibfield  {author} {\bibinfo {author} {\bibfnamefont {S.}~\bibnamefont
  {Bahamonde}}, \bibinfo {author} {\bibfnamefont {C.~G.}\ \bibnamefont
  {B\"ohmer}}, \ and\ \bibinfo {author} {\bibfnamefont {M.}~\bibnamefont
  {Wright}},\ }\href {\doibase 10.1103/PhysRevD.92.104042} {\bibfield
  {journal} {\bibinfo  {journal} {Phys. Rev. D}\ }\textbf {\bibinfo {volume}
  {92}},\ \bibinfo {pages} {104042} (\bibinfo {year} {2015})},\ \Eprint
  {http://arxiv.org/abs/1508.05120} {arXiv:1508.05120 [gr-qc]} \BibitemShut
  {NoStop}%
\bibitem [{\citenamefont {Capozziello}\ \emph {et~al.}(2020)\citenamefont
  {Capozziello}, \citenamefont {Capriolo},\ and\ \citenamefont
  {Caso}}]{Capozziello2019GW}%
  \BibitemOpen
  \bibfield  {author} {\bibinfo {author} {\bibfnamefont {S.}~\bibnamefont
  {Capozziello}}, \bibinfo {author} {\bibfnamefont {M.}~\bibnamefont
  {Capriolo}}, \ and\ \bibinfo {author} {\bibfnamefont {L.}~\bibnamefont
  {Caso}},\ }\href {\doibase 10.1140/epjc/s10052-020-7737-9} {\bibfield
  {journal} {\bibinfo  {journal} {Eur. Phys. J. C}\ }\textbf {\bibinfo {volume}
  {80}},\ \bibinfo {pages} {156} (\bibinfo {year} {2020})},\ \Eprint
  {http://arxiv.org/abs/1912.12469} {arXiv:1912.12469 [gr-qc]} \BibitemShut
  {NoStop}%
\bibitem [{\citenamefont {Xu}\ \emph {et~al.}(2019)\citenamefont {Xu},
  \citenamefont {Li}, \citenamefont {Harko},\ and\ \citenamefont
  {Liang}}]{fQB}%
  \BibitemOpen
  \bibfield  {author} {\bibinfo {author} {\bibfnamefont {Y.}~\bibnamefont
  {Xu}}, \bibinfo {author} {\bibfnamefont {G.}~\bibnamefont {Li}}, \bibinfo
  {author} {\bibfnamefont {T.}~\bibnamefont {Harko}}, \ and\ \bibinfo {author}
  {\bibfnamefont {S.-D.}\ \bibnamefont {Liang}},\ }\href {\doibase
  https://doi.org/10.1140/epjc/s10052-019-7207-4} {\bibfield  {journal}
  {\bibinfo  {journal} {The European Physical Journal C}\ }\textbf {\bibinfo
  {volume} {79}} (\bibinfo {year} {2019}),\
  https://doi.org/10.1140/epjc/s10052-019-7207-4},\ \Eprint
  {http://arxiv.org/abs/1908.04760} {1908.04760} \BibitemShut {NoStop}%
\bibitem [{\citenamefont {Sotiriou}\ and\ \citenamefont {Faraoni}(2010)}]{SF}%
  \BibitemOpen
  \bibfield  {author} {\bibinfo {author} {\bibfnamefont {T.~P.}\ \bibnamefont
  {Sotiriou}}\ and\ \bibinfo {author} {\bibfnamefont {V.}~\bibnamefont
  {Faraoni}},\ }\href {\doibase https://doi.org/10.1103/RevModPhys.82.451}
  {\bibfield  {journal} {\bibinfo  {journal} {Rev. Mod. Phys.}\ }\textbf
  {\bibinfo {volume} {82}} (\bibinfo {year} {2010}),\
  https://doi.org/10.1103/RevModPhys.82.451},\ \Eprint
  {http://arxiv.org/abs/0805.1726} {0805.1726} \BibitemShut {NoStop}%
\bibitem [{\citenamefont {{De Felice}}\ and\ \citenamefont
  {{Tsujikawa}}(2010)}]{DeFelice10}%
  \BibitemOpen
  \bibfield  {author} {\bibinfo {author} {\bibfnamefont {A.}~\bibnamefont {{De
  Felice}}}\ and\ \bibinfo {author} {\bibfnamefont {S.}~\bibnamefont
  {{Tsujikawa}}},\ }\href {\doibase 10.12942/lrr-2010-3} {\bibfield  {journal}
  {\bibinfo  {journal} {Living Reviews in Relativity}\ }\textbf {\bibinfo
  {volume} {13}},\ \bibinfo {eid} {3} (\bibinfo {year} {2010})},\ \Eprint
  {http://arxiv.org/abs/1002.4928} {arXiv:1002.4928 [gr-qc]} \BibitemShut
  {NoStop}%
\bibitem [{\citenamefont {Allemandi}\ \emph {et~al.}(2006)\citenamefont
  {Allemandi}, \citenamefont {Capone}, \citenamefont {Capozziello},\ and\
  \citenamefont {Francaviglia}}]{Allemandi:2004yx}%
  \BibitemOpen
  \bibfield  {author} {\bibinfo {author} {\bibfnamefont {G.}~\bibnamefont
  {Allemandi}}, \bibinfo {author} {\bibfnamefont {M.}~\bibnamefont {Capone}},
  \bibinfo {author} {\bibfnamefont {S.}~\bibnamefont {Capozziello}}, \ and\
  \bibinfo {author} {\bibfnamefont {M.}~\bibnamefont {Francaviglia}},\ }\href
  {\doibase 10.1007/s10714-005-0208-7} {\bibfield  {journal} {\bibinfo
  {journal} {Gen. Rel. Grav.}\ }\textbf {\bibinfo {volume} {38}},\ \bibinfo
  {pages} {33} (\bibinfo {year} {2006})},\ \Eprint
  {http://arxiv.org/abs/hep-th/0409198} {arXiv:hep-th/0409198} \BibitemShut
  {NoStop}%
\bibitem [{\citenamefont {Starobinsky}(1980)}]{STAROBINSKY80}%
  \BibitemOpen
  \bibfield  {author} {\bibinfo {author} {\bibfnamefont {A.}~\bibnamefont
  {Starobinsky}},\ }\href {\doibase
  https://doi.org/10.1016/0370-2693(80)90670-X} {\bibfield  {journal} {\bibinfo
   {journal} {Physics Letters B}\ }\textbf {\bibinfo {volume} {91}},\ \bibinfo
  {pages} {99} (\bibinfo {year} {1980})}\BibitemShut {NoStop}%
\bibitem [{\citenamefont {Maggiore}(2007)}]{Maggiore07}%
  \BibitemOpen
  \bibfield  {author} {\bibinfo {author} {\bibfnamefont {M.}~\bibnamefont
  {Maggiore}},\ }\href {\doibase 10.1093/acprof:oso/9780198570745.001.0001}
  {\emph {\bibinfo {title} {{Gravitational Waves. Vol. 1: Theory and
  Experiments}}}}\ (\bibinfo  {publisher} {Oxford University Press},\ \bibinfo
  {year} {2007})\BibitemShut {NoStop}%
\bibitem [{\citenamefont {Bogdanos}\ \emph {et~al.}(2010)\citenamefont
  {Bogdanos}, \citenamefont {Capozziello}, \citenamefont {De~Laurentis},\ and\
  \citenamefont {Nesseris}}]{Bogdanos:2009tn}%
  \BibitemOpen
  \bibfield  {author} {\bibinfo {author} {\bibfnamefont {C.}~\bibnamefont
  {Bogdanos}}, \bibinfo {author} {\bibfnamefont {S.}~\bibnamefont
  {Capozziello}}, \bibinfo {author} {\bibfnamefont {M.}~\bibnamefont
  {De~Laurentis}}, \ and\ \bibinfo {author} {\bibfnamefont {S.}~\bibnamefont
  {Nesseris}},\ }\href {\doibase 10.1016/j.astropartphys.2010.08.001}
  {\bibfield  {journal} {\bibinfo  {journal} {Astropart. Phys.}\ }\textbf
  {\bibinfo {volume} {34}},\ \bibinfo {pages} {236} (\bibinfo {year} {2010})},\
  \Eprint {http://arxiv.org/abs/0911.3094} {arXiv:0911.3094 [gr-qc]}
  \BibitemShut {NoStop}%
\bibitem [{\citenamefont {Capozziello}\ \emph {et~al.}(2008)\citenamefont
  {Capozziello}, \citenamefont {Corda},\ and\ \citenamefont {{De
  Laurentis}}}]{DeLaurentis08}%
  \BibitemOpen
  \bibfield  {author} {\bibinfo {author} {\bibfnamefont {S.}~\bibnamefont
  {Capozziello}}, \bibinfo {author} {\bibfnamefont {C.}~\bibnamefont {Corda}},
  \ and\ \bibinfo {author} {\bibfnamefont {M.~F.}\ \bibnamefont {{De
  Laurentis}}},\ }\href {\doibase
  https://doi.org/10.1016/j.physletb.2008.10.001} {\bibfield  {journal}
  {\bibinfo  {journal} {Physics Letters B}\ }\textbf {\bibinfo {volume}
  {669}},\ \bibinfo {pages} {255} (\bibinfo {year} {2008})}\BibitemShut
  {NoStop}%
\bibitem [{\citenamefont {Capozziello}\ and\ \citenamefont
  {Vignolo}(2010)}]{Vignolo09}%
  \BibitemOpen
  \bibfield  {author} {\bibinfo {author} {\bibfnamefont {S.}~\bibnamefont
  {Capozziello}}\ and\ \bibinfo {author} {\bibfnamefont {S.}~\bibnamefont
  {Vignolo}},\ }\href {\doibase 10.1002/andp.201010420} {\bibfield  {journal}
  {\bibinfo  {journal} {Annalen Phys.}\ }\textbf {\bibinfo {volume} {19}},\
  \bibinfo {pages} {238} (\bibinfo {year} {2010})},\ \Eprint
  {http://arxiv.org/abs/0910.5230} {arXiv:0910.5230 [gr-qc]} \BibitemShut
  {NoStop}%
\bibitem [{\citenamefont {Capozziello}\ \emph {et~al.}(2010)\citenamefont
  {Capozziello}, \citenamefont {Cianci}, \citenamefont {De~Laurentis},\ and\
  \citenamefont {Vignolo}}]{Vignolo10}%
  \BibitemOpen
  \bibfield  {author} {\bibinfo {author} {\bibfnamefont {S.}~\bibnamefont
  {Capozziello}}, \bibinfo {author} {\bibfnamefont {R.}~\bibnamefont {Cianci}},
  \bibinfo {author} {\bibfnamefont {M.}~\bibnamefont {De~Laurentis}}, \ and\
  \bibinfo {author} {\bibfnamefont {S.}~\bibnamefont {Vignolo}},\ }\href
  {\doibase 10.1140/epjc/s10052-010-1412-5} {\bibfield  {journal} {\bibinfo
  {journal} {Eur. Phys. J. C}\ }\textbf {\bibinfo {volume} {70}},\ \bibinfo
  {pages} {341} (\bibinfo {year} {2010})},\ \Eprint
  {http://arxiv.org/abs/1007.3670} {arXiv:1007.3670 [gr-qc]} \BibitemShut
  {NoStop}%
\bibitem [{\citenamefont {Corda}(2008)}]{Corda07}%
  \BibitemOpen
  \bibfield  {author} {\bibinfo {author} {\bibfnamefont {C.}~\bibnamefont
  {Corda}},\ }\href {\doibase 10.1142/S0217751X08038603} {\bibfield  {journal}
  {\bibinfo  {journal} {Int. J. Mod. Phys. A}\ }\textbf {\bibinfo {volume}
  {23}},\ \bibinfo {pages} {1521} (\bibinfo {year} {2008})},\ \Eprint
  {http://arxiv.org/abs/0711.4917} {arXiv:0711.4917 [gr-qc]} \BibitemShut
  {NoStop}%
\bibitem [{\citenamefont {Corda}(2007)}]{Corda2007}%
  \BibitemOpen
  \bibfield  {author} {\bibinfo {author} {\bibfnamefont {C.}~\bibnamefont
  {Corda}},\ }\href {\doibase 10.1088/1475-7516/2007/04/009} {\bibfield
  {journal} {\bibinfo  {journal} {Journal of Cosmology and Astroparticle
  Physics}\ }\textbf {\bibinfo {volume} {2007}},\ \bibinfo {pages} {009}
  (\bibinfo {year} {2007})}\BibitemShut {NoStop}%
\bibitem [{\citenamefont {Katsuragawa}\ \emph {et~al.}(2019)\citenamefont
  {Katsuragawa}, \citenamefont {Nakamura}, \citenamefont {Ikeda},\ and\
  \citenamefont {Capozziello}}]{Taishi19}%
  \BibitemOpen
  \bibfield  {author} {\bibinfo {author} {\bibfnamefont {T.}~\bibnamefont
  {Katsuragawa}}, \bibinfo {author} {\bibfnamefont {T.}~\bibnamefont
  {Nakamura}}, \bibinfo {author} {\bibfnamefont {T.}~\bibnamefont {Ikeda}}, \
  and\ \bibinfo {author} {\bibfnamefont {S.}~\bibnamefont {Capozziello}},\
  }\href {\doibase 10.1103/PhysRevD.99.124050} {\bibfield  {journal} {\bibinfo
  {journal} {Phys. Rev. D}\ }\textbf {\bibinfo {volume} {99}},\ \bibinfo
  {pages} {124050} (\bibinfo {year} {2019})}\BibitemShut {NoStop}%
\bibitem [{\citenamefont {{Yang}}\ \emph {et~al.}(2011)\citenamefont {{Yang}},
  \citenamefont {{Lee}},\ and\ \citenamefont {{Geng}}}]{Yang11}%
  \BibitemOpen
  \bibfield  {author} {\bibinfo {author} {\bibfnamefont {L.}~\bibnamefont
  {{Yang}}}, \bibinfo {author} {\bibfnamefont {C.-C.}\ \bibnamefont {{Lee}}}, \
  and\ \bibinfo {author} {\bibfnamefont {C.-Q.}\ \bibnamefont {{Geng}}},\
  }\href {\doibase 10.1088/1475-7516/2011/08/029} {\bibfield  {journal}
  {\bibinfo  {journal} {JCAP}\ }\textbf {\bibinfo {volume} {2011}},\ \bibinfo
  {eid} {029} (\bibinfo {year} {2011})},\ \Eprint
  {http://arxiv.org/abs/1106.5582} {arXiv:1106.5582 [astro-ph.CO]} \BibitemShut
  {NoStop}%
\bibitem [{\citenamefont {N\"af}\ \emph {et~al.}(2009)\citenamefont {N\"af},
  \citenamefont {Jetzer},\ and\ \citenamefont {Sereno}}]{Jetzer09}%
  \BibitemOpen
  \bibfield  {author} {\bibinfo {author} {\bibfnamefont {J.}~\bibnamefont
  {N\"af}}, \bibinfo {author} {\bibfnamefont {P.}~\bibnamefont {Jetzer}}, \
  and\ \bibinfo {author} {\bibfnamefont {M.}~\bibnamefont {Sereno}},\ }\href
  {\doibase 10.1103/PhysRevD.79.024014} {\bibfield  {journal} {\bibinfo
  {journal} {Phys. Rev. D}\ }\textbf {\bibinfo {volume} {79}},\ \bibinfo
  {pages} {024014} (\bibinfo {year} {2009})}\BibitemShut {NoStop}%
\bibitem [{\citenamefont {Li}\ \emph {et~al.}(2011)\citenamefont {Li},
  \citenamefont {Sotiriou},\ and\ \citenamefont {Barrow}}]{Barrow11}%
  \BibitemOpen
  \bibfield  {author} {\bibinfo {author} {\bibfnamefont {B.}~\bibnamefont
  {Li}}, \bibinfo {author} {\bibfnamefont {T.~P.}\ \bibnamefont {Sotiriou}}, \
  and\ \bibinfo {author} {\bibfnamefont {J.~D.}\ \bibnamefont {Barrow}},\
  }\href {\doibase 10.1103/PhysRevD.83.064035} {\bibfield  {journal} {\bibinfo
  {journal} {Phys. Rev. D}\ }\textbf {\bibinfo {volume} {83}},\ \bibinfo
  {pages} {064035} (\bibinfo {year} {2011})}\BibitemShut {NoStop}%
\bibitem [{\citenamefont {Krššák}\ and\ \citenamefont
  {Saridakis}(2016)}]{Krššák_2016}%
  \BibitemOpen
  \bibfield  {author} {\bibinfo {author} {\bibfnamefont {M.}~\bibnamefont
  {Krššák}}\ and\ \bibinfo {author} {\bibfnamefont {E.~N.}\ \bibnamefont
  {Saridakis}},\ }\href {\doibase 10.1088/0264-9381/33/11/115009} {\bibfield
  {journal} {\bibinfo  {journal} {Classical and Quantum Gravity}\ }\textbf
  {\bibinfo {volume} {33}},\ \bibinfo {pages} {115009} (\bibinfo {year}
  {2016})}\BibitemShut {NoStop}%
\bibitem [{\citenamefont {Bamba}\ \emph {et~al.}(2013)\citenamefont {Bamba},
  \citenamefont {Capozziello}, \citenamefont {{De Laurentis}}, \citenamefont
  {Nojiri},\ and\ \citenamefont {Sáez-Gómez}}]{BAMBA13}%
  \BibitemOpen
  \bibfield  {author} {\bibinfo {author} {\bibfnamefont {K.}~\bibnamefont
  {Bamba}}, \bibinfo {author} {\bibfnamefont {S.}~\bibnamefont {Capozziello}},
  \bibinfo {author} {\bibfnamefont {M.}~\bibnamefont {{De Laurentis}}},
  \bibinfo {author} {\bibfnamefont {S.}~\bibnamefont {Nojiri}}, \ and\ \bibinfo
  {author} {\bibfnamefont {D.}~\bibnamefont {Sáez-Gómez}},\ }\href {\doibase
  https://doi.org/10.1016/j.physletb.2013.10.022} {\bibfield  {journal}
  {\bibinfo  {journal} {Physics Letters B}\ }\textbf {\bibinfo {volume}
  {727}},\ \bibinfo {pages} {194} (\bibinfo {year} {2013})}\BibitemShut
  {NoStop}%
\bibitem [{\citenamefont {Bahamonde}\ and\ \citenamefont
  {Capozziello}(2017)}]{bahamonde2017}%
  \BibitemOpen
  \bibfield  {author} {\bibinfo {author} {\bibfnamefont {S.}~\bibnamefont
  {Bahamonde}}\ and\ \bibinfo {author} {\bibfnamefont {S.}~\bibnamefont
  {Capozziello}},\ }\href@noop {} {\bibfield  {journal} {\bibinfo  {journal}
  {The European Physical Journal C}\ }\textbf {\bibinfo {volume} {77}},\
  \bibinfo {pages} {1} (\bibinfo {year} {2017})}\BibitemShut {NoStop}%
\bibitem [{\citenamefont {Capozziello}\ \emph
  {et~al.}(2024{\natexlab{a}})\citenamefont {Capozziello}, \citenamefont
  {Capriolo},\ and\ \citenamefont {Lambiase}}]{Capriolo2024}%
  \BibitemOpen
  \bibfield  {author} {\bibinfo {author} {\bibfnamefont {S.}~\bibnamefont
  {Capozziello}}, \bibinfo {author} {\bibfnamefont {M.}~\bibnamefont
  {Capriolo}}, \ and\ \bibinfo {author} {\bibfnamefont {G.}~\bibnamefont
  {Lambiase}},\ }\href {\doibase 10.1103/PhysRevD.110.104028} {\bibfield
  {journal} {\bibinfo  {journal} {Phys. Rev. D}\ }\textbf {\bibinfo {volume}
  {110}},\ \bibinfo {pages} {104028} (\bibinfo {year} {2024}{\natexlab{a}})},\
  \Eprint {http://arxiv.org/abs/2407.14862} {arXiv:2407.14862 [gr-qc]}
  \BibitemShut {NoStop}%
\bibitem [{\citenamefont {Zhao}(2022)}]{Zhao21}%
  \BibitemOpen
  \bibfield  {author} {\bibinfo {author} {\bibfnamefont {D.}~\bibnamefont
  {Zhao}},\ }\href {\doibase 10.1140/epjc/s10052-022-10266-4} {\bibfield
  {journal} {\bibinfo  {journal} {Eur. Phys. J. C}\ }\textbf {\bibinfo {volume}
  {82}},\ \bibinfo {pages} {303} (\bibinfo {year} {2022})},\ \Eprint
  {http://arxiv.org/abs/2104.02483} {arXiv:2104.02483 [gr-qc]} \BibitemShut
  {NoStop}%
\bibitem [{\citenamefont {Capozziello}\ \emph
  {et~al.}(2024{\natexlab{b}})\citenamefont {Capozziello}, \citenamefont
  {Capriolo},\ and\ \citenamefont {Nojiri}}]{CAPOZZIELLO2024}%
  \BibitemOpen
  \bibfield  {author} {\bibinfo {author} {\bibfnamefont {S.}~\bibnamefont
  {Capozziello}}, \bibinfo {author} {\bibfnamefont {M.}~\bibnamefont
  {Capriolo}}, \ and\ \bibinfo {author} {\bibfnamefont {S.}~\bibnamefont
  {Nojiri}},\ }\href {\doibase https://doi.org/10.1016/j.physletb.2024.138510}
  {\bibfield  {journal} {\bibinfo  {journal} {Physics Letters B}\ }\textbf
  {\bibinfo {volume} {850}},\ \bibinfo {pages} {138510} (\bibinfo {year}
  {2024}{\natexlab{b}})}\BibitemShut {NoStop}%
\bibitem [{\citenamefont {Capozziello}\ and\ \citenamefont
  {Ferrara}(2024)}]{Ferrara24}%
  \BibitemOpen
  \bibfield  {author} {\bibinfo {author} {\bibfnamefont {S.}~\bibnamefont
  {Capozziello}}\ and\ \bibinfo {author} {\bibfnamefont {C.}~\bibnamefont
  {Ferrara}},\ }\href@noop {} {\  (\bibinfo {year} {2024})},\ \Eprint
  {http://arxiv.org/abs/2401.09737} {arXiv:2401.09737 [gr-qc]} \BibitemShut
  {NoStop}%
\bibitem [{\citenamefont {Capozziello}\ \emph {et~al.}(2014)\citenamefont
  {Capozziello}, \citenamefont {Lobo},\ and\ \citenamefont
  {Mimoso}}]{Capozziello:2013}%
  \BibitemOpen
  \bibfield  {author} {\bibinfo {author} {\bibfnamefont {S.}~\bibnamefont
  {Capozziello}}, \bibinfo {author} {\bibfnamefont {F.~S.~N.}\ \bibnamefont
  {Lobo}}, \ and\ \bibinfo {author} {\bibfnamefont {J.~P.}\ \bibnamefont
  {Mimoso}},\ }\href {\doibase 10.1016/j.physletb.2014.01.066} {\bibfield
  {journal} {\bibinfo  {journal} {Phys. Lett. B}\ }\textbf {\bibinfo {volume}
  {730}},\ \bibinfo {pages} {280} (\bibinfo {year} {2014})},\ \Eprint
  {http://arxiv.org/abs/1312.0784} {arXiv:1312.0784 [gr-qc]} \BibitemShut
  {NoStop}%
\bibitem [{\citenamefont {Capozziello}\ \emph {et~al.}(2015)\citenamefont
  {Capozziello}, \citenamefont {Lobo},\ and\ \citenamefont
  {Mimoso}}]{Capozziello:2014}%
  \BibitemOpen
  \bibfield  {author} {\bibinfo {author} {\bibfnamefont {S.}~\bibnamefont
  {Capozziello}}, \bibinfo {author} {\bibfnamefont {F.~S.~N.}\ \bibnamefont
  {Lobo}}, \ and\ \bibinfo {author} {\bibfnamefont {J.~P.}\ \bibnamefont
  {Mimoso}},\ }\href {\doibase 10.1103/PhysRevD.91.124019} {\bibfield
  {journal} {\bibinfo  {journal} {Phys. Rev. D}\ }\textbf {\bibinfo {volume}
  {91}},\ \bibinfo {pages} {124019} (\bibinfo {year} {2015})},\ \Eprint
  {http://arxiv.org/abs/1407.7293} {arXiv:1407.7293 [gr-qc]} \BibitemShut
  {NoStop}%
\bibitem [{\citenamefont {Harko}\ and\ \citenamefont
  {Lobo}(2014)}]{Harko:2014}%
  \BibitemOpen
  \bibfield  {author} {\bibinfo {author} {\bibfnamefont {T.}~\bibnamefont
  {Harko}}\ and\ \bibinfo {author} {\bibfnamefont {F.~S.~N.}\ \bibnamefont
  {Lobo}},\ }\href {\doibase 10.3390/galaxies2030410} {\bibfield  {journal}
  {\bibinfo  {journal} {Galaxies}\ }\textbf {\bibinfo {volume} {2}},\ \bibinfo
  {pages} {410} (\bibinfo {year} {2014})},\ \Eprint
  {http://arxiv.org/abs/1407.2013} {arXiv:1407.2013 [gr-qc]} \BibitemShut
  {NoStop}%
\bibitem [{\citenamefont {Bertolami}\ \emph {et~al.}(2007)\citenamefont
  {Bertolami}, \citenamefont {Boehmer}, \citenamefont {Harko},\ and\
  \citenamefont {Lobo}}]{Bertolami:2007}%
  \BibitemOpen
  \bibfield  {author} {\bibinfo {author} {\bibfnamefont {O.}~\bibnamefont
  {Bertolami}}, \bibinfo {author} {\bibfnamefont {C.~G.}\ \bibnamefont
  {Boehmer}}, \bibinfo {author} {\bibfnamefont {T.}~\bibnamefont {Harko}}, \
  and\ \bibinfo {author} {\bibfnamefont {F.~S.~N.}\ \bibnamefont {Lobo}},\
  }\href {\doibase 10.1103/PhysRevD.75.104016} {\bibfield  {journal} {\bibinfo
  {journal} {Phys. Rev. D}\ }\textbf {\bibinfo {volume} {75}},\ \bibinfo
  {pages} {104016} (\bibinfo {year} {2007})},\ \Eprint
  {http://arxiv.org/abs/0704.1733} {arXiv:0704.1733 [gr-qc]} \BibitemShut
  {NoStop}%
\bibitem [{\citenamefont {Capozziello}\ \emph
  {et~al.}(2024{\natexlab{c}})\citenamefont {Capozziello}, \citenamefont
  {Caruana}, \citenamefont {Farrugia}, \citenamefont {Levi~Said},\ and\
  \citenamefont {Sultana}}]{Caruana23}%
  \BibitemOpen
  \bibfield  {author} {\bibinfo {author} {\bibfnamefont {S.}~\bibnamefont
  {Capozziello}}, \bibinfo {author} {\bibfnamefont {M.}~\bibnamefont
  {Caruana}}, \bibinfo {author} {\bibfnamefont {G.}~\bibnamefont {Farrugia}},
  \bibinfo {author} {\bibfnamefont {J.}~\bibnamefont {Levi~Said}}, \ and\
  \bibinfo {author} {\bibfnamefont {J.}~\bibnamefont {Sultana}},\ }\href
  {\doibase 10.1007/s10714-024-03204-0} {\bibfield  {journal} {\bibinfo
  {journal} {Gen. Rel. Grav.}\ }\textbf {\bibinfo {volume} {56}},\ \bibinfo
  {pages} {27} (\bibinfo {year} {2024}{\natexlab{c}})},\ \Eprint
  {http://arxiv.org/abs/2308.15995} {arXiv:2308.15995 [gr-qc]} \BibitemShut
  {NoStop}%
\bibitem [{\citenamefont {Colg\'ain}\ \emph {et~al.}(2024)\citenamefont
  {Colg\'ain}, \citenamefont {Dainotti}, \citenamefont {Capozziello},
  \citenamefont {Pourojaghi}, \citenamefont {Sheikh-Jabbari},\ and\
  \citenamefont {Stojkovic}}]{Colgain:2024xqj}%
  \BibitemOpen
  \bibfield  {author} {\bibinfo {author} {\bibfnamefont {E.~O.}\ \bibnamefont
  {Colg\'ain}}, \bibinfo {author} {\bibfnamefont {M.~G.}\ \bibnamefont
  {Dainotti}}, \bibinfo {author} {\bibfnamefont {S.}~\bibnamefont
  {Capozziello}}, \bibinfo {author} {\bibfnamefont {S.}~\bibnamefont
  {Pourojaghi}}, \bibinfo {author} {\bibfnamefont {M.~M.}\ \bibnamefont
  {Sheikh-Jabbari}}, \ and\ \bibinfo {author} {\bibfnamefont {D.}~\bibnamefont
  {Stojkovic}},\ }\href@noop {} {\  (\bibinfo {year} {2024})},\ \Eprint
  {http://arxiv.org/abs/2404.08633} {arXiv:2404.08633 [astro-ph.CO]}
  \BibitemShut {NoStop}%
\bibitem [{\citenamefont {Odintsov}\ \emph {et~al.}(2024)\citenamefont
  {Odintsov}, \citenamefont {S\'aez-Chill\'on~G\'omez},\ and\ \citenamefont
  {Sharov}}]{Odintsov:2024woi}%
  \BibitemOpen
  \bibfield  {author} {\bibinfo {author} {\bibfnamefont {S.~D.}\ \bibnamefont
  {Odintsov}}, \bibinfo {author} {\bibfnamefont {D.}~\bibnamefont
  {S\'aez-Chill\'on~G\'omez}}, \ and\ \bibinfo {author} {\bibfnamefont {G.~S.}\
  \bibnamefont {Sharov}},\ }\href@noop {} {\  (\bibinfo {year} {2024})},\
  \Eprint {http://arxiv.org/abs/2412.09409} {arXiv:2412.09409 [gr-qc]}
  \BibitemShut {NoStop}%
\bibitem [{\citenamefont {Hashimoto}\ and\ \citenamefont
  {Tanahashi}(2017)}]{Hashimoto2016}%
  \BibitemOpen
  \bibfield  {author} {\bibinfo {author} {\bibfnamefont {K.}~\bibnamefont
  {Hashimoto}}\ and\ \bibinfo {author} {\bibfnamefont {N.}~\bibnamefont
  {Tanahashi}},\ }\href {\doibase 10.1103/PhysRevD.95.024007} {\bibfield
  {journal} {\bibinfo  {journal} {Phys. Rev. D}\ }\textbf {\bibinfo {volume}
  {95}},\ \bibinfo {pages} {024007} (\bibinfo {year} {2017})},\ \Eprint
  {http://arxiv.org/abs/1610.06070} {arXiv:1610.06070 [hep-th]} \BibitemShut
  {NoStop}%
\bibitem [{\citenamefont {Capozziello}\ \emph
  {et~al.}(2024{\natexlab{d}})\citenamefont {Capozziello}, \citenamefont
  {Lapponi}, \citenamefont {Luongo},\ and\ \citenamefont
  {Mancini}}]{Capozziello:2024mxh}%
  \BibitemOpen
  \bibfield  {author} {\bibinfo {author} {\bibfnamefont {S.}~\bibnamefont
  {Capozziello}}, \bibinfo {author} {\bibfnamefont {A.}~\bibnamefont
  {Lapponi}}, \bibinfo {author} {\bibfnamefont {O.}~\bibnamefont {Luongo}}, \
  and\ \bibinfo {author} {\bibfnamefont {S.}~\bibnamefont {Mancini}},\ }\href
  {\doibase 10.1140/epjc/s10052-024-13449-3} {\bibfield  {journal} {\bibinfo
  {journal} {Eur. Phys. J. C}\ }\textbf {\bibinfo {volume} {84}},\ \bibinfo
  {pages} {1081} (\bibinfo {year} {2024}{\natexlab{d}})},\ \Eprint
  {http://arxiv.org/abs/2406.19274} {arXiv:2406.19274 [gr-qc]} \BibitemShut
  {NoStop}%
\bibitem [{\citenamefont {Maldacena}\ \emph {et~al.}(2016)\citenamefont
  {Maldacena}, \citenamefont {Shenker},\ and\ \citenamefont {Stanford}}]{MSS}%
  \BibitemOpen
  \bibfield  {author} {\bibinfo {author} {\bibfnamefont {J.}~\bibnamefont
  {Maldacena}}, \bibinfo {author} {\bibfnamefont {S.~H.}\ \bibnamefont
  {Shenker}}, \ and\ \bibinfo {author} {\bibfnamefont {D.}~\bibnamefont
  {Stanford}},\ }\href {https://arxiv.org/pdf/1503.01409.pdf} {\bibfield
  {journal} {\bibinfo  {journal} {Journal of High Energy Physics}\ } (\bibinfo
  {year} {2016})},\ \Eprint {http://arxiv.org/abs/1503.01409} {1503.01409}
  \BibitemShut {NoStop}%
\bibitem [{\citenamefont {Addazi}\ \emph {et~al.}(2021)\citenamefont {Addazi},
  \citenamefont {Capozziello},\ and\ \citenamefont {Odintsov}}]{AC}%
  \BibitemOpen
  \bibfield  {author} {\bibinfo {author} {\bibfnamefont {A.}~\bibnamefont
  {Addazi}}, \bibinfo {author} {\bibfnamefont {S.}~\bibnamefont {Capozziello}},
  \ and\ \bibinfo {author} {\bibfnamefont {S.~D.}\ \bibnamefont {Odintsov}},\
  }\href {https://arxiv.org/pdf/2103.16856.pdf} {\bibfield  {journal} {\bibinfo
   {journal} {Physics Letters B}\ } (\bibinfo {year} {2021})},\ \Eprint
  {http://arxiv.org/abs/2103.16856} {2103.16856} \BibitemShut {NoStop}%
\bibitem [{\citenamefont {Addazi}\ and\ \citenamefont
  {Capozziello}(2023)}]{AC1}%
  \BibitemOpen
  \bibfield  {author} {\bibinfo {author} {\bibfnamefont {A.}~\bibnamefont
  {Addazi}}\ and\ \bibinfo {author} {\bibfnamefont {S.}~\bibnamefont
  {Capozziello}},\ }\href {\doibase
  https://doi.org/10.1016/j.physletb.2023.137828} {\bibfield  {journal}
  {\bibinfo  {journal} {Physics Letters B}\ } (\bibinfo {year} {2023}),\
  https://doi.org/10.1016/j.physletb.2023.137828},\ \Eprint
  {http://arxiv.org/abs/2303.01956} {2303.01956} \BibitemShut {NoStop}%
\end{thebibliography}%

\end{document}